\documentclass[a4paper,11pt]{article}
\usepackage{graphicx,amssymb,amstext,amsmath,xcolor,color}
\usepackage{color,yfonts}
\usepackage{floatflt}

\usepackage{array}
\usepackage{boldline}
\usepackage{multirow}
\usepackage{arydshln}
\usepackage{bigdelim}
\usepackage{hyperref}

\numberwithin{equation}{section}

\definecolor{cbl}{rgb}{0,0,1}                

\newcommand{\bc}{\begin{center}}
\newcommand{\ec}{\end{center}}
\def\ba#1{\begin{array}{#1}\displaystyle}
\newcommand{\ea}{\end{array}}

\newcommand{\beq}{\begin{equation}}
\newcommand{\eeq}{\end{equation}}
\newcommand{\beqa}{\begin{eqnarray}}
\newcommand{\eeqa}{\end{eqnarray}}
\newcommand{\no}{\nonumber}
\newcommand{\n}{\nonumber\\}
\newcommand{\bi}{\begin{itemize}}
\newcommand{\ei}{\end{itemize}}

 \newcommand{\Aa}{\textfrak{a}} 
 \newcommand{\Bb}{\textfrak{b}}

\def\lt#1{\left#1}
\def\rt#1{\right#1}
\def\t#1{\tilde{#1}}

\def\b#1{\bar{#1}}
\def\frc#1#2{\frac{#1}{#2}}

\newcommand{\p}{\partial}
\newcommand{\prin}{\underline{\mathrm{P}}}

\newcommand{\vac}{{\rm vac}}
\newcommand{\bra}{\langle}
\newcommand{\ket}{\rangle}
\newcommand{\Z}{{\mathbb{Z}}}
\newcommand{\N}{{\mathbb{N}}}
\newcommand{\R}{{\mathbb{R}}}
\newcommand{\C}{{\mathbb{C}}}

\newcommand{\dd}{\mathrm{d}}

\newcommand{\Or}{{\cal O}}

\newcommand{\Tr}{{\rm Tr}}

\newcommand{\TT}{{\cal T}}

\def\ta{\theta}

\newcommand{\ri}{{\rm i}}

\begin{document}
\begin{titlepage}
\vspace{0.2cm}
\begin{center}

{\large{\bf{Entanglement Content of Quantum Particle Excitations I. Free Field Theory}}}

\vspace{0.8cm} 
{\large \text{Olalla A. Castro-Alvaredo${}^{\heartsuit}$, Cecilia De Fazio{\LARGE ${}^{\bullet}$}, Benjamin Doyon{\LARGE${}^{\star}$} and Istv\'an M. Sz\'ecs\'enyi{$\,{}^{\spadesuit}$}}}

\vspace{0.8cm}
{\small
{\small ${}^{\heartsuit\,\spadesuit}$} Department of Mathematics, City, University of London, 10 Northampton Square EC1V 0HB, UK\\
\vspace{0.2cm}
{\LARGE ${}^{\bullet}$} Dipartimento di Fisica e Astronomia, Universit\`a di Bologna, Via Irnerio 46, I-40126 Bologna, Italy\\
\vspace{0.2cm}
{\LARGE${}^{\star}$}Department of Mathematics, King's College London, Strand WC2R 2LS, UK}\\

\end{center}

\vspace{1cm}

We evaluate the entanglement entropy of a single connected region in excited states of one-dimensional massive free theories with finite numbers of particles, in the limit of large volume and region length. For this purpose, we use finite-volume form factor expansions of branch-point twist field two-point functions. We find that the additive contribution to the entanglement due to the presence of particles has a simple ``qubit" interpretation, and is largely independent of momenta: it only depends on the numbers of groups of particles with equal momenta. We conjecture that at large momenta, the same result holds for any volume and region lengths, including at small scales. We  provide accurate numerical verifications.

\medskip

\vfill

\noindent 
${}^{\heartsuit}$ o.castro-alvaredo@city.ac.uk\\
{\LARGE ${}^{\bullet}$} cecilia.defazio@studio.unibo.it\\
{\LARGE${}^{\star}$} benjamin.doyon@kcl.ac.uk\\
{${}^{\spadesuit}$} istvan.szecsenyi@city.ac.uk\\

\hfill \today

\end{titlepage}


\section{Introduction}
Measures of entanglement, such as the entanglement entropy, have attracted much attention in recent years, particularly in the context of one-dimensional many body quantum systems (see e.g.~review articles in \cite{EERev1,specialissue, EERev2}).  Among such systems, those enjoying conformal invariance in the scaling limit  are of particular interest as they provide a theoretical and universal description of critical phenomena.  In their seminal work Calabrese and Cardy \cite{Calabrese:2004eu} used principles of conformal field theory (CFT) to study the entanglement entropy (EE) \cite{bennet} of quantum critical systems. Their results generalised previous work \cite{HolzheyLW94}, provided theoretical support for numerical observations in critical quantum spin chains \cite{latorre1} and highlighted the fact that the EE encodes universal information about quantum critical points, such as the central change of the corresponding CFT and, in more complex setups, about the full primary operator content of CFT \cite{disco1,negativity1,negativity2}.

{\begin{floatingfigure}[h]{7.5cm} 
 \begin{center} 
 \includegraphics[width=7cm]{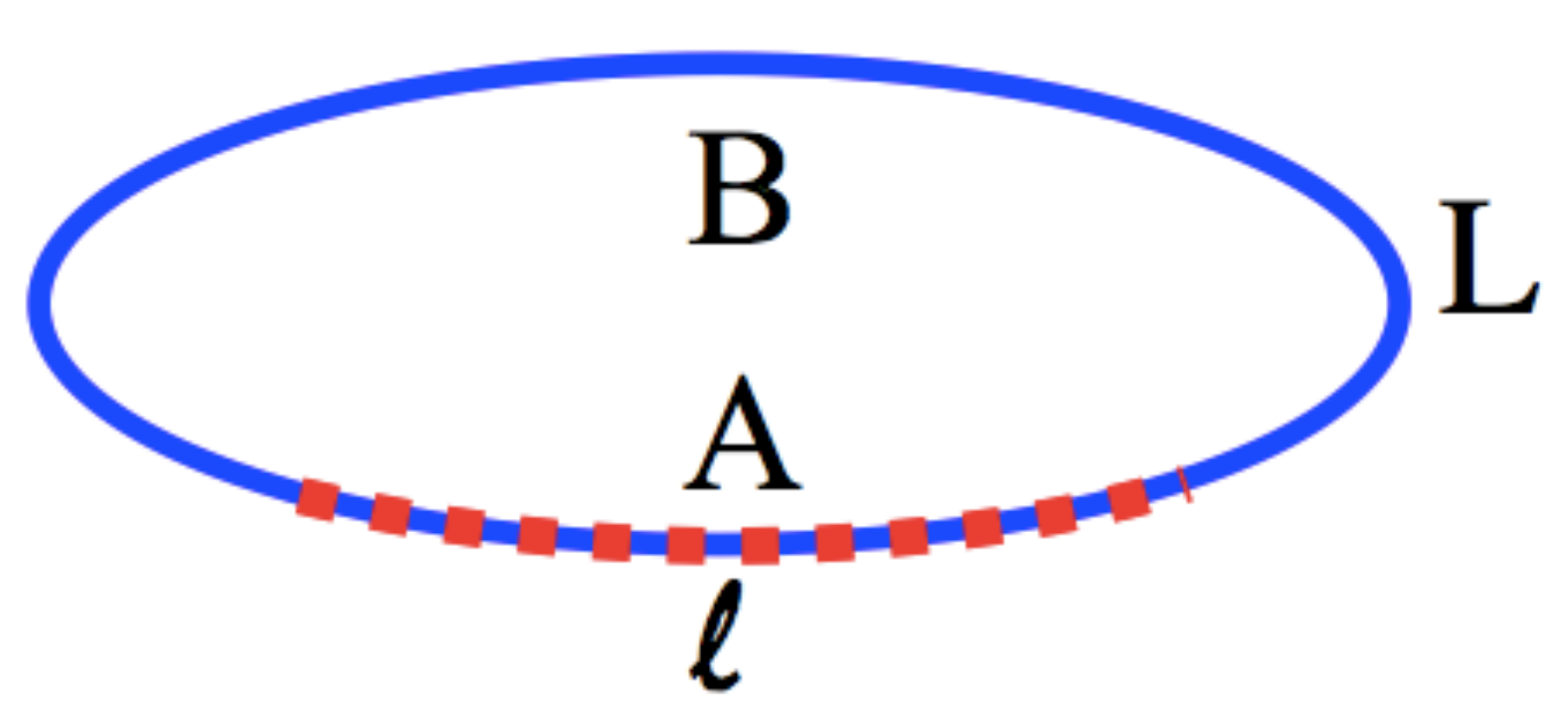} 
 \end{center} 
 \caption{Typical bipartition of a one-dimensional finite system of total length $L$ into region $A$ of length $\ell$ and region $B$ of length $L-\ell$.\vspace{1cm}} 
 \label{typical} 
 \end{floatingfigure}}
The von Neumann and R\'enyi {\it entanglement entropies} are measures of the amount of quantum
entanglement, in a pure quantum state, between the degrees of
freedom associated to two sets of independent observables whose
union is complete on the Hilbert space $\mathcal{H}=\mathcal{H}_A \otimes \mathcal{H}_B$.  In the scaling limit\footnote{Starting from a lattice system with a critical point for some value of a parameter $\lambda=\lambda_c$, the scaling limit to a critical point described by CFT may be taken by first setting $\lambda=\lambda_c$ so the correlation length $\xi\rightarrow \infty$ and then taking the thermodynamic limit $L\rightarrow \infty$. The near-critical behaviour of massive QFT is recovered by taking the limit $\lambda \rightarrow \lambda_c$ and $L, \ell \rightarrow \infty$ simultaneously, whilst keeping $L/\xi$ and $\ell/\xi$ fixed. This is the regime we consider in this paper.}, at quantum critical points, they have been widely studied using CFT \cite{CallanW94,HolzheyLW94,latorre1, Latorre2,Calabrese:2004eu,Calabrese:2005in} and in lattice realizations of critical systems such as quantum spin chains \cite{latorre3,Jin,Lambert,Keating, Weston,fabian,parity} and lattice models \cite{peschel,rava1,rava2}. In particular, the combination of a geometric description, Riemann uniformization techniques and standard expressions for CFT partition functions is very fruitful. Beyond criticality, EEs are accessible by means of the branch point twist field approach introduced in \cite{entropy} and also through numerical techniques.

Consider a bipartition where the two sets of observables correspond to the local observables in two finite-size complementary connected regions, $A$ and $B$ (see for instance Fig.~\ref{typical}). Let the system by in a state $|\Psi\ket_L$, then the von Neumann entropy associated to region $A$ is 
\beq
S_1^\Psi(\ell,L)=-\Tr(\rho_A \log \rho_A)\,,
\label{trace}
\eeq
 where $\rho_A=\Tr_B(|\Psi \rangle_L{}_L \langle \Psi|)$ is the reduced density matrix associated to subsystem $A$ and the trace (\ref{trace}) is over the degrees of freedom in subsystem $A$. One may obtain the entropy $S_1^\Psi(\ell,L)$ as a limiting case of the sequence of  $n$th R\'enyi entropies  defined as
 \beq 
 S_n^\Psi(\ell,L)=\frac{\log \Tr\rho_A^n }{1-n}\,,
 \label{renyi}
 \eeq 
thanks to the property 
\beq 
\lim_{n\rightarrow 1} S_n^\Psi(\ell,L)=S_1^\Psi(\ell,L)\,.
\label{rtrick}
\eeq
One may also consider the so-called single-copy entropy \cite{EC, PZ, DD}, defined as
\beq 
S_\infty^\Psi(\ell,L):=\lim_{n\rightarrow \infty} S_n^\Psi(\ell,L)\,.
\eeq

Much of the work carried out so far deals with the entanglement properties of the ground state (mostly, but not always, in infinite systems). In conformal field theory, universal results for certain types of excited states are known: in \cite{german1,german2}, the increment of R\'enyi entropy in an excited state $|\Upsilon\ket$ with respect to the ground state of a CFT for the configuration of Fig.~\ref{typical} was found to be
 \beq 
S_n^{\Upsilon}(r)-S^{0}_n(r)=\frac{(1+n)(h+\bar{h})}{3n}(\pi r)^2 + O\left(r^{2\Delta_\psi}\right)\,,
\label{there}
\eeq 
for small values of $r=\frac{\ell}{L}$. The excited state was defined as
\beq
|\Upsilon\ket = \lim_{\xi,\bar{\xi}\rightarrow -i\infty}\Upsilon(\xi,\bar{\xi})|0\ket\,, 
\eeq
where $\Upsilon(\xi,\bar{\xi})$ is a CFT field, $h, \bar{h}$ are its holomorphic and antiholomorphic dimensions, $\xi, \bar{\xi}$ are coordinates on the cylinder, and $\Delta_\psi=h_\psi+\bar{h}_\psi$ is the smallest scaling dimension of any field in the theory. Therefore, a measurement of the EE of a low-lying excited state in CFT at finite volume can provide information about the primary field content of the theory. The most extensive numerical study of other kinds of excited states in critical systems was performed in \cite{Vincenzo2}. In this work a very detailed study of the excited states of the XY model in a transverse field and the XXZ Heisenberg spin-chain was carried out. The authors focussed on the case when $L\gg\ell \gg  1$  and on excited states that are macroscopically different from the ground-state (we will consider instead zero-density states). The EE of excited states with finite energy density in quantum field theory (QFT) or quantum lattice models is very simple by the eigenstate thermalization hypothesis (or its extension to integrable systems): it is dominated by the thermodynamic entropy of the corresponding Gibbs (or generalized Gibbs) state and is known to satisfy a volume law \cite{EERev2}. However, little is known so far about the EE of zero-density excited states in gapped systems. The most extensive numerical study in gapped quantum spin chains was carried out in \cite{Vincenzo} where some of the results we obtain here (see 
Section~\ref{smain}) were proposed as describing the ``semi-classical" limit of the EE. Indeed the authors observe how the EE of certain excited states approaches such semi-classical limit for large enough volumes and appropriate correlation lengths. In our work \cite{excitations} we have shown that these bounds, and generalisations, provide, in fact, exact large-volume predictions that are much more widely applicable.

In the present paper, we provide full analytical computations supporting some of the results in \cite{excitations}. We consider excited states of 1+1-dimensional massive QFT with zero energy density: those formed of finite numbers of asymptotic particles, at various momenta. We consider the situation depicted in  Fig.~\ref{typical}, in the limit where both the systems size $L$ and the length $\ell$ of region $A$ are large and in fixed proportion
\beq
 \ell, L \rightarrow \infty\quad \mathrm{with} \quad r=\frac{\ell}{L} \,\, \in\,\, [0,1]\,.
\eeq
Let $|\Psi\rangle_L$ be such an excited state.  Employing the branch point twist field approach \cite{entropy}, we compute the difference between the R\'enyi entropy in the excited state and in the ground state, in this limit,
\beq
	\lim_{L\rightarrow \infty} S_n^\Psi(r L, L)-S_n^0(rL,L)=:\Delta{S}_n^\Psi(r)\,.
\eeq
This entropy increment can be formally written as a ratio of branch point twist field correlators,
\beq 
\Delta{S}_n^\Psi(r)= \lim_{L\rightarrow \infty}
\frac{1}{1-n}{\log\left[\frac{{}_L\bra \Psi|\mathcal{T}(0)\tilde{\TT}(rL)|\Psi \ket_{L}}{{}_L\bra 0|\mathcal{T}(0)\tilde{\TT}(rL)|0 \ket_L}\right]}\,,
\label{re1}
\eeq 
where $\mathcal{T}$ is the branch point twist field and $\tilde{\TT}$ is its hermitian conjugate \cite{entropy}. Recall that branch point twist fields are local fields of the $n$-copy ``replica" QFT, the theory constructed as $n$ not mutually-interacting copies of the model under study. In the replica theory, the state $|\Psi\ket_L$ has the structure
\beq
|\Psi\ket_L=  |\Psi\ket^1_L \otimes |\Psi\ket^2_L \otimes \cdots \otimes |\Psi\ket^n_L\,,
\label{replicastate}
\eeq
where $|\Psi \ket^i_L$ is an excited state of the $i$-th single-copy theory in finite volume $L$. We concentrate on the (uncompactified) massive free real boson and free Majorana fermion models. The techniques that we use --  based on form factors of branch point twist fields -- have been chosen so that they are (hopefully) generalizable to integrable models, in view of extending our results in a future work.

The results we find are very surprising, for various reasons:
\begin{itemize}
\item All results are {\em independent of the momenta} of the excitations, except for the sole condition of coincidence or not of rapidities, and are independent of the model under consideration.
\item The structure of all functions $\Delta S^\Psi_n(r)$ is extremely simple. They in fact admit a {\em combinatorial, or qubit interpretation}, related to counting all possible configurations with various numbers of excitations (particles) ``located" in the region $A$ and outside of it.
\item Our numerical analysis also suggests that the formulae above hold very precisely even for {\em arbitrary systems size $L$, no matter how small, if the momenta of the excitations are large} (even though our calculation methodology employs a large volume expansion). 
\item Additional numerical analysis presented in \cite{excitations} has shown they hold also in higher dimensional free theories and at least some states of interacting quantum spin chains. 
\end{itemize}
The paper is organized as follows: In Section~\ref{smain} we review our results for the increment of EE for states with a finite number of excitations.  The formulae presented in Section~\ref{smain} as well as their ``qubit" interpretation appeared first in \cite{excitations}. Here we present a more general discussion of the ``qubit" interpretation. In Section~\ref{renyireview} we review the connection between branch point twist fields in replica theories  and R\'enyi entropy. We also highlight the challenges of generalizing such connection to finite volume and excited states. We explain how these challenges may be resolved in the case of the massive free boson theory and introduce the ``doubling trick" in this context. In Section~\ref{itgetsserious} we derive the general formulae for the R\'enyi entropy of a single-particle excited state, a $k$-particle excited state involving distinct momenta only, and a $k$-particle excited state consisting of equal momenta. We provide concrete examples of all three cases for  the 2nd R\'enyi entropy of the massive free boson theory. We compare the analytical results to numerical results obtained by employing the wave functional method. In Section~\ref{thefreefermion} we generalize the results of the previous section to the massive free fermion. We find that the expressions for the EEs of states with distinct momenta are identical to those in the free boson theory, even if there are technical differences in the computations involved. In Section~\ref{conclusion} we present our conclusions and outlook. In Appendix A we review the wave functional method and its application to the computation of the R\'enyi entropies of the harmonic chain. In Appendix B we present a derivation of the selection rules which single out those terms in the form factor expansion that provide the leading large-volume contribution to the R\'enyi entropies. In Appendix C we prove some properties of the functions $g_p^n(r)$ in terms of which all EEs can be expressed.


\section{Summary of the Main Results}\label{smain}

The computation of the ratio (\ref{re1}) for a generic $k$-particle excited state of a massive free theory in finite volume involves the use of a considerable number of techniques we will be presenting in the next sections: the form factor programme for branch point twist fields \cite{entropy}, the generalization of this programme for finite volume correlators following the ideas of \cite{PT1,PT2}, the rewriting of the branch point twist field in terms of $U(1)$ fields of the replica free  theory by employing the ``doubling trick" introduced in \cite{doubling}. We then use a new numerical technique based on wave functionals in order to test our analytical results. This is therefore a rather technical work. However, the results that we have obtained are surprisingly simple and can be easily summarized.  They have been shown to hold more widely in \cite{excitations}.


\subsection{Main Formulae}
\label{mainformulae}

Consider a state consisting of a single particle excitation. Let us denote the entropy increments of such a state by $\Delta S_n^1(r)$. We find that
\beq
\Delta{S}_n^{1}(r)=\frac{ \log(r^n+(1-r)^n)}{1-n}\,.
\label{ren}
\eeq
The increment of von Neumann entropies is given by
\beq
 \Delta{S}_1^{1}(r)=-r \log r - (1-r)\log(1-r)\,,
\label{vn}
\eeq
and the increment of single-copy entropies has the form 
\beq
\Delta{S}_\infty^{1}(r)=\left\{\begin{array}{ll} -\log(1-r) & \mathrm{for}\quad 0\leq r <\frac{1}{2}\,,\\
 -\log r & \mathrm{for}\quad \frac{1}{2}\leq r \leq 1\,.
 \end{array} \right. 
 \label{single}
\eeq
For excited states consisting of a finite number $k$ of excitations of {\it{distinct}} momenta the results are simply as above, multiplied by $k$. 
In the free boson, we may also consider states consisting of $k$ particles of {\it{equal}} momenta. We will denote the entropy increments of such states by $\Delta{S}_n^{k}(r)$. We find
\beqa
\Delta{S}_n^{k}(r)&=&\frac{1}{1-n}{\log
\sum\limits_{q=0}^k \left[\left(\begin{array}{c}
k\\
q
\end{array}\right) r^{q} (1-r)^{k-q}\right]^n
},
\label{11}\\
 \Delta{S}_1^{k}(r)&=&-{ \sum\limits_{q=0}^k \left(\begin{array}{c}
k\\
q
\end{array}\right) r^{q} (1-r)^{k-q}\log\left[  \left(\begin{array}{c}
k\\
q
\end{array}\right) r^{q} (1-r)^{k-q}\right]\,.}
\label{2}
\eeqa
The single-copy entropy is a function which is non-differentiable at $k$  points in the interval $r\in(0,1)$ (generalizing (\ref{single}) which has one non-differentiable point).  The positions of these singularities are given by the values
\beq
r=\frac{1+q}{1+k}\,, \quad \mathrm{for} \quad q=0,\ldots,k-1\,,
\eeq
and the single-copy entropy is given by
\beq
 \Delta{S}_\infty^{k}(r)=
 -\log\left[  \left(\begin{array}{c}
k\\
q
\end{array}\right)  r^q (1-r)^{k-q}\right]\,, \quad \mathrm{for}\quad \frac{q}{1+k}\leq r < \frac{1+q}{1+k} \quad \mathrm{and} \quad q=0, \dots, k\,.  
\label{3}
\eeq
Therefore, if the rapidities are distinct, the contribution to the entanglement entropy of $k$ particles is exactly $k$ times the contribution of a single particle excitation, while if they are equal, this is not true: the contribution is in fact smaller.

\begin{figure}[h!]
\begin{center} 
    \includegraphics[width=7.943cm]{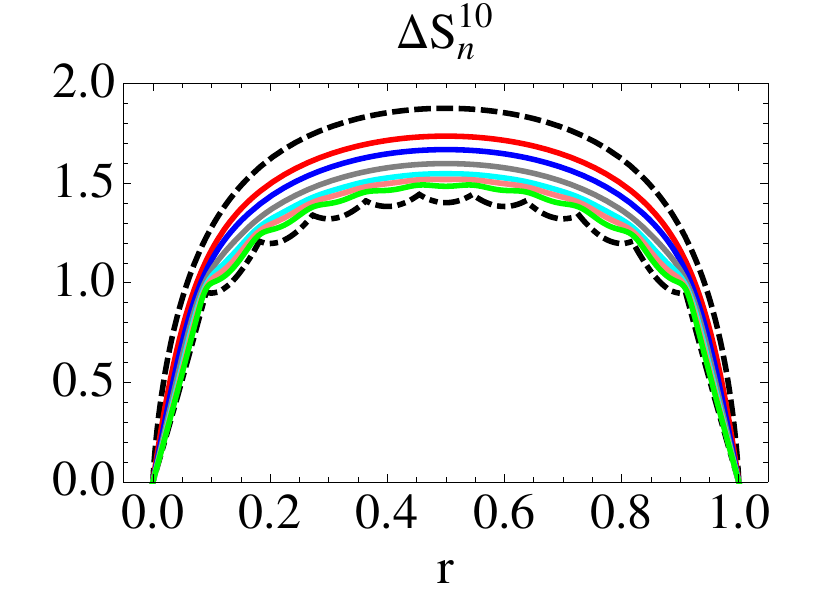} 
     \includegraphics[width=7.943cm]{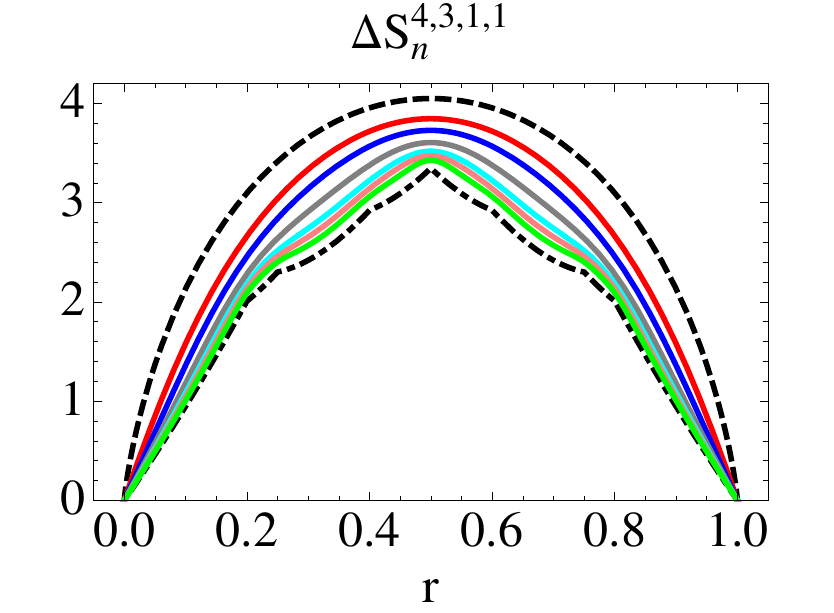} 
  \includegraphics[width=7.943cm]{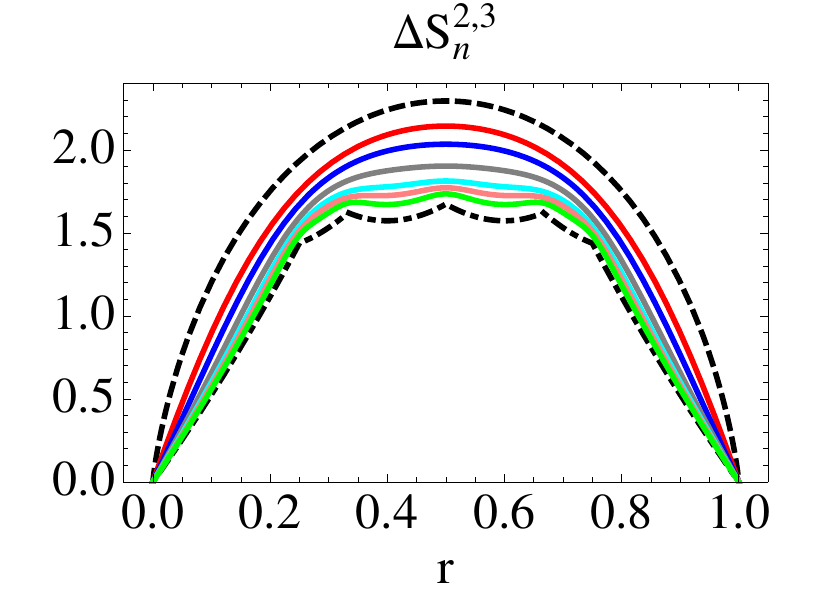} 
   \includegraphics[width=7.943cm]{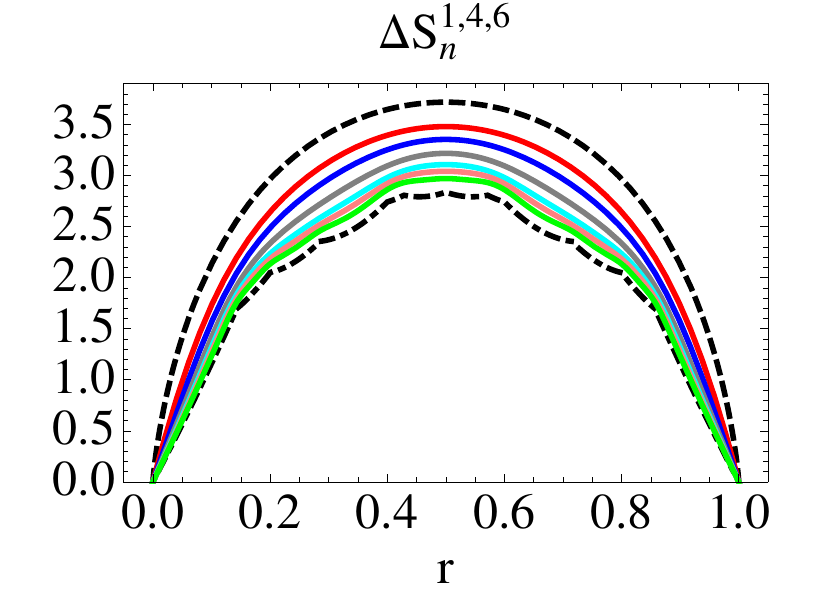} 
 \end{center} 
 \caption{The functions (\ref{11}), (\ref{2}) and (\ref{3}) for a state of 10 equal momenta and for three ``mixed" states with some equal and some distinct momenta.  We plot the R\'enyi entropies for $n=2,3,5,8, 11, 17$ and the von Neumann and single-copy entropies. In each figure, the dashed (outer-most) curve is the von Neumann entropy and the dot dashed (inner-most) curve is the single-copy entropy.} 
 \label{cuspy} 
  \end{figure}

For generic states containing a mixture of excitations with equal and distinct rapidities we find formulae which are sums of those above. Denoting by 
$\Delta S_n^{k_1,k_2,\cdots}(r)$ the R\'enyi entropies of an excited state consisting of $k_i$ particles of momentum $p_i$ with $p_i \neq p_j$ for $i \neq j$ we find
\beq
\Delta S_n^{k_1,k_2,\cdots}(r)=\sum_{i}\Delta S_n^{k_i}(r),\,\,\, \Delta S_1^{k_1,k_2,\cdots}(r)=\sum_{i}\Delta S_1^{k_i}(r), \,\,\, \Delta S_\infty^{k_1,k_2,\cdots}(r)=\sum_{i}\Delta S_\infty^{k_i}(r)\,.
\label{general}
\eeq
Note that (\ref{11}), (\ref{2}) and (\ref{3}) reduce to (\ref{ren}), (\ref{vn}) and (\ref{single}), respectively, for $k=1$. 
In Fig.~\ref{cuspy} we present several examples of the functions above for states of equal and mixed rapidities (other examples were presented in \cite{excitations}).
 
It is easy to show that all the differences of von Neumann entropies have their maximum value at $r=\frac{1}{2}$. 
For states with $k$ distinct rapidities this maximum value is given simply by $k \log 2$ so that a $k$-particle excited state may at most add  $k$ qubits to the entanglement entropy with respect to its ground state value. This fact was discussed in \cite{ln2} for one-particle excitations and shown to hold beyond free theories, for integrable and non-integrable theories. 

For states with some equal rapidities, the maximum is lower. In particular for $k$ coinciding rapidities, it is given by
\beq
\Delta S_1^{k}\left(\frac{1}{2}\right)={ \sum\limits_{q=0}^k \frac{1}{2^k}\left(\begin{array}{c}
k\\
q
\end{array}\right) \log\left[ \frac{1}{2^k} \left(\begin{array}{c}
k\\
q
\end{array}\right) \right]}< k \log 2\,,\quad \mathrm{for} \quad k>1\,.
\eeq


\subsection{Qubit Interpretation}
\label{bens_qubit}

It turns out that the general formulae \eqref{ren}-\eqref{3} have  interpretations as the entanglement entropies of simple states formed of qubits, and are easily understandable from a quasi-classical particle picture of the actual QFT states considered. This was discussed in \cite{excitations}, and we give here slightly more precision.

In order to explain this, consider a bipartite Hilbert space ${\cal H} = {\cal H}_{\rm int}\otimes {\cal H}_{\rm ext}$. Each factor ${\cal H}_{\rm int}\simeq {\cal H}_{\rm ext}$ is the Hilbert space for $N_j$ distinguishable sets each of $j$ indistinguishable qubits, for $j=1,2,3,\ldots$. Making the relation with the entanglement problem described above, we associate ${\cal H}_{\rm int}$ with the interior of the entanglement region of length $\ell$ and ${\cal H}_{\rm ext}$ with its exterior, and we identify the qubit state $1$ with the presence of a particle and $0$ with its absence. With $k$ particles lying on $(0,L)$, we construct the state  $|\Psi_{\mathrm{qb}}\ket \in {\cal H}$ by the (naive) picture according to which equal-rapidity particles are indistinguishable, and a particle can lie anywhere in $(0,L)$ with flat probability: any given particle has probability $r$ of lying in the entanglement region, and $1-r$ of lying outside. We make a linear combination of qubit states following this picture, with coefficients that are (square roots of) the total probability of a given qubit configuration, taking proper care of (in)distinguishability. Then, the R\'enyi and von Neumann entanglement entropies of $|\Psi_{\mathrm{qb}}\ket$ are given exactly by the formulae seen earlier. 
In general
\beq\label{qubit}
	S_n^{\Psi_{\mathrm{qb}}}(r) = \frc{\log\left(\Tr\rho_{{\cal H}_{\rm int}}^n\right)}{1-n}\,,\qquad
	\rho_{{\cal H}_{\rm int}} = \Tr_{{\cal H}_{\rm ext}} |\Psi_{\mathrm{qb}}\ket\bra\Psi_{\mathrm{qb}}|\,,
\eeq
and the statement is that $S_n^{\Psi_{\mathrm{qb}}}(r) =\Delta S_n^{\Psi}(r) $ for some excited state $|\Psi\ket_L$ corresponding to the probability distribution described above.

More precisely, we have ${\cal H}_{\rm int} \simeq{\cal H}_{\rm ext}\simeq  \otimes_{j\geq 1} (\C^{j+1})^{\otimes N_j}$. Here $\C^{j+1}$ is the Hilbert space of $j$ indistinguishable qubits, with basis elements $|q\ket,\,q=0,1,\ldots,j$ labelled by the number of qubits that are in their state $1$. One can also write ${\cal H}_{\rm int} \simeq{\cal H}_{\rm ext}\simeq \otimes_{i=1}^{N} \C^{j_i+1}$, where $N$ is the total number of groups, $N = \sum_{j\geq 1}N_j$, and $j_i$ take values $j_1=\cdots=j_{N_1}=1$, $j_{N_1+1}=\cdots=j_{N_1+N_2}=2$, etc.  We denote the basis of vectors in ${\cal H}_{\rm int}\simeq{\cal H}_{\rm ext}$ by $|{\bf q}\ket$ for ${\bf q} = (q_i:i=1,\ldots,N)\in \prod_{j\geq 1} \{0,1,\ldots,j\}^{N_j}$.  We use the notation $\b {\bf q}= (j_i-q_i:i=1,\ldots,N)$ for the state where the qubits are inverted.  We then construct
\beq
	|\Psi_{\mathrm{qb}}\ket = \sum_{{\bf q}\in \prod_{j\geq 1} \{0,1,\ldots,j\}^{N_j}}
	\sqrt{p_{\bf q}} \,|{\bf q}\ket \otimes |\b {\bf q}\ket\,,
\eeq
where $p_{\bf q}$ is the probability of finding the particle configuration ${\bf q}$ in the entanglement region according to the naive picture above, given by
\beq
	p_{\bf q} = \prod_{i} {j_i\choose q_i} r^{q_i}(1-r)^{j_i-{q_i}}\,.
\eeq

For instance, if a single particle is present, then the state is
\beq
	|\Psi_{\mathrm{qb}}\ket= \sqrt{r} \;|1\ket \otimes |0\ket + \sqrt{1-r}\;|0\ket \otimes |1\ket\,,
	\label{2qubit}
\eeq
as either the particle is in the region, with probability $r$, or outside of it, with probability $1-r$. If two particles of coinciding rapidities are present, then we have
\beq
	|\Psi_{\mathrm{qb}}\ket = \sqrt{r^2}\; |2\ket \otimes |0\ket + \sqrt{2r(1-r)}\;|1\ket \otimes |1\ket + \sqrt{(1-r)^2}\;|0\ket \otimes |2\ket\,,
\eeq
as either the two particles are in the region, with probability $r^2$, or one is in the region and one outside of it (no matter which one), with probability $2r(1-r)$, or both are outside the region, with probability $(1-r)^2$. For two particles of different rapidities,
\beq
	|\Psi_{\mathrm{qb}}\ket= \sqrt{r^2}\; |11\ket \otimes |00\ket + \sqrt{r(1-r)}\;(|10\ket \otimes |01\ket + |01\ket \otimes |10\ket) + \sqrt{(1-r)^2}\;|00\ket \otimes |11\ket\,,
\eeq
counting the various ways two distinct particles can be distributed inside or outside the region.

From this explicit construction, one can indeed show that \eqref{qubit} gives the formula \eqref{general}.


\section{R\'enyi Entropies and Branch Point Twist Fields}
\label{renyireview}
It has been known for some time that several entanglement measures, including the R\'enyi entropies, can be expressed in terms of correlation functions of a special class of local fields $\TT$ which have been termed {\it branch point twist fields} in \cite{entropy}. Branch point twist fields are, on the one hand, twist fields in the broader sense, that is, fields associated with an internal symmetry of the theory under consideration, and on the other hand related to branch points of multi-sheeted Riemann surfaces. They are twist fields associated to the cyclic permutation symmetry of a model composed of $n$ copies of the original model, with exchange relations
\beqa
\label{basictwistfield}
\TT(x)\Or_i(y) &=& \Or_{i+1}(y)\TT (x)  \quad \mathrm{for} \quad y^1>x^1\,,\\ 
&=& \Or_i(y)\TT(x)  \quad \mathrm{for} \quad x^1>y^1\,,
\eeqa
where $\Or_i(y)$ is any local field on copy number $i$, and with $\Or_{n+1}(y)=\Or_1(y)$.

The idea of quantum fields associated with branch points of Riemann surfaces in the context of entanglement appeared first in \cite{Calabrese:2004eu}, where their scaling dimension was evaluated in CFT 
\beq 
\Delta_\TT=\frac{c}{24}\left(n-\frac{1}{n} \right)\,,
\label{dtwist}
\eeq 
(see also \cite{kniz} for an earlier work concerned with similar ideas in the context of orbifold CFT). Here $c$ is the central charge and $n$ is the number of sheets in the Riemann surface. The general description in terms of branch point twist fields as symmetry fields associated to cyclic permutation symmetry of the $n$ Riemann surface's sheets, as per \eqref{basictwistfield}, was given in \cite{entropy}, where they were studied in integrable massive QFT. This description is however independent of integrability, and it was first used in massive QFT outside of integrability in \cite{next}. 

The missing logical link that connects the Riemann surface structure mentioned above with a computation of entanglement measures comes through a result commonly known as  the {\it replica trick}.  Mathematically speaking, the replica trick is simply the statement (\ref{rtrick}) with (\ref{renyi}). However, the word ``replica" originates from the fact that the object $\Tr \rho_A^n$ which features in (\ref{renyi}) can be interpreted as the partition function of a replica QFT understood as $n$ non-interacting copies of the original QFT.  In the limit $L\to\infty$ (for the configuration in Fig.~1 with $L\rightarrow \infty$), this partition function is evaluated precisely on a Riemann surface with $n$ sheets as described earlier, with a branch cut of length $\ell$ across which Riemann sheets are connected cyclically (when the branch cut starts at the origin and $L\rightarrow \infty$ this is exactly the structure of the Riemann surface of the function $\sqrt[n]{\frac{z}{z-\ell}}$). Hence the number of sheets and the number of replicas are both $n$. For finite volume $L$, the Riemann sheets are replaced by cylinders of circumference $L$ cyclically connected along a branch cut in the compactified (space) direction. In this picture, the  $n$th R\'enyi entropy with the partitioning protocol presented in Fig.~1 is given by:
\beq
S_n(\ell,L)=\frac{\log\left(\varepsilon^{4\Delta_\TT}{}_L\bra \Psi| \TT(0)\tilde{\TT}(\ell)|\Psi\ket_L\right)}{1-n}\,,
\label{renyi2}
\eeq 
where $|\Psi\ket_L$ is an excited state of a finite number of excitations in the finite-volume $L$, replica QFT. The structure of the state is as reported in (\ref{replicastate}), $\tilde{\TT}=\TT^\dagger$ is the hermitian conjugate of the branch point twist field $\TT$, and $\varepsilon$ is a non-universal short-distance cut-off. Notice that the dependance on the cut-off $\varepsilon$ cancels out when considering the entropy increment (\ref{re1}).


\subsection{Challenges Posed by the Treatment of Excited States}

For $L\rightarrow \infty$ in the ground state the function (\ref{renyi2}) has been extensively investigated, both from the point of view of its universal features \cite{entropy, next} and for particular models \cite{nexttonext,other,BCD,freeboson,E8toda}. The study of excited states however presents new challenges.

First, in the context of integrable models of massive QFT, it is natural to evaluate two-point functions by inserting a sum over a complete set of states and then computing the matrix elements of local operators that are the building blocks of the resulting sum. Schematically we can represent this process by writing
\beq 
{}_L\bra \Psi | \TT(0)\tilde{\TT}(\ell)|\Psi \ket_L\propto \sum_{|\Phi\ket} {}_L\bra \Psi | \TT(0)|\Phi\ket_L \times {}_L\bra \Phi| \tilde{\TT}(\ell)|\Psi\ket_L\,.
\label{renyi3}
\eeq
The advantage of this decomposition is that for integrable models at least, there exist effective methods to exactly compute the matrix elements $ {}_L\bra \Psi | \TT(0)|\Phi\ket_L$. In infinite volume such methods are usually refered to as the form factor programme \cite{KW,SmirnovBook} and they provide the most powerful and successful approach to the computation of correlation functions, both analytically and numerically. For branch point twist fields, the form factor programme was generalized in \cite{entropy}.
For finite volume and local fields $\mathcal{O}$ -- excluding twist fields -- matrix elements of the type ${}_L\bra \Psi| \mathcal{O}(0)|\Phi\ket_L$ are also well understood \cite{PT1,PT2}.

In 1+1 dimensions, the states $|\Phi\ket_L$ and $|\Psi\ket_L$ are characterized by a discrete set of rapidities (or momenta). Should any of the rapidities in one set coincide with some in the other set, the matrix element ${}_L\bra \Psi| \mathcal{O}(0)|\Phi\ket_L$ for $L\rightarrow \infty$ will develop, in the usual infinite-volume normalization of the states, $\delta$-function  singularities. Considering instead finite volume form factors, provides a natural regularization scheme to deal with such singularities. Indeed, for local operators, a systematic prescription exists to compute the ``physical part" of matrix elements such as ${}_L\bra \Psi| \mathcal{O}(0)|\Phi\ket_L$ by subtracting the contributions of any occurring singularities in a way which is controlled by the particular pole structure of the form factors of local fields.

In our case however, we face the challenge that the branch point twist fields are not local in the sense required to apply the techniques of \cite{PT1, PT2}. Although they are local with respect to the Lagrangian density of the replica model (as they implement a symmetry) they are non-local with respect to the fundamental fields of the theory (those whose associated modes create and annihilate the physical particles). It is however, still very plausible that the standard general ideas for the computation of finite-volume non-diagonal form factors will be applicable to branch point twist fields. We confirm this below by analytical and numerical results in free theories. 
\begin{figure}[h!]
\begin{center} 
 \includegraphics[width=10cm]{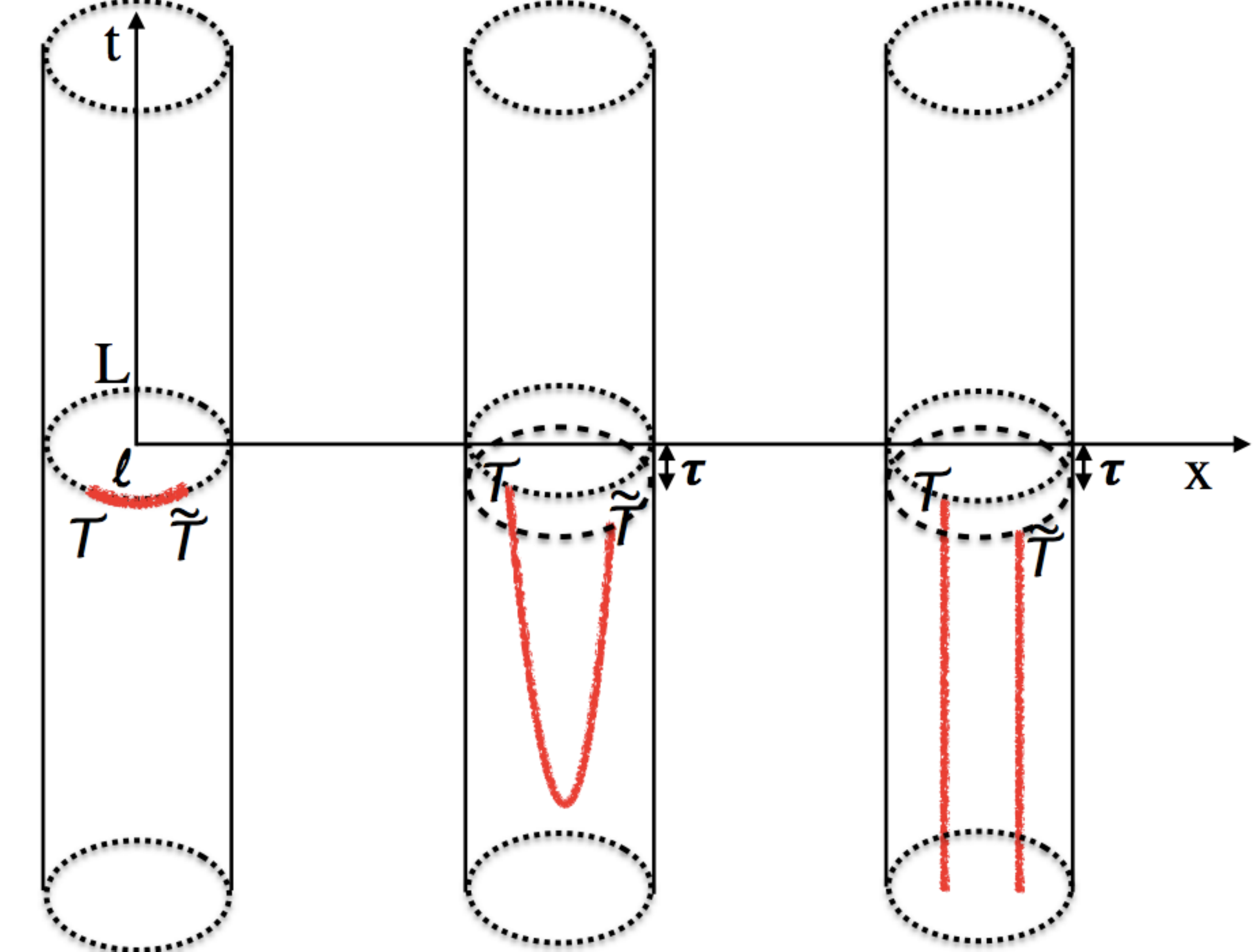} 
 \end{center} 
 \caption{Branch cut deformation along the time direction on an infinite cylinder of circumference $L$. Note that, formally, the fields are also slightly shifted in the time direction (hence the parameter $\tau$) to ensure time ordering. } 
 \label{cylinder} 
  \end{figure}
  
Second, branch point twist fields sit at the origin of branch cuts which, in the standard prescription, originate at the twist field and extend indefinitely in the space direction. For the two-point function, the two branch cuts emerging from the twist field and its hermitian conjugate combine to create a branch cut of finite length $\ell$ which is interpreted as the length of subsystem $A$. However, once we write down the expansion (\ref{renyi3}) we need to evaluate the matrix elements ${}_L\bra \Psi| \TT(0)|\Phi\ket_L$ and ${}_L\bra \Phi| \tilde{\TT}(\ell)|\Psi\ket_L$. For these matrix elements, an infinitely long branch cut extending in space is incompatible with working in finite volume $L$. This conflict can be resolved by adopting an approach which is reminiscent of that taken in \cite{FZ} for the Ising field theory and the matrix elements of its $\mathbb{Z}_2$ twist field $\sigma$. We may use the fact that the branch cut can be continuously deformed without changing the value of the correlation function. Therefore we may continuously ``stretch" the branch cut  along the time direction as indicated in Fig.~\ref{cylinder}. The result is a product of fields with branch cuts extending in the time direction. In this configuration, the fields are well defined in the quantization on the circle, where they are {\it intertwining operators}. The operator ordering of the two-point function in the quantization scheme on the circle, is implemented in the path integral by a time ordering: an infinitesimal shift $\tau$ along the cylinder, as in Fig.~\ref{cylinder}.

In parallel to the situation in \cite{FZ}, the Hilbert space of quantization on the circle is divided into sectors characterized by periodicity conditions: if an internal symmetry $\sigma$ exists, then the Hilbert space ${\cal H}_\sigma$ is that of field configurations with the quasi-periodicity condition $\Or(x+L) = \sigma\cdot \Or(x)$. For the Ising model, the $\mathbb{Z}_2$ symmetry leads to two sectors, Ramond-Ramond and Neveu-Schwarz. In the case of our replica model, we have in particular $n$ sectors labelled by cyclic elements of the permutation group. The intertwining operators corresponding to the branch point twist fields act as follows:
\beq\label{intertwine}
 	\TT: {\cal H}_\sigma\to {\cal H}_{\omega^{-1}\circ\sigma}\,,\qquad
	\t\TT:{\cal H}_\sigma\to {\cal H}_{\omega\circ\sigma}\,,
\eeq
where $\omega$ is the elementary cyclic permutation symmetry of the $n$-copy replica model, taking copy $i$ to copy $i+1\;{\rm mod}\;n$. This is seen as follows: the condition \eqref{basictwistfield} imposes continuity between $\Or_i$ below and $\Or_{i+1}$ above the branch extending towards the right. After the deformation as in Fig. \ref{cylinder}, this becomes continuity between $\Or_i$ on the left and $\Or_{i+1}$ on the right of the branch extending towards negative times. This adds a factor of $\omega$ on the Hilbert space on which $\TT$ acts, or equivalently, a factor $\omega^{-1}$ on the image Hilbert space. Therefore, in the matrix elements  ${}_L\bra \Psi | \TT(0)|\Phi\ket_L$ and ${}_L\bra \Phi| \tilde{\TT}(\ell)|\Psi\ket_L$, the state $|\Psi\ket_L$ is in a different sector than the state $|\Phi\ket_L$. In the cylinder picture of Fig.~\ref{cylinder}, the state $|\Phi\ket$ lies between the twist fields, in the time slice of extent $\tau$ introduced by the operator ordering.

Finally, the question arises as to how the matrix elements of branch point twist fields with states in different sectors can be computed. Answering this question in general integrable QFT is somewhat complicated and will be discussed in a future work. However, for free theories there are additional resources at our disposal. More precisely, for free theories, it is possible to express the branch point twist fields in terms of simpler $U(1)$ twist fields, where the permutation symmetry has been diagonalized. This is achieved by employing the so-called {\it doubling trick} introduced in \cite{doubling} and employed successfully in the branch point twist field context in \cite{entropy,VEVFB}, where it allowed for the computation of the vacuum expectation value of the branch point twist field. A similar idea was also used in \cite{CH} in the study of the EE of free theories.

The doubling trick is the simple idea that a real free fermion (Majorana) and a real free boson theory can be doubled to construct a complex free fermion (Dirac) and a complex free boson theory. This doubling induces a $U(1)$ symmetry in the new theory to which a $U(1)$ twist field is associated. In a replica theory whose fundamental building blocks are doubled free theories, the $U(1)$ symmetry on each individual copy is extended to a $U(n)$ symmetry, which includes cyclic permutation of the copies. Diagonalizing the cyclic permutation, in the new basis the branch point twist field is then expressed as a product of $n$ individual $U(1)$ twist fields $\TT_{p}$ for $p=1,\ldots,n$.

Having summarized the main challenges and techniques involved in the computation of R\'enyi entropies of excited states in finite volume, we proceed now to present these techniques in some detail for the case at hand.


\subsection{Doubling Trick and Replica Free Boson Model}
\label{doublingtrick}

In this and the remaining subsections, we concentrate on the free boson model. We then generalize the construction to the free fermion in Section \ref{thefreefermion}.

In \cite{doubling} Fonseca and Zamolodchikov introduced the ``doubling trick". There, it was employed to find differential equations that are satisfied by certain combinations of correlation functions in the Ising model. This technique was later used in order to obtain vacuum expectation values $\bra \TT\ket$ in infinite volumes  in the works \cite{entropy} (free fermion) and \cite{VEVFB} (free boson).

The idea is to ``double" the free theory in order to have an additional continuous symmetry. Let $\phi_a$ and $\phi_b$ be two independent free massive real bosons. We construct a free massive complex boson as:
\beq
\Phi=\frac{\phi_a+i\phi_b}{\sqrt{2}}\quad \mathrm{and} \quad \Phi^\dagger=\frac{\phi_a-i\phi_b}{\sqrt{2}}\,,
\eeq
which has an internal continuous $U(1)$ symmetry. This symmetry can then be exploited in order to obtain information about the original (not doubled) theory. In the context of the branch point twist field, the doubling trick is used as follows. In the doubled replica model, the combination of the $U(1)$ symmetry of the complex field on each replica, and of the permutation symmetry of the replica, implies the existence of a $U(n)$ symmetry of the model. Cyclic permutations form a subgroup of the  $U(1)$ symmetry group of rotations amongst the copies, which can be diagonalized. The diagonal basis is a new set of $n$ independent complex free bosons, each of which has its own $U(1)$ symmetry, and the cyclic permutation action is expressed as a product of $U(1)$ actions on each of these bosons. Therefore, the branch-point twist field acts as a product of $U(1)$ twist fields in the diagonal basis.

In the replica theory we have $n$ copies of the complex free boson, $\Phi_j$ with $j=1,\ldots, n$. Since the components $\phi_{a,j}, \phi_{b,j}$ are commuting fields and the permutation symmetry $\omega$ acts in a factorized way as $\omega_a\times\omega_b$, the branch point twist field also factorises,
\beq 
\TT = \TT_{\rm{complex}}=\TT_a \otimes \TT_b\,.
\eeq 
Therefore, correlators of $\TT$ in any state that is factorized into the copies $a$ and $b$, also factorize into those of $\TT_a$ and $\TT_b$ in the real boson theory. The idea is to perform calculations in the replica complex free boson theory and interpret the results in terms of the real free boson using this factorization.

In matrix form, the permutation symmetry $\omega$ acts as
\beq
\omega\left(
\begin{array}{c}
\Phi_1\\
\Phi_2\\
\vdots\\
\Phi_{n-1}\\
\Phi_n
\end{array}\right)=\left(
\begin{array}{c}
\Phi_2\\
\Phi_3\\
\vdots\\
\Phi_n\\
\Phi_1\end{array}\right)\,,\quad\mbox{that is\,,}\quad
\omega=\left(\begin{array}{ccccc}
0&1&0&\cdots &0\\
0&0&1&\cdots&0\\
\vdots&\vdots&\vdots&\ddots& \vdots\\
0&0&0&\cdots&1\\
1&0&0&\cdots&0
\end{array} 
\right)\,.
\label{matrixw}
\eeq
The eigenvalues of this matrix are exactly the $n$th roots of unity $\lambda_p= e^{\frac{2\pi \ri p}{n}}$ for $p=1,\ldots,n$. The cyclic permutation action is diagonalized by the fields
\beq
\tilde{\Phi}_p=\frac{1}{\sqrt{n}} \sum_{j=1}^{n}  e^{-\frac{2\pi \ri j p}{n}}  \Phi_{j}\,,
\label{rel}
\eeq
which are themselves canonically normalized complex free boson fields. 
Since $\omega$ acts diagonally on the basis $\tilde{\Phi}_p$, it can be factorized into a product of $U(1)$ fields. We denote by $\TT_{p}$ the $U(1)$-field acting nontrivially on sector $p$, and $\t\TT_p$ its hermitian conjugate. The field $\TT_p$ has exchange relations
 \beqa
\label{exu1}
\TT_{p}(x)\tilde{\Phi}_q(y) &=& e^{\frac{2\pi \ri p\,}{n}\delta_{qp}} \tilde{\Phi}_q(y)\TT_{p} (x)  \quad \mathrm{for} \quad y^1>x^1\,,
\\ &=& \tilde{\Phi}_q(y)\TT_{p}(x)  \quad \mathrm{for} \quad x^1>y^1\,,\no
\eeqa
for $q, p=1,\ldots,n$ with $q\equiv q+n$ and $p\equiv p+n$, and $\t\TT_p$ has similar exchange relations with complex conjugate phase. Then,
\beq \label{TTfactorizesector}
\TT=\prod_{p=1}^{n} \TT_{p}\,,
\eeq 
where, by definition, the field $\TT_n$ is the identity field.  For free bosons, such $U(1)$ fields have been studied and it is known that they have scaling dimensions \cite{orbifold}
\beq
\Delta_p= \frac{p}{2n}\left(1-\frac{p}{n}\right)\,,
 \eeq
so that
\beq
\Delta_\TT=\sum_{p=1}^{n}\Delta_p= \frac{1}{12}\left(n-\frac{1}{n}\right)\,,
\eeq
which coincides with (\ref{dtwist}) for $c=2$ (the central charge of the complex free boson).

In order to study the entanglement entropy of excited states in finite volume $L$, we will consider states of the complex boson theory which are $k$-particle states in copy $a$ times the vacuum in copy $b$,
\beq\label{k-part-state}
	|k\ket_L = |k\ket^a_L \otimes |0\ket^b_L
	= |k\ket^{a,1}_L\otimes\cdots\otimes|k\ket^{a,n}_L
	\otimes |0\ket^{b,1}_L\otimes\cdots\otimes|0\ket^{b,n}_L\,.
\eeq
In this factorized state, we have
\beq
	{}_L\bra k|\TT(0)\t\TT(\ell)|k\ket_L=
	{}^a_L\bra k|\TT_a(0)\t\TT_a(\ell)|k\ket^a_L\, \times \, {}^b_L\bra 0|\TT_b(0)\t\TT_b(\ell)|0\ket^b_L\,.
\eeq
The second factor is the vacuum expectation value, which is known. We therefore obtain the required real free boson result as
\beq
	{}^a_L\bra k|\TT_a(0)\t\TT_a(\ell)|k\ket^a_L = \frc{{}_L\bra k|\TT(0)\t\TT(\ell)|k\ket_L}{{}^{b}_L\bra 0|\TT_b(0)\t\TT_b(\ell)|0\ket^b_L}\,.
\eeq

In order to describe the many-particle states $|k\ket_L$ more precisely, we introduce the creation and annihilation operators $(a_j^\pm)^\dag(\theta)$ and $a_j^\pm(\theta)$, respectively, of the complex free boson $\Phi_j$; these create / annihilate a particle of rapidity $\theta$ and $U(1)$ charge $\pm$ in replica copy $j$. The creation operator on doubling-trick copy $a$ and replica copy $j$ is expressed as
\beq
(a_j^{a})^\dagger(\theta)=\frac{1}{\sqrt{2}}((a_j^{+})^\dagger(\theta)+(a_{j}^{-})^\dagger(\theta))\,.
\label{copya}
\eeq
Therefore, the normalized $k$-particle state \eqref{k-part-state} is, explicitly in the case of distinct rapidities,
\beq
|k\ket_L = |\theta_1,\ldots,\theta_k \ket^a_L\otimes |0\ket^b_L=\frac{1}{2^\frac{kn }{2}} \prod_{j=1}^n\prod_{i=1}^k \Big((a_j^+)^\dagger(\theta_i)+(a_j^-)^\dagger(\theta_i)\Big)|0\ket_L\,. 
\label{statekp_basic}
\eeq
In the free boson theory, one may consider states with some coinciding rapidities, in which case the normalization of the state needs to be slightly modified. This will be discussed in more detail in Subsection~\ref{coincide}.

On the other hand, in the diagonal basis \eqref{rel}, the operators $\tilde{a}^\pm_p(\theta)$ (and hermitian conjugate) are given by
\beq 
(\tilde{a}_p^{\pm})^\dagger(\theta)=\frac{1}{\sqrt{n}}\sum_{j=1}^{n}e^{\pm\frac{2\pi \ri j p}{n}} (a_j^{\pm})^\dagger(\theta) \quad \mathrm{or} \quad 
({a}_j^{\pm})^\dagger(\theta)=\frac{1}{\sqrt{n}}\sum_{p=1}^{n}e^{\mp\frac{2\pi \ri j p}{n}} (\tilde{a}_p^{\pm})^\dagger(\theta)\,. 
\label{newb}
\eeq
Expressing the operators $a_j^{\pm}(\theta)$ in terms of the tilde operators through (\ref{newb}) leads, after some manipulations, to
 \beqa
 |k\ket_L =  \frac{1}{(2n)^\frac{n k}{2}}\prod_{j=1}^n \sum_{\epsilon_1,\ldots,\epsilon_k=\pm} \sum_{p_1,\ldots, p_{k}=1}^{n} e^{-\frac{2\pi \ri j}{n} \sum_{i=1}^{k} p_i \epsilon_i} 
( \tilde{a}_{p_1}^{\epsilon_1})^\dagger(\theta_1)\cdots ( \tilde{a}_{p_k}^{\epsilon_k})^\dagger(\theta_k)|0\ket_L\,.\label{dec}
\eeqa
This is a useful expression, because thanks to \eqref{TTfactorizesector}, correlation functions of twist fields in the diagonalized basis factorize into correlations on the sectors $p=1,\ldots,n$. Let us introduce the short-hand notation 
\beq
\tilde{a}_j^{+}(\theta):=\textfrak{a}_j(\theta)\quad \mathrm{and} \quad  \tilde{a}_j^{-}(\theta):=\textfrak{b}_j(\theta)\,.
\eeq
Then,  the following state factorizes as
\beq
 \textfrak{a}_{1}^\dagger(\theta)\textfrak{a}_{2}^\dagger(\theta) \textfrak{b}_{3}^\dagger(\theta) \textfrak{b}_{2}^\dagger(\theta)|0\ket_L=  \textfrak{a}_{1}^\dagger(\theta)|0\ket_{1;L} \otimes  \textfrak{a}_{2}^\dagger(\theta)  \textfrak{b}_{2}^\dagger(\theta)|0\ket_{2;L} \otimes  \textfrak{b}_{3}^\dagger(\theta)|0\ket_{3;L}\,,
\eeq
where we write $|0\ket_L = \otimes_{j=1}^n|0\ket_{j;L}$. Using this, for $n=3$ we have for instance
\beqa
&&{}_L \bra 0 |\textfrak{a}_{1}(\theta)\textfrak{a}_{2}(\theta) \textfrak{b}_{3}(\theta) \textfrak{b}_{2}(\theta)\TT(0) \tilde{\TT}(\ell) \textfrak{a}_{1}^\dagger(\theta)\textfrak{a}_{2}^\dagger(\theta) \textfrak{b}_{3}^\dagger(\theta) \textfrak{b}_{2}^\dagger(\theta)|0\ket_L\nonumber\\
&& ={}_{1;L}\bra 0 |\textfrak{a}_{1}(\theta) \TT_1(0) \tilde{\TT}_1(\ell) \textfrak{a}_{1}^\dagger(\theta)|0\ket_{1;L} \times 
{}_{2;L} \bra 0 |\textfrak{a}_{2}(\theta) \textfrak{b}_{2}(\theta)\TT_2(0) \tilde{\TT}_2(\ell) \textfrak{a}_{2}^\dagger(\theta) \textfrak{b}_{2}^\dagger(\theta)|0\ket_{2;L}\,, 
\label{suchas}
\eeqa
where we used the fact that 
\beq {}_{3;L} \bra 0 | \textfrak{b}_{3}(\theta)  \TT_3(0) \tilde{\TT}_3(\ell) \textfrak{b}_{3}^\dagger(\theta) |0\ket_{3;L}=1\,,
\eeq
 since $\TT_3$ is the identity field for $n=3$. In this way, any two-point function can be expressed as a sum of factorized correlators involving only particles and $U(1)$ twist fields acting on a particular sector of the theory. A detailed computation for $k$-particle states of equal and distinct momenta will be presented below.
 
 The computation of matrix elements such as (\ref{suchas}) requires two additional ingredients: first, the introduction of finite volume form factors, and second, the understanding of how particle rapidities are quantized in finite volume. We  address these questions  in the next two subsections. 

\subsection{Infinite Volume Form Factors of $U(1)$ Fields}

As explained in the previous subsection, explicit computations of the R\'enyi entropy may be obtained by computing matrix elements of $U(1)$ twist fields. Let us review here some of the properties of these form factors in the free boson theory.  The form factors have been known in the literature for quite some time \cite{SMJ, VEVFB}. We define the two particle form factors of the $p$-th $U(1)$ field as
\beqa 
&&F^{p|+-}(\theta_1-\theta_2):={}_p\bra 0| \TT_p(0) \textfrak{a}_{p}^\dag(\theta_1) \textfrak{b}_{p}^\dag(\theta_2)|0\ket_p =F^{p|-+}(\theta_2-\theta_1)\,,\nonumber\\
&&F^{p|++}(\theta_1-\theta_2):={}_p\bra 0| \TT_p(0) \textfrak{a}_{p}^\dag(\theta_1) \textfrak{a}_{p}^\dag(\theta_2)|0\ket_p=0\,, \nonumber\\
&&F^{p|--}(\theta_1-\theta_2):={}_p\bra 0| \TT_p(0) \textfrak{b}_{p}^\dag(\theta_1) \textfrak{b}_{p}^\dag(\theta_2)|0\ket_p=0\,. \eeqa
The last two form factors are vanishing  for symmetry reasons (the twist field preserves the total $U(1)$ charge).
The form factor programme for quasi-local fields \cite{KW,SmirnovBook,YZam} tells us that the nonvanishing form factors may be computed as the solutions to a set of three equations. First, 
Watson's equations
\beq 
F^{p|\pm \mp}(\theta)=F^{p|\mp \pm}(-\theta) \quad \mathrm{and} \quad F^{p|\pm \mp}(\theta+2\pi i) =\gamma^{\pm}_p F^{p|\mp \pm}(-\theta)= \gamma_p^{\pm} F^{p|\pm \mp}(\theta)\,,
\eeq 
where $\gamma^{\pm}_p$ are the factors of local commutativity associated to the bosons $\pm$. From the exchange relations (\ref{exu1}) we expect that $\gamma^{+}_p=(\gamma^{-}_p)^{-1}=e^{\frac{2\pi i p}{n}}$. Finally, the kinematic residue equation is
\beq 
\mathrm{Res_{\theta=0}}F^{p|\pm \mp}(\theta+i \pi) = i (1-\gamma^{\pm}_p) \tau_p\,,
\label{kinu}
\eeq 
where 
\beq
\tau_p={}_p\bra0|\TT_{p}(0)|0\ket_p\,,
\eeq
 is the vacuum expectation value. 
Based on the equations above it is easy to make a general ansatz:
\beq
F^{p|+ -}(\theta)=\frac{A e^{a \theta}}{\cosh\frac{\theta}{2}}\,,
\eeq 
where $A$ and $a$ are constants to be determined. It is then easy to show that the equations are satisfied if 
\beq 
a=\frac{p}{n} - \frac{1}{2} \quad \mathrm{and} \quad A=- \tau_p \sin \frac{\pi p}{n}\,. 
\eeq 
This gives the solution
\beq 
F^{p|+-}(\theta)=-\tau_p \sin\frac{\pi p}{n} \frac{e^{\left(\frac{p}{n}-\frac{1}{2} \right)\theta}}{\cosh \frac{\theta}{2}}\,. \label{minus}
\eeq 
Another solution can be obtained by shifting $j\mapsto j+n$ but if we assume $p\leq n$ the solution above is singled out. Since the theory is free, higher particle form factors can be obtained by simply employing Wick's theorem. For the complex free boson they have the structure
\beqa 
\!\!\!\!\!\!F_{2m}^{p,n}(\theta_1,\ldots,\theta_m; \beta_1,\ldots, \beta_m)&=&
 {}_p\bra 0|\TT_p(0)\textfrak{a}_{p}^\dag(\theta_1)\cdots \textfrak{a}_{p}^\dag(\theta_m) \textfrak{b}_{p}^\dag(\beta_1)\cdots \textfrak{b}_{p}^\dag(\beta_m)|0\ket_p \nonumber\\
&=&
 \tau_p \sum_{\sigma \in S_m} f_p^n(\theta_{\sigma(1)}-\beta_1)\cdots 
f_p^n(\theta_{\sigma(m)}-\beta_{m})\,, \label{suma}
\eeqa 
where we introduced the normalized two-particle form factor
\beq
f^{n}_p(\theta):=\frac{{F}^{p|+-}(\theta)}{\tau_p}\,,
\eeq
and $\sigma$ are all elements of the permutation group $S_m$ of $m$ symbols.

In what follows we will require the form factors (\ref{suma}) as well as slightly more general matrix elements.
These can be related to form factors as
\beqa
&&  {}_{p}\bra 0|\prod_{i_1=1}^s \textfrak{a}_p(\theta_{i_1})\prod_{i_2=1}^q \textfrak{b}_p(\beta_{i_2})\TT_p(0)\prod_{i_4=1}^{q'} \textfrak{b}^\dagger_p(\beta'_{i_4})\prod_{i_3=1}^{s'} \textfrak{a}_p^\dagger(\theta'_{i_3})|0\ket_{p} =\label{inf_vol}\\
&& \quad  F_{s+s'+q+q'}^{p,n}(\theta'_1,\ldots,\theta'_{s'},\beta_1+i\pi,\ldots, \beta_q+i\pi; \beta'_1,\ldots, \beta'_{q'},\theta_1+i\pi,\ldots, \theta_s + i\pi) \delta_{s-q, s'-q'}\,,\nonumber
\eeqa
as long as $\theta_i \neq \theta'_i$ and $\beta_i \neq \beta'_i$ for all $i$. That is, any matrix element can be written in terms of form factors as long as there are no repeated rapidities leading to additional singularities \cite{SmirnovBook}.


\subsection{Finite Volume Matrix Elements: A Simple Example}
Once the correlation function has been expressed in terms of correlators acting on a particular sector, the latter can be computed by the insertion of a complete set of states. In finite volume both the rapidities of the excited state and  intermediate states are quantized. We will use the following simple example to explain what these quantization conditions are in general. 

Consider a simple matrix element on sector $p$ of the form
\beqa
 && {}_{p;L}\bra 0|\prod_{i=1}^k \textfrak{a}_p(\theta_i)\TT_p(0)\tilde{\TT}_p(\ell) \prod_{i=1}^k \textfrak{a}_p^\dagger(\theta_i)|0\ket_{p;L}=\nonumber\\
 && \qquad \qquad \sum_{|q\ket_p} {}_{p;L}\bra 0|\prod_{i=1}^k \textfrak{a}_p(\theta_i)\TT_p(0)|q\ket_{p;L} \times {}_{p;L}\bra q|
 \tilde{\TT}_p(\ell) \prod_{i=1}^k \textfrak{a}_p^\dagger(\theta_i)|0\ket_{p;L}\,.
 \label{exa1}
\eeqa 
We will think of this matrix element as a particular building block of a more complicated two-point function. This means that the external state $\prod_{i=1}^k \textfrak{a}_p^\dagger(\theta_i)|0\ket_{p;L}$ depends on rapidities $\{\theta_i\}$ which are the same rapidities of the original excited state $|k \ket_L$ in (\ref{statekp_basic}). 
Here $|q\ket_{p;L}$ are $q$-particle states of the form 
\beq
|q\ket_{p;L}=\prod_{i=1}^s \textfrak{a}_p^\dagger (\beta_i)\prod_{i=s+1}^q \textfrak{b}_p^\dagger (\beta_i)|0\ket_{p;L}\,,
\label{qstate}
\eeq
 and the sum over intermediate states is a sum over $q=0,\ldots,\infty$ and over $\beta_i$s. Charge conservation requires that
\beq 
2s-q=k.
\label{condition}
\eeq

In finite volume $L$  one must choose a quantization sector in order to determine the set of values the rapidities $\{\theta_i\}$ and $\{\beta_i\}$ may take. Below we choose the state $|k\ket_L$ to be in the trivial quantization sector, where the field is periodic, $\Phi_j(x+L) = \Phi_j(x)$ for all $j$.  In each copy this generates the Hilbert space ${\cal H}_{\bf 1}$. According to \eqref{intertwine}, the twist fields $\TT$ and $\t\TT$ change quantization sector as follows:
\beq
	\t\TT : {\cal H}_{\bf 1} \to {\cal H}_{\omega}\,,\qquad
	\TT  : {\cal H}_{\omega} \to {\cal H}_{\bf 1}\,,
\eeq
where ${\cal H}_{\omega}$ is the Hilbert space with quasi-periodicity condition $\Phi_i(x+L) = \Phi_{i+1}(x)$. Therefore, in the two-point function \eqref{exa1}, the intermediate states are in the quantization sector ${\cal H}_\omega$. As per \eqref{exu1}, in the diagonal basis, ${\cal H}_\omega$ has quasi-periodicity condition $\t \Phi_p(x+L) = e^{\frc{2\pi \ri p}n}\t\Phi_p(x)$. This means that the quantization of momenta (rapidities) is as follows:
\beq 
P(\theta_i)=m L \sinh \theta_i= 2 \pi I_i\quad \mathrm{with}\quad I_i\in \mathbb{Z} \quad \mathrm{and} \quad i=1, \ldots, k\,.
\label{Is}
\eeq 
for the external state (as these are the rapidities of the excited state), and
\beqa
P(\beta_i)&=&m L \sinh \beta_i = 2 \pi J_i^+ + \frac{2\pi p}{n}\quad \mathrm{with}\quad J_i^+\in \mathbb{Z} \quad \mathrm{and} \quad i=1, \ldots, s\,,
\label{Ja}\\
P(\beta_i)&=&m L \sinh \beta_i = 2 \pi J_i^- - \frac{2\pi p}{n}\quad \mathrm{with}\quad J_i^-\in \mathbb{Z} \quad \mathrm{and} \quad i=s+1, \ldots, q\,,
\label{Jb}
\eeqa
for the intermediate states (\ref{qstate}). Note that the different signs in (\ref{Ja})-(\ref{Jb}) are associated with particles created by operators $\textfrak{a}_p^\dagger(\beta_i)$ and $\textfrak{b}_p^\dagger(\beta_i)$, respectively.

These quantization conditions provide the generalization of the Bethe-Yang equations  \cite{Bethe,YangYang} (in the free case) in the presence of the branch cut induced by the $U(1)$ twist field $\TT_p$
and can be naturally extended to more general external states. 

With this information the finite-volume correlator can be expanded as (the full details of this expansion will be discussed in Section 4)
\beqa
&& {}_{p;L}\bra 0|\prod_{i=1}^k \textfrak{a}_p(\theta_i)\TT_p(0)\tilde{\TT}_p(\ell) \prod_{i=1}^k \textfrak{a}_p^\dagger(\theta_i)|0\ket_{p;L}  \nonumber\\
&&=\sum_{s=k}^\infty \frac{1}{s! (s-k)!}\sum_{\{J_i^+\}} \sum_{\{J_i^-\}} {}_{p;L}\bra 0|\prod_{i=1}^k \textfrak{a}_p(\theta_i)\TT_p(0)\prod_{r=1}^s \textfrak{a}^\dagger(\beta_r)\prod_{r=s+1}^{2s-k} \textfrak{b}^\dagger(\beta_r)|0\ket_{p;L} \nonumber\\
&&  \qquad\qquad \qquad \qquad \qquad \times\; 
{}_{p;L}\bra 0|\prod_{r=1}^s \textfrak{a}(\beta_r)\prod_{r=s+1}^{2s-k} \textfrak{b}(\beta_r)
 \tilde{\TT}_p(\ell) \prod_{i=1}^k \textfrak{a}_p^\dagger(\theta_i)|0\ket_{p;L}\,.
 \label{exa2}
\eeqa 

Although (\ref{exa2}) only shows the form factor expansion of a particular correlator, the above analysis easily extends to any other cases.  We note that the expansion (\ref{exa2}) may alternatively be expressed by replacing the sums $\sum_{\{J^{\pm}_i\}} $ by a set of contour integrals such that the sum over residues enclosed by the contours reproduces the original sum. This technique turns out to be rather useful in order to generalize the computation above to any external state. We will make full use of it in Subsection~\ref{istvan_contours}.

The final ingredient needed to evaluate (\ref{exa2}) are the finite-volume non-diagonal form factors inside the sums.
Fortunately, it is known \cite{PT1,PT2} that such matrix elements can generically be related to the infinite-volume form factors (\ref{suma}) 
simply as 
\beqa
&&  {}_{p;L}\bra 0|\prod_{i_1=1}^s \textfrak{a}_p(\theta_{i_1})\prod_{i_2=1}^q \textfrak{b}_p(\beta_{i_2})\TT_p(0)\prod_{i_4=1}^{q'} \textfrak{b}^\dagger_p(\beta'_{i_4})\prod_{i_3=1}^{s'} \textfrak{a}_p^\dagger(\theta'_{i_3})|0\ket_{p;L} = \label{ftofL}\\
&& \quad  \frac{F_{s+s'+q+q'}^{p,n}(\theta'_1,\ldots,\theta'_{s'},\beta_1+i\pi,\ldots, \beta_q+i\pi; \beta'_1,\ldots, \beta'_{q'},\theta_1+i\pi,\ldots, \theta_s + i\pi) \delta_{s-q', s'-q'}}{\sqrt{ \rho(\theta_1,\ldots,\theta_s; \beta_1,\ldots,\beta_q)\rho(\theta'_1,\ldots,\theta'_{s'}; \beta'_1,\ldots,\beta'_{q'})}} \,,
\nonumber
\eeqa
up to exponentially decaying corrections $O(e^{-\mu L})$. The functions in the denominator are the so-called density functions of the left- and right-states, respectively. In general, these  can be computed from the Bethe-Yang equations \cite{Bethe,YangYang}. However, for free theories they are simply products over the particle energies times the volume, that is,
\beq
 \rho(\theta_1,\ldots,\theta_s; \beta_1,\ldots,\beta_q)=\prod_{i_1=1}^{s} L E(\theta'_i)\prod_{i_2=1}^{q} L E(\beta'_i),
\eeq
\beq
\rho(\theta'_1,\ldots,\theta'_{s'}; \beta'_1,\ldots,\beta'_{q'})=\prod_{i_3=1}^{s'} L E(\theta_i) \prod_{i_4=1}^{q'} L E(\beta'_i)\,,
\eeq
with $E(\theta)=m\cosh\theta$.
The form factor in the numerator is exactly the same function as in the infinite volume expression (\ref{inf_vol}) up to the quantization conditions on the rapidities discussed earlier. 

We now know that finite-volume form factors are proportional to infinite-volume ones up to quantization of the rapitidities. It is worth noting here an important property of the form factor (\ref{minus}), namely its leading behaviour near the kinematic singularity. Consider the form factor $f_p^n(\beta_1-\theta_1+ i\pi)$ and suppose that the rapidites are quantized through Bethe-Yang equations of the form (\ref{Is}) for $\theta_1$ and (\ref{Ja}) for $\beta_1$. Then the leading contribution for $\theta_1 \approx \beta_1$ can be expressed as
\beq 
f_p^n(\beta_1-\theta_1+i\pi) \underset{\theta_1\approx \beta_1}{=}\frac{m L\sin\frac{\pi p}{n}\cosh\theta_1 \,e^{\frac{i\pi p}{n}}}{\pi (J_1^+-I_1+\frac{p}{n})}\,.
\label{z12}
\eeq 
Later computations will often involve the evaluation of the modulus square of $f_p^n(\theta)$ near a kinematic pole, giving rise to sums of the form 
\beq 
g_{p}^n(r)=\frac{\sin^2\frac{\pi p}{n}}{\pi^2} \sum_{J\in \mathbb Z} \frac{e^{2 \pi i r (J+ \frac{p}{n})}}{(J+ \frac{p}{n})^2}=1-(1-e^{\frac{ 2\pi i p}{n}})r\,.
\label{gjn}
\eeq 
A proof of the equality (\ref{gjn}) and a discussion of some other properties of the functions $g_p^n(r)$ is presented in Appendix \ref{thefunctionsg}.

\section{R\'enyi Entropy in the Massive Free Boson}
\label{itgetsserious}

We have now reviewed all the techniques necessary to perform a computation of the R\'enyi entropy of the configuration in Fig.~1 in a generic zero density excited state of the massive free boson theory.  Below, we describe in detail the computation for the cases of a single-particle excitation, a $k$-particle excitation with distinct rapidities, and  a $k$-particle excitation with equal rapidities. In each case we illustrate our method for $n=2$ and, for many-particle excitations, we choose the simplest state with $k=2$.
 
\subsection{Single-Particle Excited States}
\label{contours}

We will start by considering the simplest type of excited state, namely a one-particle excited state of rapidity $\theta$, that satisfies the Bethe-Yang equation \eqref{Is} with quantum number $I$. The excited state (\ref{statekp_basic}) for $k=1$ has the form 
\beqa
 |1\ket_L = \frac{1}{2^\frac{n}{2}} \prod_{j=1}^n \Big((a_j^+)^\dagger(\theta)+(a_j^-)^\dagger(\theta)\Big)|0\ket_L\,.
 \label{state1p}
\eeqa
As explained in the previous section, such a state admits a more intuitive expression after changing to the new basis of creation operators \eqref{newb}, as per \eqref{dec}. Here we write it as
\beq
|1 \ket_L=\sum_{ \{N^\pm\}} C_n\left( \{N^\pm\} \right)  \prod_{p=1}^n \left[\textfrak{a}_p^\dagger(\theta)\right]^{N_p^+}\left[\textfrak{b}_p^\dagger(\theta)\right]^{N_p^-}|0\ket_L\,, 
\label{state1p_exp}
\eeq
where the $C_n\left( \{N^\pm\} \right) $ coefficients contain all the phase factors from the transformation \eqref{newb}, and the summation runs over the integer sets $\{N^{\pm}\}=\{N^+_1,N^-_1,\dots,N^+_n,N^-_n\}$ subject to the condition 
\beq 
\sum_{p=1}^n\sum_{\epsilon=\pm}N_p^\epsilon=n\,.
\eeq 
These are the boson occupation numbers of particles/antiparticles in each sector. 
As seen before, both the branch point twist fields and generic states factorize into sectors so that the two-point function of branch point twist fields in the excited state (\ref{state1p}) at finite volume may be expressed using \eqref{state1p_exp} as 
\beqa
\,{}_{L}\bra 1 | \TT (0)\tilde{\TT}(\ell)|1 \ket_L & =& \sum_{ \{ N^{\pm}\} }\sum_{\{ \tilde{N}^{\pm}\} }[C_n(\{ N^{\pm}\} )]^*C_n(\{ \tilde{N}^{\pm}\}) \prod_{p=1}^n\mathcal{F}_{p}\left(N_{p}^{\pm},\tilde{N}_{p}^{\pm}\right)  \,,
\eeqa
where $*$ denotes complex conjugation, and
\beq
\mathcal{F}_{p}\left(N_{p}^{\pm},\tilde{N}_{p}^{\pm}\right) =\,{}_{p;L}\bra 0| (\textfrak{a}_p(\theta))^{N_p^+}(\textfrak{b}_p(\theta))^{N_p^-}\TT_p (0) \tilde{\TT}_p(\ell) (\textfrak{b}^\dagger_p(\theta))^{\tilde{N}_p^-} (\textfrak{a}^\dagger_p(\theta))^{\tilde{N}_p^+}|0\ket_{p;L}\,,
\label{eq:2pt_sector}
\eeq
is the finite-volume two-point function in sector $p$. Note that both here and later, the order of the creation and annihilation operators is irrelevant as they all commute in the free boson case.

In sector $n$, the $U(1)$ twist-fields  coincide with the identity, hence the two-point function is only nonzero if $N_n^\pm=\tilde{N}_n^\pm$, and its value is just the normalization of the finite-volume states
\beq
\mathcal{F}_{n}\left(N_{n}^{\pm},{N}_{n}^{\pm}\right)= N_n^+! N_n^-!\,.
\eeq
For other sectors however, the matrix elements (\ref{eq:2pt_sector}) are non-trivial.  As standard, they can be obtained by inserting a complete set of states between the two fields so that (\ref{eq:2pt_sector}) becomes a sum over products of the form factors (\ref{suma}). Explicitly,
\beq
I= \sum_{m^\pm=0}^\infty \sum_{J^\pm_1\le J^\pm_2\le\dots\le J^\pm_{m^\pm}} \frac{\prod_{i=1}^{m^+}\textfrak{a}_p^\dagger(\theta_i)\prod_{j=1}^{m^-}\textfrak{b}_p^\dagger(\beta_j)
 |0  \ket_{p;L} \,{}_{p;L}\bra 0| \prod_{j=1}^{m^-} \textfrak{b}_p(\beta_j) \prod_{i=1}^{m^+}
 \textfrak{a}_p(\theta_i)}{\mathcal{N}(\{J_i^+\})\mathcal{N}(\{J_i^-\})}\,,
\label{eq:complete_set_basic}
\eeq
where the rapidity sets $\{\theta_i\}$, $\{\beta_i\}$ satisfy the Bethe-Yang equations \eqref{Ja} or \eqref{Jb} with the quantum numbers $\{J_i^\pm\}$. The numbers $\mathcal{N}(\{J_i^\pm\})$ are the norms of the finite-volume states. They are different from $1$ only if there are coinciding rapidities, and every group of $s$ coinciding rapidities contributes an $s!$ factor to the norm. The restriction in the sums over quantum numbers prevents us from over-counting states in the finite-volume Hilbert-space. Alternatively, combinatorial considerations allow us to rewrite \eqref{eq:complete_set_basic} in the following simpler
form 
\beq
I= \sum_{m^{\pm}=0}^\infty \sum_{\{J^\pm \} }\frac{1}{m^+! m^- !} \prod_{i=1}^{m^+}\textfrak{a}_p^\dagger(\theta_i)\prod_{j=1}^{m^-}\textfrak{b}_p^\dagger(\beta_j)
 |0  \ket_{p;L} \,{}_{p;L}\bra 0| \prod_{j=1}^{m^-} \textfrak{b}_p(\beta_j) \prod_{i=1}^{m^+} \textfrak{a}_p(\theta_i)\,,
 \label{eq:complete_set_newform}
\eeq
without any restriction. We can now insert the complete set of states \eqref{eq:complete_set_newform}  into the two-point function (\ref{eq:2pt_sector}). Employing the  action of the translation operator on energy eigenstates, and the finite-volume form factor formulae \eqref{ftofL}, we arrive to 
\beqa
&&\mathcal{F}_{p}\left(N_{p}^{\pm},\tilde{N}_{p}^{\pm}\right) = \sum_{m^\pm=0}^\infty \sum_{\{J^\pm\}} \frac{1}{m^+!m^-!} \frac{e^{i \ell \left[\sum_{i=1}^{m^+} P(\theta_i)+\sum_{i=1}^{m^-}P(\beta_i)-(\tilde{N}_{p}^++\tilde{N}_p^-)P(\theta)\right]}}{\left[\sqrt{LE(\theta)}\right]^{{N_p^++N_p^-+\tilde{N}_p^++\tilde{N}_p^-}} \prod_{i=1}^{m^+} L E({\theta_i}) \prod_{i=1}^{m^-} L E(\beta_i)} \nonumber \\
&& \qquad \qquad \qquad \times F^{p,n}_{N_p^-+N_p^++m^-+m^+}(\underbrace{\hat{\theta},\dots,\hat{\theta}}_{N_p^-},\theta_1,\dots,\theta_{m^+};\underbrace{\hat{\theta},\dots,\hat{\theta}}_{N_p^+},\beta_1,\dots,\beta_{m^-}) \nonumber \\
&&\qquad \qquad \qquad \times F^{n-p,n}_{\tilde{N}_p^-+{\tilde{N}}_p^++m^-+m^+}(\underbrace{\theta,\dots,\theta }_{\tilde{N}_p^+},\hat{\beta}_{1},\dots,\hat{\beta}_{m^-}; \underbrace{\theta,\dots,\theta }_{\tilde{N}_p^-} , \hat{\theta}_{1} ,\dots,\hat{\theta}_{m^+})\,,
\label{eq:2pt_sector_ff}
\eeqa 
where $\hat{x}:=x+i\pi$. As seen earlier in (\ref{suma})  the form factors above are only non vanishing if 
\beq
N_p^-+m^+=N_p^++m^- \qquad \mathrm{and} \qquad \tilde{N}_p^-+m^+=\tilde{N}_p^++m^-\,,
\label{istvan_condition}
\eeq
which is equivalent to $N_p^+-{N_p}^-=m^+-m^-=\t{N}_p^+-\tilde{N_p}^-$. Note that the order of rapidities is chosen as in the definition (\ref{suma}) (this is just for convenience as for free bosons the order is irrelevant).

We will now take the expression (\ref{eq:2pt_sector_ff}) and evaluate its leading large-volume behaviour.  There are two equivalent ways of doing this which we present below. 


\subsubsection{Computation by Exact Summation over Quantum Numbers}
\label{sec:summation}

For large volume, the density factors in the denominator of (\ref{eq:2pt_sector_ff}) become large. However, if some rapidity of the intermediate states approaches the rapidity of the excited state, the kinematic poles of the form factors will give rise, due to (\ref{z12}), to positive powers of the volume. The powers in the numerator and denominator will combine to give an overall power of the volume $L$. In this section we will show that the largest such power is zero. Therefore, as $L \rightarrow \infty$ the two-point function (\ref{eq:2pt_sector_ff}) tends to a volume-independent value. There are three different cases we should investigate for a given rapidity $\theta_i$ or $\beta_i$. Recall that from (\ref{suma}) each of the form factors above consists of a large sum of products over two-particle form factors.

The first case of interest occurs when we consider the contribution to (\ref{eq:2pt_sector_ff}) of  those terms where the same rapidity $\theta_i$ is paired up with the rapidity $\theta$ (in the Wick-contraction sense of (\ref{suma})) in a two-particle form factor coming from each of the form factors in (\ref{eq:2pt_sector_ff}) . If $\theta_i \sim \theta$, then the form factor product above will be dominated by the contribution around the corresponding kinematic poles and we can write 
\beqa
&& F^{p,n}_{N_p^-+N_p^++m^-+m^+}(\ldots,\theta_1,\dots,\theta_{m^+};\underbrace{\hat{\theta},\dots,\hat{\theta}}_{N_p^+},\dots) \sim  \nonumber\\
&& N^+_p f_p^n (\theta_i-\hat{\theta})F^{p,n}_{N_p^-+N_p^++m^-+m^+-2}(\ldots,\theta_1,\dots, \check{\theta}_i,\ldots, \theta_{m^+};\underbrace{\hat{\theta},\dots,\hat{\theta}}_{N_p^+-1},\dots)\,,
\label{eq:ff_expansion_T}
\eeqa
and, similarly 
\beqa
&& F^{n-p,n}_{\tilde{N}_p^-+{\tilde{N}}_p^++m^-+m^+}(\underbrace{\theta,\dots,\theta }_{\tilde{N}_p^+}, \dots; \dots,\hat{\theta}_1, \dots, \hat{\theta}_{m^+}) \sim  \nonumber\\ 
&& \t N^+_p f_{n-p}^n (\theta-\hat{\theta}_i) 
F^{n-p,n}_{\tilde{N}_p^-+{\tilde{N}}_p^++m^-+m^+-2}(\underbrace{\theta,\dots,\theta }_{\tilde{N}_p^+-1}, \dots; \dots,\hat{\theta}_1, \dots, \check{\hat{\theta}}_i,\ldots \hat{\theta}_{m^+}, \ldots)\,,
\label{eq:ff_expansion_Ttilde}
\eeqa
where $\check{x}$ means that the variable $x$ is no longer present in the form factor. Above we kept implicit the dependence of the form factors on sets of rapidities not involved in the contraction.
The  combinatorial factors $N^+_p$ and $\t N^+_p$ come from the many pairings of $\theta_i$ with $\theta$ as per the permutation in \eqref{suma}.

The leading large-volume term from  the summation over the quantum number $J_i^+$, pertaining to the rapidity $\theta_i$, is
\beqa
\sum_{J_i^+ \in \mathbb{Z}} \frac{f_p^n (\theta_i-\hat{\theta})f_{n-p}^n (\theta -\hat{\theta}_i)e^{i\ell (P(\theta_i)-P(\theta))}}{\cosh\theta\cosh \theta_i}&\sim&(mL)^2\sum_{J_i^+ \in \mathbb{Z}} \frac{\sin^2\frac{\pi p}{n}}{\pi^2}\frac{ e^{2\pi i r (J_i^+ -I +\frac{p}{n}) }}{
(J_i^+ -I+\frac{p}{n})^2}\nonumber\\
&=&(mL)^2 g_p^n(r)\,,
\label{eq:magic_sum_+}
\eeqa
where, as before, $r=\frac{\ell}{L}$ and we used the Bethe-Yang equations \eqref{Is} and \eqref{Ja} to express the rapidites in terms of the associated quantum numbers. Here $g_p^n(r)$ are the functions (\ref{gjn}). Note that since the sum is over all integers, the value of the integer $I$ has no effect on the outcome of the sum. In other words, the result is independent of the value of the rapidity $\theta$. Similarly, for the case when some $\beta_i$ is paired up with $\theta$ in both the form factors we get
\beqa
\sum_{J_i^- \in \mathbb{Z}} \frac{f_p^n (\hat{\theta}-\beta_i)f_{n-p}^n (\hat{\beta}_i -\theta) e^{i\ell (P(\beta_i)-P(\theta))}}{\cosh\theta\cosh \beta_i}&\sim&(mL)^2\sum_{J_i^- \in \mathbb{Z}} \frac{\sin^2\frac{\pi p}{n}}{\pi^2}\frac{ e^{2\pi i r (J_i^- -I -\frac{p}{n}) }}{
(J_i^--I -\frac{p}{n})^2} \nonumber\\
&=& (mL)^2 g_{-p}^n(r)\,.
\label{eq:magic_sum_-}
\eeqa
As a consequence, if a rapidity is paired up with $\theta$ in both the form factors, the summation gives an $(mL)^2$ factor. 

The second case of interest occurs when none of the rapidities $\theta_i$, $\beta_i$ are paired up in any of the form factors with $\theta$. In this case, the large volume limit is regular, there is no kinematic singularity playing a role, and we can replace the summation over quantum numbers by integration
\beq
\sum_{J_i^{+}\in \mathbb{Z}}\sim mL \int \mathrm{d}\theta_i \qquad \mathrm{and}\qquad  \sum_{J_i^{-}\in \mathbb{Z}}\sim mL \int \mathrm{d}\beta_i\,.
\eeq
This operation generates additional factors of order $mL$ for each integral. 

Finally, there is a third case which is a mixture of the previous two, namley when $\theta_i$ or $\beta_i$ is paired up with $\theta$ in one of the form factors but with a different rapidity in the other. Due to the shifts in the Bethe-Yang equations \eqref{Is}, \eqref{Ja} and \eqref{Jb}, the summation is not singular at any value of the volume, and it can be rewritten with principal value integral
 \beq
\sum_{J_i^+}\sim mL \,  \prin \int_\theta \mathrm{d}\theta_i\qquad \mathrm{and}\qquad  \sum_{J_i^-}\sim mL \,  \prin \int_\theta \mathrm{d}\beta_i\,,
\eeq 
giving once more an $mL$ factor. 

By successively using the expansion \eqref{eq:ff_expansion_T}, \eqref{eq:ff_expansion_Ttilde} with the summations \eqref{eq:magic_sum_+} and \eqref{eq:magic_sum_-} we can calculate the overall leading large-volume contribution to the two-point function. Indeed, in Appendix \ref{bens_rules} we show that this leading large-volume contribution is of order $L^0$ and is obtained exactly when  $N^\pm_p=\tilde{N}^\pm_p$ with $N^\pm_p\leq m^\pm$, and $N^+_p$ ($N^-_p$) intermediate rapidities $\theta_i$ ($\beta_i$) are paired up with $\theta$ in both form factors. Each pairing of the rapidities gives rise to a sum of the type (\ref{gjn}) with the remaining, unpaired rapidities giving rise to form factors dependant on a smaller set of variables. Explicitly
\beqa
&& \mathcal{F}_{p}\left(N_{p}^{\pm},N_{p}^{\pm}\right) = \sum_{q^\pm=0}^\infty \sum_{\{J^\pm\} \in \mathbb{Z}} \frac{1}{(q^+ +N^+_p)!(q^- +N^-_p)!} \frac{e^{i \ell \left[\sum_{i=1}^{q^+}P(\theta_i)+\sum_{i=1}^{q^-}P(\beta_i)\right]}}{\prod_{i=1}^{q^+} L E(\theta_i) \prod_{i=1}^{q^-} L E(\beta_i)} \nonumber \\
&& \qquad \qquad \times F^{p,n}_{q^+ + q^-}(\theta_1,\dots,\theta_{q^+};\beta_1,\dots,\beta_{q^-})  F^{n-p,n}_{q^++q^-}(\hat{\beta}_{1},\dots,\hat{\beta}_{q^-};  \hat{\theta}_{1},\dots,\hat{\theta}_{q^+}) \nonumber \\
&&  \qquad \qquad \times (N^+_p!)^2 (N^-_p!)^2 \binom{q^++N^+_p}{N^+_p}\binom{q^-+N^-_p}{N^-_p} \left[g_{p}^n(r)\right]^{N^+_p}\left[g_{-p}^n(r)\right]^{N^-_p}\,,
\label{eq:2pt_sector_ff_res}
\eeqa 
where $q^\pm=m^\pm-N^\pm_p$ is the number of remaining intermediate state rapidities after the contractions. The factorials in the denominator are just $m^\pm!$, that came from the complete set of state insertion.  Out of $m^\pm$ original intermediate rapidities,  $N^\pm_p$ are paired up with the rapidity $\theta$ in the sense described earlier. All  particular pairing choices are equivalent to each other under relabelling of the rapidities, as they are all integrated over, that is counted by the binomial factors.   The $N^\pm_p!$  combinatorial factors arise from the pairing of the chosen intermediate rapidities to $\theta$ in the form factors, as explained in \eqref{eq:ff_expansion_T} and \eqref{eq:ff_expansion_Ttilde}. Once all possible contractions with a rapidity $\theta$ have been carried out, two form factors will still remain depending on $q^++q^-$ rapidities. In addition, we know from (\ref{suma}) that only form factors with $q^+=q^-=q$ are non-vanishing. Simplifying we obtain
  \beqa
\mathcal{F}_{p}\left(N_{p}^{\pm},N_{p}^{\pm}\right) &=& N^+_p! N^-_p! \left[g_{p}^n(r)\right]^{N^+_p}\left[g_{-p}^n(r)\right]^{N^-_p} \sum_{p=0}^\infty \frac{1}{(q!)^2} \sum_{\{J^\pm\} \in \mathbb{Z}}  \frac{e^{i \ell \sum_{i=1}^{q}\left(P(\theta_i)+P(\beta_i)\right)}}{\prod_{i=1}^{q} L^2 E(\theta_i) E(\beta_i)} \nonumber \\
&&  \times F^{p,n}_{2q}(\theta_1,\dots,\theta_{q};\beta_1,\dots,\beta_{q})  F^{n-p,n}_{2q}(\hat{\beta}_{1},\dots,\hat{\beta}_{q};  \hat{\theta}_{1},\dots,\hat{\theta}_{q})\,.
\label{eq:2pt_sector_ff_res2}
\eeqa 
Aside from the prefactor $N^+_p! N^-_p! \left[g_{p}^n(r)\right]^{N^+_p}\left[g_{-p}^n(r)\right]^{N^-_p}$, the expression above exactly reproduces the finite-volume 
vacuum two-point function in the given sector, i.e. $\,{}_{p,L}\bra 0 | \TT_p (0)\tilde{\TT}_p(\ell)|0\ket_{p,L} $. As a consequence, our end result for the finite-volume two-point function can be expressed as
\beq
\frac{\,{}_{L}\bra 1 | \TT (0)\tilde{\TT}(\ell)|1 \ket_L }{\,{}_{L}\bra 0 | \TT (0)\tilde{\TT}(\ell)|0\ket_L }=\sum_{ \{ N^{\pm}\} }|C_n(\{ N^{\pm}\} )|^2\prod_{p=1}^n\prod_{\epsilon=\pm} (N^\epsilon_p!) \left(g_{{\epsilon p}}^n(r)\right)^{N^\epsilon_p}+\mathcal{O}(L^{-1})\,.
\label{eq:2pt_full_result}
\eeq
In particular, for $p=n$, the factor reproduces the norm of the finite-volume state as expected, since $g_{\pm n}(r)=1$ and $\,{}_{n;L}\bra 0 | \TT_n (0)\tilde{\TT}_n(\ell)|0\ket_{n;L}=1$.

 
\subsubsection{Computation by Contour Integration} 
\label{istvan_contours}

Another way of calculating the leading large-volume term of the two-point function in a given sector \eqref{eq:2pt_sector_ff} is to transform the summation over quantum numbers of the intermediate states into contour integrals. This approach not only leads to the same result (\ref{eq:2pt_full_result}) but seems more amenable to generalization to interacting theories, something we would like to attempt in future work. Consider generic sums of the form
\beq
\sum_{J^+_i \in \mathbb{Z}} \frac{h^+(\theta_i, \ldots )}{L E(\theta_i)} =\sum_{J^+_i} \int_{\mathcal{C}_{J^+_i}}\frac{\mathrm{d} \tilde{\theta}_i }{2\pi} \frac{h(\tilde{\theta}_i,\dots )}{e^{i(L P(\tilde{\theta}_i)-\frac{2\pi p}{n}) }-1}\,, 
\eeq
and
\beq
\sum_{J^-_i \in \mathbb{Z}} \frac{h^-(\beta_i, \ldots )}{L E(\beta_i)} =\sum_{J^-_i} \int_{\mathcal{C}_{J^-_i}}\frac{\mathrm{d} \tilde{\beta}_i }{2\pi} \frac{h(\tilde{\beta}_i,\dots )}{e^{i(L P(\tilde{\beta}_i)+\frac{2\pi p}{n}) }-1}\,, 
\eeq
where $h^{\pm}$ are functions that are regular at the positions $\theta_i$, $\beta_i$, respectively and  $\mathcal{C}_{J^\pm_i}$ is a small contour encircling $\theta_i, \beta_i$ with positive orientation, and the denominators inside the integrals are the exponential form of the Bethe-Yang equations \eqref{Ja} and \eqref{Jb}, that is zero at every solution of the equations. From now onwards we will omit the tilde on the integration variables. 

Transforming every sum in  \eqref{eq:2pt_sector_ff} into a contour integral we obtain the expression
\beqa
\mathcal{F}_{p}\left(N_{p}^{\pm},\tilde{N}_{p}^{\pm}\right) &= &\sum_{m^\pm=0}^\infty  \frac{1}{m^+!m^-!} \frac{1}{{\left[\sqrt{LE(\theta)}\right] }^{N_p^++N_p^-+\tilde{N}_p^++\tilde{N}_p^-} } \left[\prod_{i=1}^{m^+}  \sum_{J^+_i\in \mathbb{Z}}\int_{\mathcal{C}_{J^+_i}}\frac{\mathrm{d} \theta_i }{2\pi}\right]
\nonumber \\
&&\times  \left[ \prod_{k=1}^{m^-}\sum_{J^-_k\in \mathbb{Z}}\int_{\mathcal{C}_{J^-_k}}\frac{\mathrm{d} \beta_k }{2\pi}\right]\frac{e^{i \ell \left[\sum_{i=1}^{m^+}P({\theta_i})+\sum_{i=1}^{m^-}P({\beta_i})-(\tilde{N}_{p}^++\tilde{N}_{p}^-)P(\theta)\right]}}{\prod_{i=1}^{m^+} [e^{i(L P(\theta_i) -\frac{2\pi p}{n}) }-1] \prod_{i=1}^{m^-} [e^{i(L P(\beta_i) +\frac{2\pi p}{n}) }-1]} \nonumber \\
&& \times F^{p,n}_{N_p^++N_p^-+m^++m^-}(\underbrace{\hat{\theta},\dots,\hat{\theta}}_{N_p^-},\theta_1,\dots,\theta_{m^+};\underbrace{\hat{\theta},\dots,\hat{\theta}}_{N_p^+},\beta_1,\dots,\beta_{m^-}) \nonumber \\
&&\times F^{n-p,n}_{\tilde{N}_p^++\tilde{N}_p^-+m^++m^-}(\underbrace{\theta,\dots,\theta }_{\tilde{N}_p^+},\hat{\beta}_{1},\dots,\hat{\beta}_{m^-}; \underbrace{\theta,\dots,\theta }_{\tilde{N}_p^-} , \hat{\theta}_{1} ,\dots, \hat{\theta}_{m^+})\,. 
\label{eq:2pt_sector_ff_contour}
\eeqa
Our next step is to combine the small contours around the Bethe-Yang solutions into a contour encircling the real axis for each variable. While doing so, the contour will cross the kinematic poles of the form factors, whenever $\theta_i=\theta$ or $\beta_i=\theta$ for some $i$, and we need to account for the residues of these poles.

It is easy to see from \eqref{kinu}, that the contribution from residues at $\theta$ coming from a single kinematic singularity is of order $L^0$ in the volume and therefore they will be strongly suppressed by the power of $L$ in the denominator of (\ref{eq:2pt_sector_ff_contour}). However, if we consider terms where both form factors have a kinematic pole at the same location $\theta_i=\theta$ or $\beta_i=\theta$, then we have to calculate the residue of a second order pole, and this can change the order in the volume. Let us calculate this residue for a particular rapidity $\theta_i$
\beqa
&&-\int_{\mathcal{C}_{J^+_i}}\frac{\mathrm{d} \theta_i }{2\pi}\frac{e^{i \ell \left(P({\theta_i})-P({\theta})\right)}}{ e^{i(L P(\theta_i) -\frac{2\pi p}{n}) }-1}  F^{p,n}_{N_p^++N_p^-+m^++m^-}(\dots,\theta_i,\dots;\underbrace{\hat{\theta},\dots,\hat{\theta}}_{N_p^+},\dots) \nonumber \\
&&\qquad \qquad \qquad\qquad \qquad \qquad \times F^{n-p,n}_{\tilde{N}_p^++\tilde{N}_p^-+m^++m^-}(\underbrace{\theta,\dots,\theta }_{\tilde{N}_p^+},\dots; \dots, \hat{\theta}_{i},\dots)\,.
\eeqa
Recall that here, as earlier hatted variables are variables shifted by $i\pi$. From the kinematic residue equation (\ref{kinu}) it follows that near the kinematic poles the integrand may be approximated as
\beqa
&&-\int_{\mathcal{C}_{J^+_i}}\frac{\mathrm{d} \theta_i }{2\pi}\frac{e^{i \ell \left(P({\theta_i})-P({\theta})\right)}}{ e^{i(L P(\theta_i) -\frac{2\pi p}{n}) }-1}  \frac{-i N_p^+\left(1-e^{-\frac{2\pi i p}{n}}\right)}{\theta_i-\theta} \frac{-i\tilde{N}_p^+\left(1-e^{\frac{2\pi i p}{n}}\right)}{\theta-\theta_i}  \nonumber \\
&&\quad \qquad \qquad\qquad \qquad \qquad \times  F^{p,n}_{N_p^++N_p^-+m^++m^--2}(\dots,\check{\theta}_i,\dots;\underbrace{\hat{\theta},\dots,\hat{\theta}}_{N_p^+-1},\dots)\nonumber\\
&&\qquad \quad \qquad\qquad \qquad \qquad \times 
F^{n-p,n}_{\tilde{N}_p^++\tilde{N}_p^-+m^++m^--2}(\underbrace{\theta,\dots,\theta }_{\tilde{N}_p^+-1},\dots; \dots, \check{\hat{\theta}}_{i},\dots)\,.
\eeqa
Evaluating the corresponding residue we obtain
\beqa
&& -i N_p^+ \tilde{N}_p^+\left(1-e^{\frac{2\pi i p}{n}}\right)\left(1-e^{-\frac{2\pi i p}{n}}\right)\frac{\mathrm{d}}{\mathrm{d} \theta_i}\left(\frac{e^{i\ell \left(P{\theta_i)}-P({\theta})\right)}}{e^{i(LP(\theta_i) -\frac{2\pi p}{n}) }-1}\right)_{\theta_i=\theta}\nonumber \\
&&\times F^{p,n}_{N_p^++N_p^-+m^++m^--2}(\dots,\check{\theta}_i,\dots;\underbrace{\hat{\theta},\dots,\hat{\theta}}_{N_p^+-1},\dots) 
\nonumber \\
&&\times F^{n-p,n}_{\tilde{N}_p^++\tilde{N}_p^-+m^++m^--2}(\underbrace{\theta,\dots,\theta }_{\tilde{N}_p^+-1},\dots;\dots, \check{\hat{\theta}}_{i},\dots)\,,
 \eeqa
 where the checked variables are absent.  Simplifying, the final result is
 \beqa
 && L E(\theta) N_p^+ \tilde{N}_p^+ g_p^n(r)  F^{p,n}_{N_p^++N_p^-+m^++m^--2}(\dots,\check{\theta}_i,\dots;\underbrace{\hat{\theta},\dots,\hat{\theta}}_{N_p^+-1},\dots) \nonumber\\
 &&\qquad \qquad \qquad\qquad  \times 
  F^{n-p,n}_{\tilde{N}_p^++\tilde{N}_p^-+m^++m^--2}(\underbrace{\theta,\dots,\theta }_{\tilde{N}_p^+-1},\dots; \dots, \check{\hat{\theta}}_{i},\dots)\,,
\eeqa
where we also used the Bethe-Yang equation \eqref{Is}, and the $N_p^+$, $\tilde{N}_p^+$ combinatorial factors are the result of the pairing of $\theta_i$ with the $\theta$s. It is important to note, that the result is proportional  to the volume, and also to the function $g_p^n(r)$ introduced in \eqref{gjn}. An entirely similar computation, for a rapidity $\beta_i$ gives the result
\beqa
&&-\int_{\mathcal{C}_{J_i^-}}\frac{\mathrm{d} \beta_i }{2\pi}\frac{e^{i \ell \left(P({\beta_i})-P({\beta})\right)}}{ e^{i(L P(\beta_i) +\frac{2\pi p}{n}) }-1}  F^{p,n}_{N_p^++N_p^-+m^++m^-}(\underbrace{\hat{\theta},\dots, \hat{\theta}}_{N_p^-},\dots;\dots,\beta_i,\dots) \nonumber \\
&&\qquad \qquad \qquad\qquad \qquad \qquad  \times  F^{n-p,n}_{\tilde{N}_p^++\tilde{N}_p^-+m^++m^-}(\dots,\hat{\beta_{i}},\dots; \underbrace{\theta,\dots,\theta }_{\tilde{N}_p^-} ,\dots) \nonumber \\
&=& LE(\theta) N_p^- \tilde{N}_p^- g_{-p}^n(r)  F^{p,n}_{N_p^++N_p^-+m^++m^--2}(\underbrace{\hat{\theta},\dots, \hat{\theta}}_{N_p^--1},\dots;\dots,\check{\beta}_i,\dots) \nonumber\\
&& \qquad \qquad \qquad\qquad   \times F^{n-p,n}_{\tilde{N}_p^++\tilde{N}_p^-+m^++m^--2}(\dots,\check{\hat{\beta}}_{i},\dots; \underbrace{\theta,\dots,\theta }_{\tilde{N}_p^--1} ,\dots)\,.
\eeqa
As a consequence of these residues, we get the leading large-volume contribution to the two-point function, if we pick up  the largest possible number of residues of second order poles which are enveloped as the contour is deformed. The maximum number of  second order poles is $\min(N_p^\pm,\tilde{N}_p^\pm)$, that implies, that $m^\pm \geq \max(N_p^\pm,\tilde{N}_p^\pm)$. These terms have an $e^{mL R}$ dependence on the volume with
\beq
R=\min(N^+_p,\tilde{N}^+_p)+\min(N^-_p,\tilde{N}^-_p)-\frac{N^+_p+N^-_p+\tilde{N}^+_p+\tilde{N}^-_p}{2}\,.
\eeq
As argued more generally in Appendix \ref{bens_rules} (the formula above can be seen as an especialization of equation (\ref{b4}) in  Appendix B)
the leading contribution is obtained  when $N^\pm_j=\tilde{N}^\pm_j$, and in that case $R=0$. The leading large-volume term of the two-point function then becomes
\beqa
\!\!\!\mathcal{F}_{p}\left(N_{p}^{\pm},N_{p}^{\pm}\right) &= &\sum_{q^\pm=0}^\infty  \frac{(N^+_p!)^2 (N^-_p!)^2 \binom{q^++N^+_p}{N^+_p}\binom{q^-+N^-_p}{N^-_p}}{(q^++N^+_p)!(q^-+N^-_p)!}  \left[g_{p}^n(r)\right]^{N^+_p}\left[g_{-p}^n(r)\right]^{N^-_p} \left[\prod_{i=1}^{q^+} \int_{\mathcal{C}_{\leftrightarrows}}\frac{\mathrm{d} \theta_i }{2\pi}\right] \nonumber \\
&&\times \left[\prod_{i=1}^{q^-} \int_{\mathcal{C}_{\leftrightarrows}}\frac{\mathrm{d} \beta_i }{2\pi}\right] \frac{e^{i \ell \left(\sum_{i=1}^{q^+}P(\theta_i)+\sum_{i=1}^{q^-}P(\beta_i)\right)}}{\prod_{i=1}^{p^+} (e^{i(L P(\theta_i) -\frac{2\pi p}{n}) }-1) \prod_{i=1}^{q^-} (e^{i(L P(\beta_i) +\frac{2\pi p}{n}) }-1)}\label{eq:2pt_sector_ff_contour_res}
\\
&& \times F^{p,n}_{q^+ + q^-}(\theta_1,\dots,\theta_{q^+};\beta_1,\dots,\beta_{q^-})  F^{n-p,n}_{q^++q^-}(\hat{\beta}_{1},\dots,\hat{\beta}_{q^-};  \hat{\theta}_{1},\dots,\hat{\theta}_{q^+})\,,\nonumber
\eeqa
where $\mathcal{C}_{\leftrightarrows}$ denotes the contour encircling the real axis, $q^\pm=m^\pm-N_p^\pm$, and the combinatorial factors came from counting the various choices of intermediate rapidities giving rise to double pole residue integrals. Simplifying the combinatorial factors and noticing that $q^+=q^-=q$ for the form factors above to be non-vanishing, we can easily factor out the vacuum two-point function from the expression above and we obtain once more 
the result \eqref{eq:2pt_full_result}. 

 
\subsubsection{Example: 2nd R\'enyi Entropy of a Single-Particle Excitation}

Let us illustrate the general methods above with the simplest example: we compute the 2nd R\'enyi Entropy, i.e $n=2$, of a single-particle excited state.  From \eqref{state1p} we can easily write down the state 
\beqa
|1\ket_L &=& \frac{1}{4}  \textfrak{a}_2^\dagger(\theta)\textfrak{a}_2^\dagger (\theta)|0\ket_{2;L} + \frac{1}{4}  \textfrak{b}_2^\dagger(\theta)\textfrak{b}_2^\dagger (\theta)|0\ket_{2;L}  +\frac{1}{2}   \textfrak{a}_2^\dagger(\theta)\textfrak{b}_2^\dagger (\theta)|0\ket_{2;L} \nonumber\\
&& - \frac{1}{4}  \textfrak{a}_1^\dagger(\theta)\textfrak{a}_1^\dagger (\theta)|0\ket_{1;L}- \frac{1}{4}  \textfrak{b}_1^\dagger(\theta)\textfrak{b}_1^\dagger (\theta)|0\ket_{1;L}- \frac{1}{2}  \textfrak{a}_1^\dagger(\theta)\textfrak{b}_1^\dagger (\theta)|0\ket_{1;L} \nonumber\\
&=& \frac{1}{4}\left[(\textfrak{a}_2^\dagger(\theta)+\textfrak{b}_2^\dagger(\theta))^2-(\textfrak{a}_1^\dagger(\theta)+\textfrak{b}_1^\dagger(\theta))^2\right]|0\ket_L\,,
\label{eq:excited_state_n2}
\eeqa 
and identify the nonzero coefficients $C_2(N_1^+,N_1^-,N_2^+,N_2^-)$ of the expansion \eqref{state1p_exp} as
\beqa
C_2(2,0,0,0)=-\frac{1}{4}\,, &&C_2(0,0,2,0)=\frac{1}{4}\,, \nonumber \\
C_2(0,2,0,0)=-\frac{1}{4}\,,&&C_2(0,0,0,2)=\frac{1}{4}\,, \nonumber \\
C_2(1,1,0,0)=-\frac{1}{2}\,,&&C_2(0,0,1,1)=\frac{1}{2}\,.
\label{coefficients}
\eeqa
These can be directly plugged into \eqref{eq:2pt_full_result} 
\beqa
\lim_{L\rightarrow \infty} \frac{\,{}_{L}\bra 1 | \TT (0)\tilde{\TT}(rL)|1\ket_L }{\,{}_{L}\bra 0 | \TT (0)\tilde{\TT}(\ell)|0\ket_L }&=&\frac{2!}{16} \left[g_{1}^2(r)\right]^2+\frac{2!}{16} \left[g_{-1}^2(r)\right]^2+ \frac{1}{4} g_{1}^2(r) g_{-1}^2(r)\nonumber\\
&+& \frac{2!}{16}\left[g_{2}^2(r)\right]^2+\frac{2!}{16}\left[g_{-2}^2(r)\right]^2+\frac{1}{4}g_{2}^2(r)g_{-2}^2(r)\nonumber \\
&=&\frac{1}{2}+\frac{1}{2} [g_1^2(r)]^2=r^2+(1-r)^2\,,
\label{example1p}
\eeqa
where we used the fact that $g_2^2(r)=g_{-2}^{2}(r)=1$ and $g_1^2(r)=g_{-1}^2(r)=1-2r$. 
Therefore the difference of R\'enyi entropies is
\beq
 \Delta S_2^1(r)=-\log(r^2+(1-r)^2)\,,
\eeq 
which agrees with the expression (\ref{ren}) for $n=2$. This is also exactly the second R\'enyi entropy of the two qubit state (\ref{2qubit}).

 
\subsection{Multi-Particle Excited States}
\label{sec:multi_part}
In this section we adapt the techniques presented for the one-particle excited state case to more general states involving both distinct and equal rapidities. As we will see the essential ideas are the same but the state is more involved which makes the combinatorics of the problem more complicated.

 
\subsubsection{Distinct Rapidities}

Let us denote a general $k$-particle state \eqref{statekp_basic}  involving only distinct rapidity excitations as $ |\underbrace{1,1,\ldots,1}_k\ket_L$. It can be expressed similarly as \eqref{state1p_exp} in the form
\beqa
 |\underbrace{1,1,\ldots,1}_k\ket_L =\prod_{q=1}^k \sum_{ \{N^{q,\pm}\} } C_n\left( \{N^{q,\pm}\} \right)  \prod_{p=1}^n \left[\textfrak{a}_p^\dagger(\theta_q)\right]^{N_{p}^{q,+}}\left[\textfrak{b}_p^\dagger(\theta_q)\right]^{N_{p}^{q,-}}|0\ket_L\,, 
\label{eq:multi_k_part_state}
\eeqa
where the $C_n\left( \{N^{q,\pm}\} \right) $ coefficients are all identical for each value of $q$ (the state is invariant under relabelling of the rapidities). For fixed $q$ they are exactly the same as for the one-particle state.  We have the following restrictions for the integers
\beq 
\sum_{p=1}^n\sum_{\epsilon=\pm}N_p^{q,\epsilon}=n\,,
\label{eq:multiplicity_restriction}
\eeq 
for all $q$. The two-point function takes the form 
\beqa
&&\,{}_{L}\bra {1,1,\ldots,1}| \TT (0)\tilde{\TT}(\ell) |1,1,\ldots,1\ket_L  \\
&&=\lt[\prod_{q=1}^k\sum_{ \{ N^{q,\pm}\} }\sum_{\{ \tilde{N}^{q,\pm}\} }[C_n(\{ N^{q,\pm}\} )]^*C_n(\{ \tilde{N}^{q,\pm}\})\rt]  \prod_{p=1}^n\mathcal{F}_{p}\left(\{N_{p}^{q,\pm}\},\{\tilde{N}_{p}^{q,\pm}\}\right) \,,
\label{eq:2pt_multi}
\eeqa
where
\beqa
&&\mathcal{F}_{p}\left(\{N_{p}^{q,\pm}\},\{\tilde{N}_{p}^{q,\pm}\}\right)\\
&& = \,{}_{p;L}\bra 0| \left[\prod_{q=1}^k (\textfrak{a}_p(\theta_q))^{N_p^{q,+}}(\textfrak{b}_p(\theta_q))^{N_p^{q,-}}\right]\TT_p (0) 
 \tilde{\TT}_p(\ell) \left[\prod_{q=1}^k (\textfrak{b}^\dagger_p(\theta_q))^{\tilde{N}_p^{q,-}} (\textfrak{a}^\dagger_p(\theta_q))^{\tilde{N}_p^{q,+}}\right]|0\ket_{p;L}\,.\nonumber
\label{eq:2pt_sector_multi}
\eeqa
To find the leading contribution in the volume to $\mathcal{F}_{p}\left(\{N_{p}^{q,\pm}\},\{\tilde{N}_{p}^{q,\pm}\}\right)$, we follow the same steps as in Section  \ref{contours}. As seen in Subsections~\ref{sec:summation} and \ref{istvan_contours}, we need to focus on the contributions arising when some  intermediate rapidity approaches one of the rapidities of the excited state in both of the form factors. In other words, we need to pair up the intermediate rapidities with the same rapidity of the excited state from the in- and out-states. This mechanism singles out the leading large-volume contribution as corresponding to $N_p^{q,\pm}=\tilde{N}_p^{q,\pm}$ for every $q$. Carrying out the calculation, the combinatorial factors simplify to yield the  result 
\beqa
&&\mathcal{F}_{p}\left(\{N_{p}^{q,\pm}\},\{\tilde{N}_{p}^{q,\pm}\}\right)\nonumber \\
&&  =\prod_{q=1}^k N^{q,+}_p! N^{q,-}_p! \left[g_{p}^n(r)\right]^{N^{q,+}_p}\left[g_{-p}^n(r)\right]^{N^{q,-}_p}  \,{}_{p;L}\bra 0 | \TT (0)\tilde{\TT}(\ell)|0\ket_{p;L} + O(L^{-1})\,. \label{factor}
\eeqa
 As a consequence, in the infinite volume limit, the result for a state involving $k$ distinct rapidities factorizes into $k$ single-particle state contributions. That is
\beqa
 \lim_{L\rightarrow \infty} \frac{\,{}_{L}\bra 1,1,\ldots | \TT (0)\tilde{\TT}(rL)|1,1,\ldots \ket_L }{\,{}_{L}\bra 0 | \TT (0)\tilde{\TT}(\ell)|0 \ket_L } &=&\prod_{q=1}^k\left[\sum_{ \{ N^{q,\pm}\} }|C_n(\{ N^{q,\pm}\} )|^2\prod_{p=1}^n\prod_{\epsilon=\pm} N^{q,\epsilon}_p! \left[g_{\epsilon p}^n(r)\right]^{N^{q,\epsilon}_p}\right]  \nonumber \\
&=&\lim_{L\rightarrow \infty}  \left[\frac{\,{}_{L}\bra 1 | \TT (0)\tilde{\TT}(rL)|1 \ket_L }{\,{}_{L}\bra 0 | \TT (0)\tilde{\TT}(\ell)|0 \ket_L }\right]^k\,.
\label{eq:2pt_multi_fin}
\eeqa
This in turn leads to the relation 
\beq
\Delta S^{1,1,\ldots}_n(r)=\sum_{q=1}^k \Delta S^1_n(r)=k \Delta S^1_n(r)\,,
\label{facds}
\eeq
which is a special case of the formula (\ref{general}). 


\subsubsection{Coinciding rapidities}
\label{coincide}

The simple result (\ref{facds}) no longer holds if all or some rapidities of the excited state coincide. Let us consider a $k$-particle excited state where all the rapidities coincide, and are denoted by $\theta$. In this case the norm of the $k$-particle state  as written in \eqref{eq:multi_k_part_state} is $k!^n$, thus the normalization needs to be appropriately modified. The properly normalized state can then be written as
\beqa
 |k\ket_L =\frac{1}{\sqrt{k!}^n} \sum_{ \{N^\pm\}} D_n^k\left( \{N^\pm\} \right)  \prod_{p=1}^n \left[\textfrak{a}_p^\dagger(\theta)\right]^{N_p^+}\left[\textfrak{b}_p^\dagger(\theta)\right]^{N_p^-}|0\ket_L\,,
\eeqa
which looks very much like the one-particle state (\ref{state1p_exp}). This is not too surprising as both states depend on a single rapidity variable. The coefficients $D_n^k\left( \{N^\pm\} \right)$ are related to the coefficients $C_n\left( \{N^\pm\} \right)$ of the previous subsections by
\beq
D_n^k(\{N^\pm\})=\prod_{q=1}^k\sum_{ \{ N^{q,\pm}\}  } C_n(\{N^{q,\pm}\})\prod_{p=1}^n\prod_{\epsilon=\pm}\delta_{N^{\epsilon}_p,\sum_{q=1}^k N^{q,\epsilon}_p}\,.
\label{eq:new_coeff}
\eeq
This relation is of practical use when evaluating our formulae with the help of algebraic manipulation software. 
 The two point function is then
\beqa
\,{}_{L}\bra k | \TT (0)\tilde{\TT}(\ell)|k \ket_L  =\frac{1}{(k!)^n}\sum_{ \{ N^{\pm}\} }\sum_{\{ \tilde{N}^{\pm}\} }[D^k_n(\{ N^{\pm}\} )]^*D^k_n(\{ \tilde{N}^{\pm}\}) \prod_{p=1}^n\mathcal{F}_{p}\left(N_{p}^{\pm},\tilde{N}_{p}^{\pm}\right) \,,
\label{eq:2pt_multi_coincide}
\eeqa
where $\mathcal{F}_p$ is the same function as for the one-particle case \eqref{eq:2pt_sector}, but now the integers $N^{\pm}_p$ obey the selection rule
\beq
\prod_{p=1}^n\prod_{\epsilon=\pm}N^{\epsilon}_p=nk\,,
\eeq
which depends on the number of excitations $k$, and the same condition holds for $\tilde{N}^{\pm}_p$. 
The leading large-volume term of the two-point function then becomes
\beqa
\!\!\lim_{L\rightarrow \infty}\frac{\,{}_{L}\bra k | \TT (0)\tilde{\TT}(rL)|k \ket_L }{\,{}_{L}\bra 0 | \TT (0)\tilde{\TT}(rL)|0 \ket_L } = \frac{1}{(k!)^n}\sum_{ \{ N^{\pm}\} }|D_n^k(\{ N^{\pm}\} )|^2  \prod_{p=1}^n\prod_{\epsilon=\pm} \left(N^{\epsilon}_p\right)! \left[g_{\epsilon p}^n(r)\right]^{N^{\epsilon}_p}\,.
\label{eq:2pt_multi_coincide_res}
\eeqa
Explicit evaluation of this product for specific values of $k$ and $n$ then leads to the result (\ref{11}).

\subsubsection{The General Case}
\label{general_case}
The techniques we have just presented for states of distinct and equal rapidities can be easily adapted to deal with more general states: states where some rapidities are equal and other distinct. As expected, the EE difference for a multi-particle mixed state is a sum over the EEs of simpler states associated with groups of coinciding rapidities. This result is expressed by the formula (\ref{general}).

Regarding the results of this section overall, it is worth noting that we do not yet have closed formula for  coefficients $C_n(\{N^\pm\})$ and $D_n^k(\{N^\pm\})$ for general $n$, however it is straightforward to calculate them systematically on the computer and we have done this up to $n=6$ for two coinciding rapidities and up to smaller values of $n$ as the number of coinciding rapidities was increased to $k=6$. Once the coefficients are known we can easily evaluate formula \eqref{eq:2pt_multi_coincide_res} for several values  of $k$, and we observe that the results are always polynomials that have  $r\leftrightarrow 1-r$ symmetry as expected. It was by working out such particular examples that we were eventually able to establish the general pattern  \eqref{ren}-\eqref{3}.

\subsubsection{Example: 2nd R\'enyi Entropy of a Two-Particle Excitation}
In order to make the results above more concrete, we will now consider the EE of two-particle excited states both with distinct and with equal rapidities. 
The non-trivial part of the computation is in the characterization of the states, namely the computation of the coefficients $C_n(\{N^\pm\})$ and $D_n^k(\{N^\pm\})$ as arising in the formulae (\ref{eq:multi_k_part_state}) and (\ref{eq:2pt_multi_coincide_res}). Once these are know the EEs can be systematically obtained for any state. 

Let us consider a two-particle excited state with distinct rapidities which we represent as $|1,1\ket_L$. From the general expression (\ref{dec}) it is easy to see that
\beqa 
\!\!\!\!\!\!|1,1\ket_L
&=& \frac{1}{4}\left[(\textfrak{a}^\dagger_2(\theta_1)+\textfrak{b}^\dagger_2(\theta_1))^2-(\textfrak{a}^\dagger_1(\theta_1)+\textfrak{b}^\dagger_1(\theta_1))^2\right]\nonumber\\
& \times &
\frac{1}{4}\left[(\textfrak{a}^\dagger_2(\theta_2)+\textfrak{b}^\dagger_2(\theta_2))^2-(\textfrak{a}^\dagger_1(\theta_2)+\textfrak{b}^\dagger_1(\theta_2))^2\right]|0\ket_L\,.
\label{2dis2}
\eeqa 
The state can be fully characterized by the coefficients $C_2(\{N^{q,\pm}\})$ with $q=1,2$
 and these give two copies of the coefficients (\ref{coefficients}) of the one-particle state (\ref{eq:excited_state_n2}).
Substituting these values into the formula we obtain exactly the square of (\ref{example1p}), that is
\beqa
 \lim_{L\rightarrow \infty} \frac{{}_L\bra 1,1|\TT(0)\tilde{\TT}(r L)|1,1\ket_L}{\bra 0|\TT(0)\tilde{\TT}(\ell)|0\ket}  =\left[\frac{1}{2}+ \frac{1}{2} [g_1^2(r)]^2\right]^2=\left[r^2+(1-r)^2\right]^2\,.
  \label{fun1}
\eeqa 
 Consider  instead a two-particle excited state of equal rapidities. The state may be written as 
\beqa 
\!\!\!\!\!\!|2\ket_L
&=& \frac{1}{2!}\left[\frac{1}{4}\left[(\textfrak{a}^\dagger_2(\theta)+\textfrak{b}^\dagger_2(\theta))^2-(\textfrak{a}^\dagger_1(\theta)+\textfrak{b}^\dagger_1(\theta))^2\right]\right]^2|0\ket_L\,.
\label{dec3}
\eeqa 
The coefficients $D_2^2(N_1^+,N_1^-,N_2^+,N_2^-)$ entering the formula (\ref{eq:2pt_multi_coincide_res}) can be read off by either expanding (\ref{dec3}) and looking at the coefficients of all distinct states in the ensuing linear combination, or by using (\ref{eq:new_coeff})
\begin{align}
 D_2^2(4,0,0,0)&=\frac{1}{16}\,, &  D_2^2(0,4,0,0)&=\frac{1}{16}\,, & D_2^2(0,0,4,0)&=\frac{1}{16}\,, & D_2^2(0,0,0,4)&=\frac{1}{16}\,,\nonumber \\  
  D_2^2(2,0,2,0)&=-\frac{1}{8}\,, & D_2^2(2,0,0,2)&=-\frac{1}{8}\,, & D_2^2(0,2,2,0)&=-\frac{1}{8}\,, & D_2^2(0,2,2,0)&=-\frac{1}{8}\,,\nonumber \\ 
 D_2^2(3,1,0,0)&=\frac{1}{4}\,, &D_2^2(1,3,0,0)&=\frac{1}{4}\,, & D_2^2(0,0,3,1)&=\frac{1}{4}\,, & D_2^2(0,0,1,3)&=\frac{1}{4}\,,\nonumber \\ 
  D_2^21,1,2,0)&=-\frac{1}{4}\,,& D_2^2(1,1,0,2)&=-\frac{1}{4}\,, &D_2^2(2,0,1,1)&=-\frac{1}{4}\,, &D_2^2(0,2,1,1)&=-\frac{1}{4}\,,\nonumber \\ 
 D_2^2(2,2,0,0)&=\frac{3}{8}\,, &D_2^2(0,0,2,2)&=\frac{3}{8}\,, &D_2^2(1,1,1,1)&=-\frac{1}{2}\,. &
\label{eq:2nd_two_part_coincide_coeff}
\end{align}
Plugging the coefficients into \eqref{eq:2pt_multi_coincide_res} leads to 
\beqa
\lim_{L\rightarrow \infty} \frac{\,{}_{L}\bra 2 | \TT (0)\tilde{\TT}(\ell)|2\ket_L }{\,{}_{L}\bra 0 | \TT (0)\tilde{\TT}(\ell)|0\ket_L }&=&\frac{1}{2!^2}\left\{ \left(\frac{1}{16}\right)^2 4!\left( \left[g_{1}^2(r)\right]^4 +\left[g_{-1}^2(r)\right]^4+\left[g_{2}^2(r)\right]^4+\left[g_{-2}^2(r)\right]^4 \right) \right.  \nonumber \\
&&+\left(\frac{3}{8}\right)^2 2! 2! \left( \left[g_{1}^2(r)\right]^2 \left[g_{-1}^2(r)\right]^2 + \left[g_{2}^2(r)\right]^2 \left[g_{-2}^2(r)\right]^2 \right)  \nonumber \\
&&+\left(\frac{1}{8}\right)^2 2! 2! \left( \left[g_{1}^2(r)\right]^2+ \left[g_{-1}^2(r)\right]^2\right)\left(  \left[g_{2}^2(r)\right]^2+ \left[g_{-2}^2(r)\right]^2 \right)  \nonumber \\
&&+\left(\frac{1}{4}\right)^2 3!  \left( \left[g_{1}^2(r)\right]^3 g_{-1}^2(r) + g_{1}^2(r)\left[g_{-1}^2(r)\right]^3   \right) \nonumber \\
&&+\left(\frac{1}{4}\right)^2 3!  \left( \left[g_{2}^2(r)\right]^3 g_{-2}^2(r) + g_{2}^2(r)\left[g_{-2}^2(r)\right]^3  \right) \nonumber \\
&&+\left(\frac{1}{4}\right)^2 2!   g_{1}^2(r) g_{-1}^2(r) \left(  \left[g_{2}^2(r)\right]^2+ \left[g_{-2}^2(r)\right]^2 \right) \nonumber \\
&&+\left(\frac{1}{4}\right)^2 2!  \left( \left[g_{1}^2(r)\right]^2+ \left[g_{-1}^2(r)\right]^2\right)  g_{2}^2(r) g_{-2}^2(r)  \nonumber \\
&&+ \left.\left(\frac{1}{2}\right)^2  g_{1}^2(r) g_{-1}^2(r)g_{2}^2(r) g_{-2}^2(r)   \right\} \nonumber \\
&=&\frac{3}{8}+ \frac{3}{8}\left[g_1^2(r)\right]^4+ \frac{1}{4}\left[g_1^2(r)\right]^2=r^4+4 r^2 (1-r)^2 +(1-r)^4\,,
\eeqa
where the last line follows from noting once more that $g_2^2(r)=g_{-2}^2(r)=1$ and $g_1^2(r)=g_{-1}^2(r)=1-2r$.
This then gives the expression
\beq
\Delta S_2^2(r)=-\log(r^4+4r^2(1-r)^2+(1-r)^4)\,.
\label{fun2}
\eeq
\subsection{Numerical results: the harmonic chain}
\label{ssnumerics}

The formulae   \eqref{ren}-\eqref{3} are somewhat surprising for their simplicity and their qubit and semiclassical interpretations, especially as that they emerge from an exact, involved QFT computation. It is therefore important to convince ourselves that this is indeed the behaviour of entanglement that emerges when explicitly carrying out the scaling and thermodynamic limit of a discrete quantum mechanical system. In the free boson case the ideal model on which to test our formulae is the harmonic chain. 

\begin{figure}[h!]
\begin{center} 
\includegraphics[width=7.943cm]{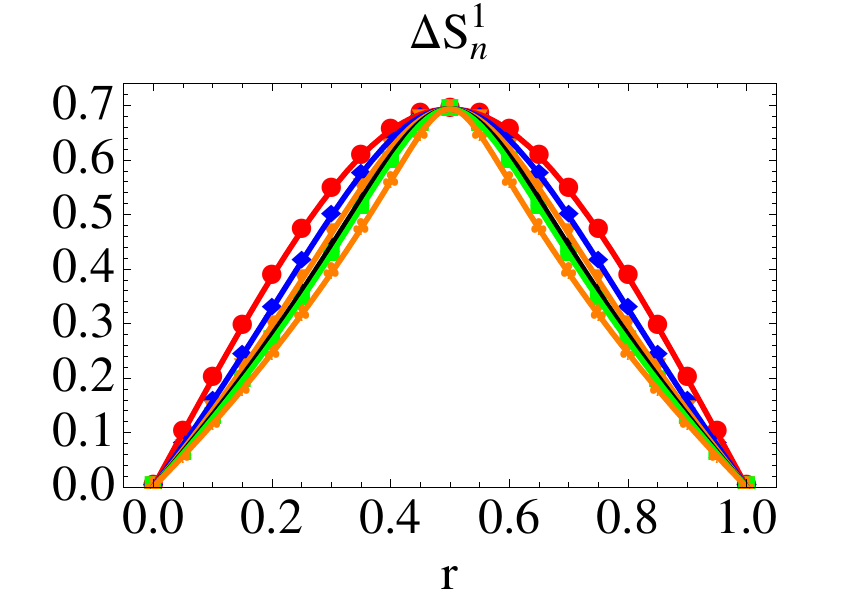} 
\includegraphics[width=7.943cm]{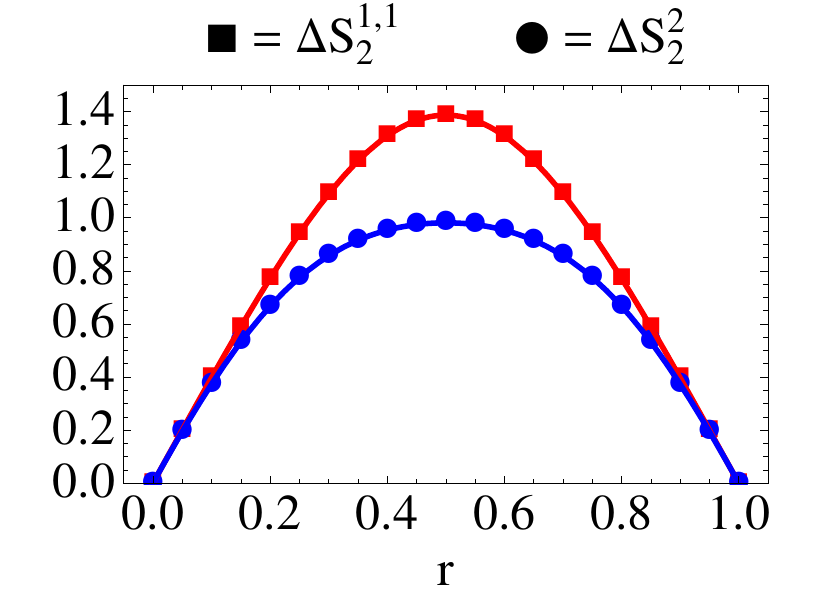} 
 \end{center} 
 \caption{Comparison between analytic results (continuous curves) and numerical values (dots) of the R\'enyi entropies. Left: a single particle, R\'enyi entropies from $n=2$ (red) to $n=11$ (orange), with momentum $P=100 \pi$. Right: two particle states, $n=2$, with distinct momenta given by $P_1\approx 30,\;P_2 \approx 45$ (squares, red curve)  and with equal momenta $P_1=P_2\approx 50$  (dots, blue curve). Additional choices of the momenta are explored in Tables 1 and 2.}
\label{numerics}
 \end{figure}

\begin{figure}[h!]
\begin{center} 
\includegraphics[width=7.943cm]{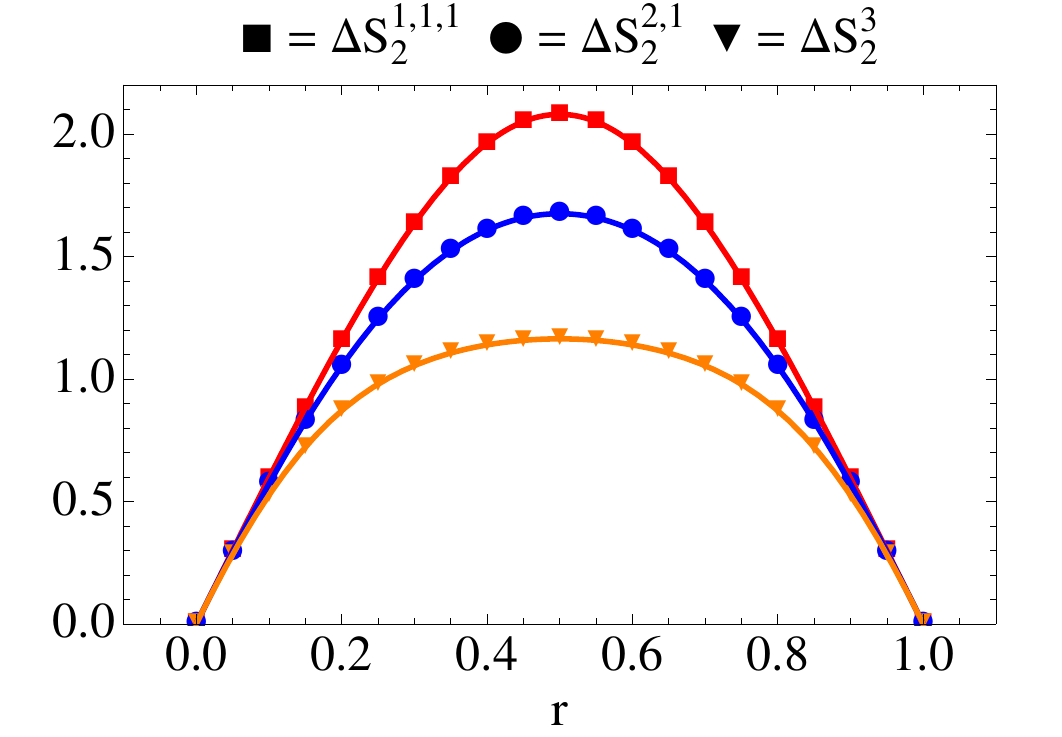} 
\includegraphics[width=7.943cm]{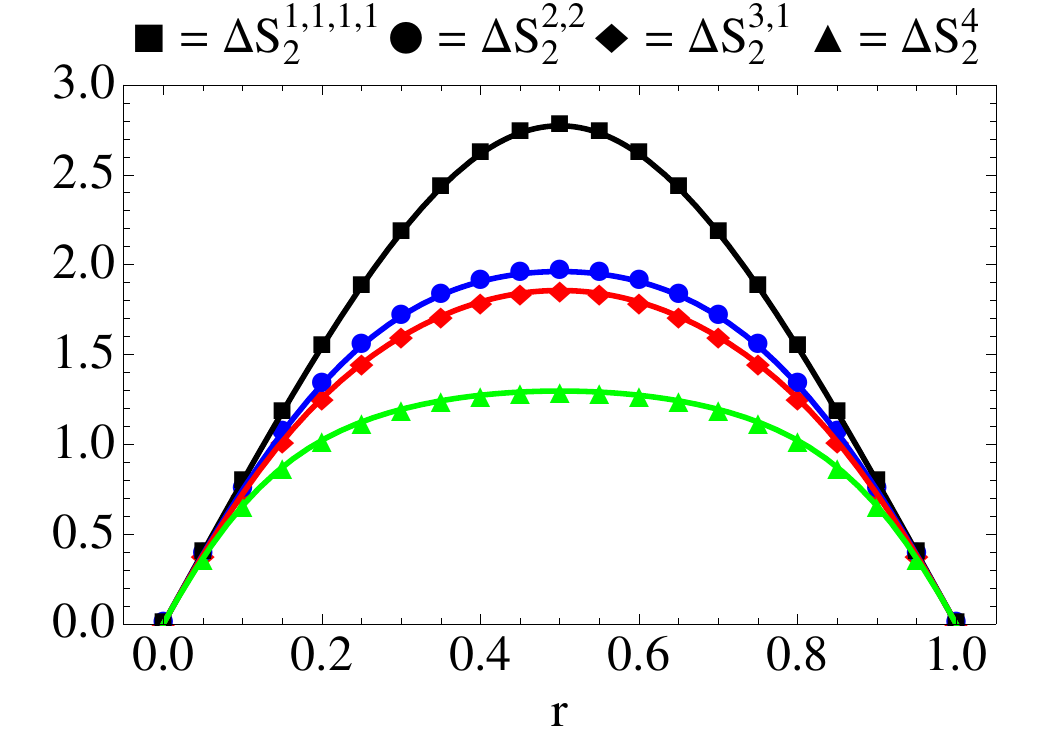} 
 \end{center} 
 \caption{Left: three particle states, $n=2$, with momenta $P_1\approx 10,\, P_2\approx 20,\, P_3\approx 30$ (squares, red curve), with momenta $P_1=P_2\approx 30,\,P_3\approx 50$ (circles, blue curve) and with momenta $P_1=P_2=P_3\approx 50$ (triangles, light brown curve). Right: four particle states, with momenta $P_1\approx 10, P_2 \approx 20, P_3\approx 30, P_4 \approx 40$ (squares, black curve), with momenta $P_1=P_2\approx 30,\,P_3=P_4\approx 50$ (circles, blue curve), with momenta $P_1=P_2=P_3\approx 30,\,P_4=50$ (diamonds, red curve), and with momenta $P_1=P_2=P_3=P_4\approx 50$ (triangles, green curve). In all cases $m=1,\,L=10$ and $\Delta x = 0.01$. Agreement with analytic expressions is excellent in all cases. This is expected as the momenta chosen are well within the QFT regime and  comparable to the mass. For instance with $P=50$ we have $\sin(P\Delta x/2) \approx P\Delta x/2=0.25$ to within 1\%.}
\label{numerics2}
 \end{figure}

The numerical method that we have employed is a wave functional method and we present the details in Appendix A. This is a method based on the exact inversion of a matrix, and gives machine-precision results for the EE of excited states in the harmonic chain in finite volume. Some of the results are presented in this section, see Figs.~\ref{numerics}, \ref{numerics2}.  In all cases, there is excellent agreement between the numerical computation in the limit of large volume and region length $L, \ell$ and small lattice spacing $\Delta x$ (the large-volume scaling limit $L,\ell \gg m^{-1}\gg \Delta x$) and the analytical large volume results  (\ref{ren})-(\ref{3}).

As was explained in \cite{excitations}, the results are in fact expected to be correct in a regime of parameters that goes beyond the universal scaling regime of QFT. The condition, expressed in full generality in \cite{excitations}, is that the minimum of the maximal De Broglie wavelength $2\pi/P_i$ of all particles, and the correlation length $\xi=1/m$, must be much smaller than the minimum of $\ell$ and $L-\ell$. This include large momenta regions, beyond the low-energy QFT regime, and holds independently of the value of $\Delta x$. Below we present some large-momenta results that confirm this.

The results presented in Fig.~\ref{numerics} are as follows.  On the left panel a series of the R\'enyi entropies (\ref{ren}) is presented in the case of a single particle, $k=1$. In the cases $n=2, 3, 4, 5, 6, 11$, both the analytic (continuous curves) and numerical (dots, squares, triangles etc.) results are presented. All curves have a single maximum at $r=\frac{1}{2}$. The numerical results are in perfect agreement with the analytic results, with relative errors less than $10^{-7}$. Numerical results are obtained for $mL=5$ and with the largest momentum allowed by the chosen lattice spacing ($\Delta x= 0.01$), which is in the middle of the Brillouin zone.

The right panel in Fig.~\ref{numerics} shows the 2nd R\'enyi entropy for a two-particle excited state. The outer-most curve is twice the function \eqref{ren}, that is 
\beq
\Delta S_2^{1,1}(r) =-2\log(r^2+(1-r)^2)\,.
\label{exam1}
\eeq
This is twice the second R\'enyi entropy of a single excitation. The squares exactly fitting this curve are the numerical values for volume $L=10$, $m=1$ and a particular choice of (relatively large) distinct momenta. It is interesting to investigate how the chosen values of the momenta affect the accuracy of the fit. Table~1 shows an additional example for distinct (small) momenta $P_1\approx 0.6$ and $P_2 \approx 2$. 

The inner-most curve (with the lowest maximum) is the function \eqref{11} with $k=2$, that is
 \beq
\Delta S_2^{2} =  -\log(r^4 + 4r^2(1-r)^2 + (1-r)^4)\,.
\label{exam2}
 \eeq
This describes the entanglement of a two-particle excited state with particles of the same momentum. Numerical results are presented with $L = 10$, $m=1$ and $P_1=P_2=50$. Table~2 shows additional values for the same quantity and momenta $P_1=P_2=2$ and $P_1=P_2=10$. High precision is obtained even for relatively small momenta.

\begin{table}[h!]
\begin{center}
\begin{tabular}{|l||c|c|c|c|c|c|c|c|c|c|c|}
\hline
$r$ &0& 0.05&0.1&0.15&0.2&0.25&0.3&0.35&0.4&0.45&0.5\\
\hline
\hline
$\Delta S_2^{1,1}(r)$ &0&0.20&0.40&0.59&0.77&0.94&1.09&1.21&1.31&1.37&1.39\\
\hline
$P_1\approx 0.6, P_2\approx 2 $&0&0.21&0.37&0.53&0.70&0.87&1.03&1.18&1.29&1.35&1.37\\
\hline
\end{tabular}
\caption{The difference of 2nd R\'enyi entropies of a two-particle excited state with distinct momenta. The second row shows the exact values of the function (\ref{exam1}). The third row shows the numerical values for the given momenta. The other parameters are $m=1$, $L=10$ and $\Delta x=0.01$. We see that agreement is not as good as for the data in Fig.~\ref{numerics} (top right), especially for small $\ell$. This is due to momenta being too small. More precisely $\min(2\pi/P_1,2\pi/P_2, \xi)=1$ which is larger than some of the values of $\ell$ considered, a regime in which we do not expect our formulae to hold. However, even for such small momenta the disagreement with (\ref{exam1}) is at worse around 10\%.}
\end{center}
\end{table}
\begin{table}[h!]
\begin{center}
\begin{tabular}{|l||c|c|c|c|c|c|c|c|c|c|c|}
\hline
$r$ &0& 0.05&0.1&0.15&0.2&0.25&0.3&0.35&0.4&0.45&0.5\\
\hline \hline
$\Delta S_2^{2}(r)$ &0&0.19&0.37&0.53&0.67&0.77&0.86&0.91&0.95&0.97&0.98\\
\hline
$P_1=P_2\approx 2 $&0&0.18&0.35&0.51&0.66&0.78&0.85&0.91&0.95&0.97&0.98\\
\hline
$P_1=P_2\approx 10 $&0&0.20&0.37&0.53&0.67&0.77&0.86&0.91&0.95&0.97&0.98\\
\hline
\end{tabular}
\caption{The difference of 2nd R\'enyi entropies of two-particle excited states with equal momenta. The second row shows the exact values of the function (\ref{exam2}). The third and fourth rows show numerical values for the given momenta. The other parameters are $m=1$, $L=10$ and $\Delta x=0.01$. For $P_1=P_2=2$ agreement is poorer, especially for small $\ell$ due to the momenta being too small. More precisely $\min(2\pi/P_1, \xi)=1$ which is larger than some of the values of $\ell$ considered, a regime where we do not expect our formulae to hold. However the disagreement with (\ref{exam2}), even for such small momenta is relatively small. For $P_1=P_2=10$ (as for 50, in Fig.~\ref{numerics}) agreement is excellent for all values of $r$.}
\end{center}
\end{table}

The left panel in Fig.~\ref{numerics2} presents the 2nd R\'enyi entropy of three kinds of three-particle excited states. The outer-most curve is three times the function \eqref{ren} with $n=2$,
 \beq
 \Delta S_2^{1,1,1}(r)=-3\log(r^2+(1-r)^2)\,,
 \eeq 
The middle curve is the function
\beq
  \Delta S_2^1(r)+\Delta S_2^{2}(r)\,,
\eeq
which describes the entanglement of a three-particle excited state with two particles of the same momentum and one of a different momentum. Finally, the innner-most curve is the function \eqref{11} with $k=3$,
\beq
   \Delta S_2^{3}=  -\log(r^6 + 9r^4(1-r)^2 + 9r^2(1-r)^4 + (1-r)^6)\,,
\eeq
which is the second R\'enyi entropy of a three-particle excited state with equal momenta.

Finally, the right panel of Fig.~\ref{numerics2} is the 2nd R\'enyi entropy of a four-particle excited state. Here four cases are shown: the outer-most curve is the case where all momenta are distinct, corresponding to the function $4\Delta S_2^1(r)$; the curve with the second highest maximum is the case where particles are divided into two distinct-momentum groups of two equal-momentum particles, corresponding to the function $2\Delta S_2^{2}(r)$; the curve with the third highest maximum is the case where three particles have equal momenta and the fourth particle has a different momentum, corresponding to the function $\Delta S_2^1(r)+\Delta S_2^{3}(r)$; finally, the inner-most curve is the second R\'enyi entropy of a four-particle excited state with all rapidities equal. This is given by the function  
\beqa
   \Delta S_2^{4}(r)= -\log(r^8 + 16r^6(1-r)^2 + 36 r^4(1-r)^4 + 16 r^2(1-r)^6 + (1-r)^8)\,.
\eeqa 
In all cases the volume is again $mL=10$ and momenta are chosen high enough.

As discussed earlier we observe that as the momentum increases, the agreement between numerics and analytical functions becomes better. At large volume, it is possible to reach very high precision with momenta that are large enough while still within the QFT regime, where the dispersion relation is relativistic. We also studied momenta beyond the QFT regime see Fig.~\ref{numerics} (left), towards the middle of the Brillouin zone (where the energy is maximal), $P\approx \pi/\Delta x=100 \pi$. There, we not only observed near machine precision, but also, the condition of the volume $L$ being much larger than the correlation length $m^{-1}$ is no longer necessary: results keep near machine precision for any values of $L,\ell,\Delta x$ with $L,\ell\gg \Delta x$, even with $m^{-1}\gg L,\ell$ (large correlation lengths). We do not currently have a derivation of this result. Intuitively, this indicates that when the wave function of the excited state presents a large number of oscillations within each subregion, then the entanglement behaves as that of the qubit system  explained in Subsection~\ref{bens_qubit}: the large number of oscillations guarantees that the particle is ``evenly distributed" within the subregions. 

It is interesting to numerically study the finite-volume corrections to our formulae (\ref{ren})-(\ref{3}) and to compare the results to a QFT computation. We expect to investigate this problem in a future work. Some results were reported in the supplementary material of \cite{excitations} which, for the harmonic chain, where compatible with integer power law corrections in $L$.


\section{Excited State Entropies of the Massive Free Fermion}
\label{thefreefermion}
Technically speaking the computations presented in the previous few sections follow through with few but important changes for the free fermion theory. 
Interestingly however, the results (\ref{ren})-(\ref{single}) hold unchanged for free fermions. For free fermions states involving two identical creation operators have zero norm and therefore the more involved cases (\ref{11})-(\ref{3}) do not arise in this case. Instead, for a state $|1,1,\ldots\ket_L$ of $k$ particles of distinct rapidities the results (\ref{ren})-(\ref{single}) hold as well upon multiplication by $k$ (as for free bosons). As we will see later, in some respects, the free fermion theory is easier to treat by the techniques outlined in this paper simply because states have a simpler structure. In this section we review those technical features that are different for free fermions and present a detailed computation of the case of a one-particle excitation. 

 
\subsection{Doubling Trick and Replica Free Fermion Model}
In this section we develop similar ideas as in Section~\ref{doublingtrick}. 
Consider two copies of a real (Majorana) fermion labeled by $a$ and $b$. This gives us our ``doubled theory" which we can now regard as a single complex (Dirac) fermion. The suitably normalized spinor components of this complex fermion are
\beq 
\Psi_{R}=\frac{1}{\sqrt{2}}(\psi_{a}+ i \psi_{b}) \quad \mathrm{and}  \quad \Psi_{L}=\frac{1}{\sqrt{2}}(\bar{\psi}_{a}- i\bar{\psi}_{b})\,,
\eeq 
and, if $\psi_{a,b}, \bar{\psi}_{a,b}$ are real, then
\beq 
\Psi^\dagger_{R}=\frac{1}{\sqrt{2}}(\psi_{a}- i \psi_{b}) \quad \mathrm{and}  \quad \Psi^\dagger_{L}=\frac{1}{\sqrt{2}}(\bar{\psi}_{a}+ i\bar{\psi}_{b})\,, 
\eeq 
so $\psi_a=\frac{1}{\sqrt{2}}(\Psi_{R}+\Psi_{R}^\dagger)$. At the level of creation (annihilation) operators there exists a similar relation:
\beq
(a^{(a)})^\dagger(\theta)=\frac{1}{\sqrt{2}}((a^{+})^\dagger(\theta)+  (a^{-})^\dagger(\theta))\,,
\label{copya_fermion}
\eeq
and, considering now $n$-copies of such a real fermion in the replica theory, labelled by an index $k$ we similarly have
\beq
(a_j^{(a)})^\dagger(\theta)=\frac{1}{\sqrt{2}}((a_j^{+})^\dagger(\theta)+  (a_j^{-})^\dagger(\theta)), \quad \mathrm{for} \quad j=1, \ldots ,n\,.
\label{copyaj}
\eeq
As noted in \cite{entropy,CH} where the ground state entanglement of free fermions was studied by employing similar ideas, it is possible to diagonalize the branch point twist field as well but it is important to make a distinction between $n$ even and $n$ odd. More precisely, the relation (\ref{matrixw}) generalizes to
\beq
\omega\left(
\begin{array}{c}
\Psi_{R,1}\\
\Psi_{R,2}\\
\vdots\\
\Psi_{R,n-1}\\
\Psi_{R,n}
\end{array}\right)=\left(
\begin{array}{c}
\Psi_{R,2}\\
\Psi_{R,3}\\
\vdots\\
\Psi_{R,n}\\
\Psi_{R,1}
\end{array}\right)\,,\quad\mbox{that is\,,}\quad
\omega=\left(\begin{array}{ccccc}
0&1&0&\cdots &0\\
0&0&1&\cdots&0\\
\vdots&\vdots&\vdots&\ddots& \vdots\\
0&0&0&\cdots&1\\
(-1)^{n+1}&0&0&\cdots&0
\end{array} 
\right)\,.
\eeq
and similarly for the fields $\Psi_{L,j}$. Note that, unlike for the free boson case, the matrix above is different depending on whether $n$ is even or odd, a feature that has been discussed in \cite{entropy, CH}.
The eigenvalues of this matrix are $\lambda_p= e^{\frac{2\pi i p}{n}}$ for $p=-\frac{n-1}{2}, \cdots, \frac{n-1}{2}$, that is the $n$th roots of unity for $n$ odd the $n$th roots of $-1$ for $n$ even. The cyclic permutation action is diagonalized by the fields
\beq
\tilde{\Psi}_{R,p}=\frac{1}{\sqrt{n}} \sum_{j=1}^{n}  e^{-\frac{2\pi \ri j p}{n}}  \Psi_{R,j}, \quad \mathrm{with} \quad p=-\frac{n-1}{2}, \cdots, \frac{n-1}{2}\,,
\label{relff}
\eeq
and the creation operators satisfy the relations
 \beq 
(\tilde{a}_p^{\pm})^\dagger(\theta)=\frac{1}{\sqrt{n}}\sum_{j=1}^{n}e^{\pm\frac{2\pi i j p}{n}} (a_j^{\pm})^\dagger(\theta), \quad \mathrm{with} \quad p=-\frac{n-1}{2}, \cdots, \frac{n-1}{2}\,,
\label{dia1}
\eeq 
and $\{a_{j_1}(\theta),a^\dagger_{j_2}(\beta)\}=\delta_{j_1 j_2}\delta(\theta-\beta)$, $\{a_{j_1}(\theta),a_{j_2}(\beta)\}=0$ for all $j_1,j_2=1, \ldots, n$. The relation can also be inverted to
\beq
({a}_j^{\pm})^\dagger(\theta)=\frac{1}{\sqrt{n}}\sum_{p=-\frac{n-1}{2}}^{\frac{n-1}{2}}e^{\pm\frac{2\pi i p j}{n}} (\tilde{a}_p^{\pm})^\dagger(\theta), \quad \mathrm{with} \quad j=1, \ldots, n\,,
\label{dia2}
\eeq
and 
$\{\tilde{a}_{p_1}(\theta),\tilde{a}^\dagger_{p_2}(\beta)\}=\delta_{p_1 p_2}\delta(\theta-\beta), \{\tilde{a}_{p_1}(\theta),\tilde{a}_{p_2}(\beta)\}=0$ for all $p_1,p_2=-\frac{n-1}{2},\cdots, \frac{n-1}{2}$.
For free fermions, the $U(1)$ fields associated to these generators have been also studied (see e.g. \cite{BL}) and it is known that they have scaling dimensions
\beq
\Delta_p=\frac{p^2}{2n^2}\,,
\label{dk}
 \eeq
 so that
\beq
\Delta_\TT=\sum_{p=\frac{1-n}{2}}^{\frac{n-1}{2}}\Delta_p= \frac{1}{24}\left(n-\frac{1}{n}\right)\,,
\eeq
note that for the massless Dirac fermion $c=1$. The form factors of these $U(1)$ fields are also discussed in \cite{BL} and they are very similar to those found for free bosons.
The two particle form factors have the same structure:
\beq
F_{2}^{p|+-}(\theta)=\frac{A e^{a \theta}}{\cosh\frac{\theta}{2}}\,, 
\eeq 
and satisfy
\beqa 
&&F^{p|+-}(\theta_1-\theta_2):={}_p\bra 0| \TT_p(0)| \textfrak{a}^\dagger_{p}(\theta_1) \textfrak{b}^\dagger_{p}(\theta_2)|0\ket_p=-F^{p|-+}(\theta_2-\theta_1)\,,\nonumber \\
&&F^{p|++}(\theta_1-\theta_2):={}_p\bra 0| \TT_p(0)| \textfrak{a}^\dagger_{p}(\theta_1) \textfrak{a}^\dagger_{p}(\theta_2)|0\ket_p=0\,, \nonumber\\
&&F^{p|--}(\theta_1-\theta_2):={}_p\bra 0| \TT_p(0)| \textfrak{b}^\dagger_{p}(\theta_1) \textfrak{b}^\dagger_{p}(\theta_2)|0\ket_p=0\,. 
\eeqa
The two last form factors are vanishing  for symmetry reasons.
The form factor programme for quasi-local fields \cite{KW,SmirnovBook,YZam} tells us that these form factors are solutions to a set of three equations. First, 
Watson's equations
\beq 
F^{p|\pm \mp}(\theta)=-F^{p|\mp \pm}(-\theta) \quad \mathrm{and} \quad F^{p|\pm \mp}(\theta+2\pi i) =\gamma^{\pm}_p F^{p|\mp \pm}(-\theta)=- \gamma^{\pm}_p F^{p|\pm \mp}(\theta)\,,
\eeq 
where $\gamma^{\pm}_p=e^{\pm\frac{2\pi i p}{n}}$ are the factors of local commutativity associated to the fermions $\pm$. Finally, the kinematic residue equation tells us that 
\beq 
\mathrm{Res_{\theta=0}}F^{p|\pm \mp}(\theta+i \pi) = i (1-\gamma^{\pm}_p) \tau_p\,,
\eeq 
where 
\beq
\tau_p={}_p\bra0|\TT_{p}(0)|0\ket_p\,,
\eeq
 is the vacuum expectation value. 
It is then easy to show that the equations are satisfied if 
\beq 
a=\frac{p}{n}  \quad \mathrm{and} \quad A= i \tau_p \sin \frac{\pi p}{n}\,. 
\eeq 
This gives the solution
\beq 
F^{p|+-}(\theta)=i\tau_p \sin\frac{\pi p}{n} \frac{e^{\frac{p}{n}\theta}}{\cosh \frac{\theta}{2}}\,. \label{minusfer}
\eeq 
 Since the theory is free, higher particle form factors can be obtained by simply employing Wick's theorem. For the Dirac fermion they have the structure
\beqa 
\!\!\!\!\!\!F_{2m}^{p,n}(\theta_1,\ldots,\theta_m; \beta_1,\ldots, \beta_m)&=&
 {}_p\bra 0|\TT_p(0)|\textfrak{a}^\dagger_{p}(\theta_1)\cdots \textfrak{a}^\dagger_{p}(\theta_m) \textfrak{b}^\dagger_{p}(\beta_1)\cdots \textfrak{b}^\dagger_{p}(\beta_m)|0\ket_p  \label{suma_fermion}
\\
&=&
 \tau_p \sum_{\sigma \in S_m} \mathrm{sign}(\sigma) f_p^n(\theta_{\sigma(1)}-\beta_{1})\cdots 
f_p^n(\theta_{\sigma(m)}-\beta_{m})\,, \nonumber
\eeqa 
where once again $f^{n}_p(\theta)$  is the normalized two-particle form factor and $\sigma$ is an element of the permutation group $S_m$ of $m$ symbols and $\mathrm{sign}(\sigma)$ is the sign of the permutation $\sigma$. 

An important property of the form factor (\ref{minus}) is its leading behaviour near the kinematic singularity. Consider the form factor $f_p^n(\theta_1-\beta_1+i\pi)$ and suppose that the rapidites are quantized through  Bethe-Yang equations of the form 
\beq
mL\sinh\beta_1=2\pi I\,, \quad m L \sinh\theta_1=2 \pi \left(J \pm \frac{p}{n}\right)\,, \quad \mathrm{with} \quad I, J \in \mathbb{Z}\,. \label{quanfer}
\eeq
 Then the leading contribution for $\theta_1 \approx \beta_1$ can be expressed as
\beq 
f_p^n(\beta_1-\theta_1+i\pi)\underset{{\theta_1\approx \beta_1} }{=}\frac{m L\sin\frac{\pi p}{n}\cosh\theta_1\,e^{\frac{i\pi p}{n}}}{\pi (J-I\pm \frac{p}{n})}\,.
\label{z12fermion}
\eeq 
Note that for free fermions it is common to distinguish between periodic and anti-periodic boundary conditions for the Bethe wave function. These lead to quantization conditions (\ref{quanfer}) which either require $I,J \in \mathbb{Z}$ or $I, J \in \mathbb{Z}+\frac{1}{2}$. In our particular computation this choice makes no difference to the final result as we will obtain expressions such as (\ref{z12}) which only depend on quantum number differences. In addition, the $U(1)$ twist fields do not change the $\mathbb{Z}_2$ sector (contrary to $\sigma$ field in the Ising model). For this reason  and without loss of generality we consider the quantization condition (\ref{quanfer}) only. 


\subsection{EE of Single-Particle Excitations}
Given the relations (\ref{copyaj}) we can represent a replica one-particle excited state in a free fermion theory as
\beq
|1\ket_L=\frac{1}{2^{\frac{n}{2}}}\prod_{j=1}^n ((a_j^+)^\dagger(\theta)+(a_j^-)^\dagger(\theta))|0\ket_L\,. 
\label{states2}
\eeq
In the basis of the generators $\textfrak{a}_{j}(\theta)= \tilde{a}_j^{+}(\theta)$ and $\textfrak{b}_{j}(\theta)= \tilde{a}_j^{-}(\theta)$ this state becomes
\beqa
|1\ket_L&=&\frac{1}{2^{\frac{n}{2}}}\prod_{j=1}^n \frac{1}{\sqrt{n}}\left(\sum_{p=-\frac{n-1}{2}}^{\frac{n-1}{2}}
\omega^{jp} \textfrak{a}_p^\dagger(\theta)+\sum_{p=-\frac{n-1}{2}}^{\frac{n-1}{2}}
\omega^{-jp} \textfrak{b}_p^\dagger(\theta)\right)|0\ket_L\nonumber\\
&=& \frac{1}{2^{\frac{n}{2}}}\prod_{j=1}^n \frac{1}{\sqrt{n}}\sum_{p=-\frac{n-1}{2}}^{\frac{n-1}{2}}
\omega^{jp} \left(\textfrak{a}_p^\dagger(\theta)+\textfrak{b}_{-p}^\dagger(\theta)\right)|0\ket_L\,,
\label{omega}
\eeqa  
where $\omega=e^{-\frac{2\pi i}{n}}$. For instance, for $n=2$
it is easy to show that the state takes simply the form
\beq
    |1\ket_L =  -\frac{i}{2} \left( \Aa_{-\frac{1}{2}}^\dag (\ta)+ \Bb_{\frac{1}{2}}^\dag (\ta)\right) \left(\Aa_{\frac{1}{2}}^\dag (\ta) +\Bb_{-\frac{1}{2}}^\dag (\ta)\right)  |0\ket_L=: -\frac{i}{2}\mathcal{S}(2)|0\ket_L\,,
\eeq
where we introduced the notation $\mathcal{S}(n)$ to denote the sum over creation operators. For $n=3$ we have instead 
\beq
|1\ket_L= -\frac{i}{2^{\frac{3}{2}}} \left( \Aa_{-1}^\dag (\ta)+ \Bb_{1}^\dag (\ta)\right) \left(\Aa_{0}^\dag (\ta) +\Bb_{0}^\dag (\ta)\right) \left( \Aa_{1}^\dag (\ta)+ \Bb_{-1}^\dag (\ta)\right)|0\ket_L
= -\frac{i}{2^{\frac{3}{2}}} \mathcal{S}(3)|0\ket_L\,.
\eeq
We note that the main difference between $n$ even and $n$ odd is that for $n$ even there is no ``trivial" sector with index 0.

These particular examples illustrate the general structure of the states. For both $n$ even and odd, they can be constructed recursively starting from the two simple examples just discussed. The state for a given $n$ can be obtained from the state for $n-2$ as follows
\beqa
\mathcal{S}(n)|0\ket_L&=& \frac{e^{i\alpha}}{{2^{\frac{n}{2}}}}  \left(\textfrak{a}_{-\frac{n-1}{2}}^\dagger(\theta)+\textfrak{b}_{\frac{n-1}{2}}^\dagger(\theta)\right)\mathcal{S}(n-2) \left(\textfrak{a}_{\frac{n-1}{2}}^\dagger(\theta) +\textfrak{b}_{-\frac{n-1}{2}}^\dagger(\theta)\right)|0\ket_L\,,
\label{Sn}
\eeqa 
where $\alpha$ is a phase which can be determined for every $n$. Its determination is actually a rather non-trivial problem but, as the states (\ref{omega}) have norm one by construction, we know it must be a real number. Its value has no effect on subsequent computations as only the norm of $e^{i\alpha}$ will be involved. 

 
\subsection{Leading Contribution to the R\'enyi Entropy}
The leading contribution to the R\'enyi entropy can be easily evaluated as all correlators emerging from the states above have a very simple factorized structure. For instance, for $n=2$ the leading contribution will come from the matrix elements
\beqa
\label{corrfuncf2}
&& {}_L\bra1| \TT (0) \Tilde{\TT} (\ell) |1\ket_L =  \prod_{p=-\frac{1}{2}}^{\frac{1}{2}} {}_L\bra 1|\TT_p(0)\tilde{\TT}_p(\ell) |1 \ket_L \nonumber\\
&&=\frac{1}{4} \left[ 
{}_{-\frac{1}{2};L}\bra 0|\Aa_{-\frac{1}{2}} (\theta) \TT (0) \Tilde{\TT} (\ell)  \Aa^\dagger_{-\frac{1}{2}}(\theta)|0\ket_{-\frac{1}{2};L} \times  {}_{\frac{1}{2};L}\bra 0|  \Aa_{\frac{1}{2}}(\theta)  \TT (0) \Tilde{\TT} (\ell)   \Aa_\frac{1}{2}(\theta) |0 \ket_{\frac{1}{2};L}\right.\nonumber\\
&&\qquad \left.+ \;
{}_{-\frac{1}{2};L}\bra 0| \Bb_{-\frac{1}{2}} (\theta)  \TT (0) \Tilde{\TT} (\ell)   \Bb^\dagger_{-\frac{1}{2}}(\theta) |0\ket_{-\frac{1}{2};L} \times  {}_{\frac{1}{2};L}\bra 0|   \Bb_{\frac{1}{2}}(\theta)  \TT (0) \Tilde{\TT} (\ell)   \Bb^\dagger_\frac{1}{2}(\theta) |0 \ket_{\frac{1}{2};L} \right.\nonumber\\
&&\qquad \left.+
{}_{-\frac{1}{2};L}\bra 0| \Aa_{-\frac{1}{2}}\Bb_{-\frac{1}{2}} (\theta)  \TT (0) \Tilde{\TT} (\ell)   \Bb^\dagger_{-\frac{1}{2}}(\theta) \Aa^\dagger_{-\frac{1}{2}}(\theta) | 0\ket_{-\frac{1}{2};L} \right.\nonumber\\
&&\qquad \left.+{}_{\frac{1}{2};L}\bra 0|\Bb_{\frac{1}{2}} (\theta) \Aa_{\frac{1}{2}}(\theta) \TT (0) \Tilde{\TT} (\ell)   \Aa^\dagger_{\frac{1}{2}}(\theta)\Bb^\dagger_{\frac{1}{2}}(\theta)|0\ket_{\frac{1}{2};L} \right]\,,
\eeqa
whereas for $n=3$ we have instead 
\beqa
&&{}_L\bra 1|\TT(0)\tilde{\TT}(\ell) | 1 \ket_L= {}_L\bra 1|\prod_{p=-1}^{1}\TT_{p}(0)\tilde{\TT}_{p}(\ell) | 1 \ket_L\nonumber\\
&&= \frac{1}{4}\left[ {}_{-1;L}\bra 0| \textfrak{a}_{-1}(\theta) \TT_{-1}(0)\tilde{\TT}_{-1}(\ell)  \textfrak{a}_{-1}^\dagger(\theta)|0\ket_{-1;L}\times
{}_{1;L}\bra 0| \textfrak{a}_1(\theta) \TT_1(0)\tilde{\TT}_1(\ell)  \textfrak{a}_1^\dagger(\theta)|0\ket_{1;L}\right.\nonumber\\
&&\qquad\left.+ {}_{-1;L}\bra 0| \textfrak{b}_{-1}(\theta) \TT_{-1}(0)\tilde{\TT}_{-1}(\ell)  \textfrak{b}_{-1}^\dagger(\theta)|0\ket_{-1;L}\times
{}_{1;L}\bra 0| \textfrak{b}_1(\theta) \TT_1(0)\tilde{\TT}_1(\ell)  \textfrak{b}_1^\dagger(\theta)|0\ket_{1;L}\right.\nonumber\\
&&\qquad\left. + {}_{-1;L}\bra 0| \textfrak{a}_{-1}(\theta) \textfrak{b}_{-1}(\theta) \TT_{-1}(0)\tilde{\TT}_{-1}(\ell)   \textfrak{b}^\dagger_{-1}(\theta)\textfrak{a}_{-1}^\dagger(\theta)|0\ket_{-1;L}\right. \nonumber\\
&&\qquad \left.+ {}_{1;L}\bra 0| \textfrak{a}_{1}(\theta) \textfrak{b}_{1}(\theta) \TT_{1}(0)\tilde{\TT}_{1}(\ell)   \textfrak{b}^\dagger_{1}(\theta)\textfrak{a}_{1}^\dagger(\theta)|0\ket_{1;L}\right]\,.
\label{2podd}
\eeqa 
By leading contribution we mean here that non-diagonal matrix elements (involving different states on the right and left) have been neglected  as the arguments presented in Appendix B show that these, even when non-vanishing, will give sub-leading contributions in the volume. 

As can be seen from these examples, the building blocks of the correlation function are generally matrix elements of the form 
\beq 
{}_{p;L}\bra 0| \textfrak{a}_{p}(\theta) \TT_{p}(0)\tilde{\TT}_{p}(\ell)  \textfrak{a}_{p}^\dagger(\theta)|0\ket_{p;L}={}_{-p;L}\bra 0| \textfrak{b}_{-p}(\theta) \TT_{-p}(0)\tilde{\TT}_{-p}(\ell)  \textfrak{b}_{-p}^\dagger(\theta)|0\ket_{-p;L}\,.
\label{abequal}
\eeq 
Matrix elements of the form $ {}_{p;L}\bra 0| \textfrak{a}_{p}(\theta) \textfrak{b}_{p}(\theta) \TT_{p}(0)\tilde{\TT}_{p}(\ell)   \textfrak{b}^\dagger_{p}(\theta)\textfrak{a}_{p}^\dagger(\theta)|0\ket_{p;L}$ have leading large $L$ behaviours which are identical to those of 
\beq
{}_{p;L}\bra 0| \textfrak{a}_{p}(\theta)  \TT_{p}(0)\tilde{\TT}_{p}(\ell)  \textfrak{a}_{p}^\dagger(\theta)|0\ket_{p;L} \times {}_{p;L}\bra 0| \textfrak{b}_{p}(\theta) \TT_{p}(0)\tilde{\TT}_{p}(\ell)   \textfrak{b}^\dagger_{p}(\theta)|0\ket_{p;L}\,,
\eeq
so they involve once more matrix elements of the type (\ref{abequal}).

The leading large volume contribution to such correlators can be evaluated along the same lines presented for the free boson theory. For instance, 
let us take one particular example:
\beqa
&&{}_{p;L}\bra 0| \textfrak{a}_{p}(\theta) \TT_{p}(0)\tilde{\TT}_{p}(\ell)  \textfrak{a}_{p}^\dagger(\theta)|0\ket_{p;L}=\nonumber\\
&&\sum_{s=0}^\infty \sum_{\{J_i^\pm\}} \frac{1}{s! (s+1)!}
 {}_{p;L}\bra 0| \textfrak{a}_p(\theta) \TT_p(0)  \textfrak{a}^\dagger_p(\theta_1)\ldots \textfrak{a}^\dagger_p(\theta_{s+1}) 
 \textfrak{b}^\dagger_p(\theta_{s+2})\ldots \textfrak{b}^\dagger_p(\theta_{2s+1})|0\ket_{p;L}\nonumber\\
 &&  \times {}_{p;L}\bra 0| \textfrak{a}_p(\theta_1)\ldots \textfrak{a}_p(\theta_{s+1}) 
 \textfrak{b}_p(\theta_{s+2})\ldots \textfrak{b}_p(\theta_{2s+1}) \tilde{\TT}_p(0)  \textfrak{a}_p^\dagger(\theta)|0\ket_{p;L}\, e^{i\ell \left[\sum_{i=1}^{2s+1} P(\theta_i)-P(\theta)\right]}\,.
\eeqa 
Recall that the sets $\{J_i^{\pm}\}$ are integers corresponding to the quantization of rapidities $\{\theta_i\}$. In finite (large) volume we can write as usual
\beqa
&&{}_{p;L}\bra 0| \textfrak{a}_{p}(\theta) \TT_{p}(0)\tilde{\TT}_{p}(\ell)  \textfrak{a}_{p}^\dagger(\theta)|0\ket_{p;L}=\nonumber\\
&&\sum_{s=0}^\infty \sum_{\{J_i^\pm\}}
 \frac{ |F_{2s+2}^{p,n}(\theta_1,\ldots,\theta_{s+1}; \theta+i\pi, \theta_{s+2},\ldots,\theta_{2s+2};L)|^2}{s!(s+1)! L E(\theta) \prod_{i=1}^{2s+1} LE(\theta_i)}e^{i\ell \left[\sum_{i=1}^{2s+1} P(\theta_i)-P(\theta)\right]}\,.
\eeqa 
From here, once more the leading contribution will come from terms in the form factor squared such that the rapidity $\theta+i\pi$ is ``contracted" with the same rapidity $\theta_{1}, \ldots,\theta_s$ 
in both form factors. Such terms (there are $s+1$ such choices) contribute a two-particle form factor squared times the vacuum two-point function, which once more factors out.
This gives
\beq
 \frac{{}_{p;L}\bra 0| \textfrak{a}_p(\theta)|\TT_p(0)\tilde{\TT}_p(\ell)| \textfrak{a}_p^\dagger(\theta)|0\ket_{p;L}}{{}_{p;L}\bra 0| \TT_p(0)\tilde{\TT}_p(\ell)|0\ket_{p;L}}=g_p^{n}(r)\,,
\eeq
where $g_n^p(r)$ are the functions discussed in Appendix \ref{thefunctionsg}.
Due to the relations (\ref{abequal}), states of the type $|1\ket_L=\mathcal{S}(n)|0\ket_L$ give
\beqa
\lim_{L\rightarrow \infty}\frac{{}_L\bra 1|\TT(0)\tilde{\TT}(\ell) | 1 \ket_L}{{}_L\bra 0|\TT(0)\tilde{\TT}(\ell) | 0 \ket_L}&=&\prod_{p=-\frac{n-1}{2}}^{\frac{n-1}{2}}g_p^{n}(r)=r^n+(1-r)^n\,,
\eeqa
both for $n$ even and odd. 
The fact that this gives the same entanglement entropy as the free boson is mathematically very interesting in the sense 
that in this case it comes from a single product of functions $g_p^{n}(r)$ whereas for the free boson it was the result of 
adding together a constant plus various powers and products of these
same functions. It is also not difficult to see that this same structure is recovered when considering multi-particle states of distinct rapidities.  


\section{Conclusions and Outlook}
\label{conclusion}

In this paper we have studied the $n^{\rm{th}}$ R\'enyi entropy increment 

\begin{minipage}{8cm}
$$
\Delta S_n^\Psi (r):=\lim_{L\rightarrow \infty} \left[S_n^\Psi(rL,L)-S_n^0(rL,L)\right]\,,
$$
\end{minipage}
\begin{minipage}{8cm}
 \begin{center} 
 \includegraphics[width=5cm]{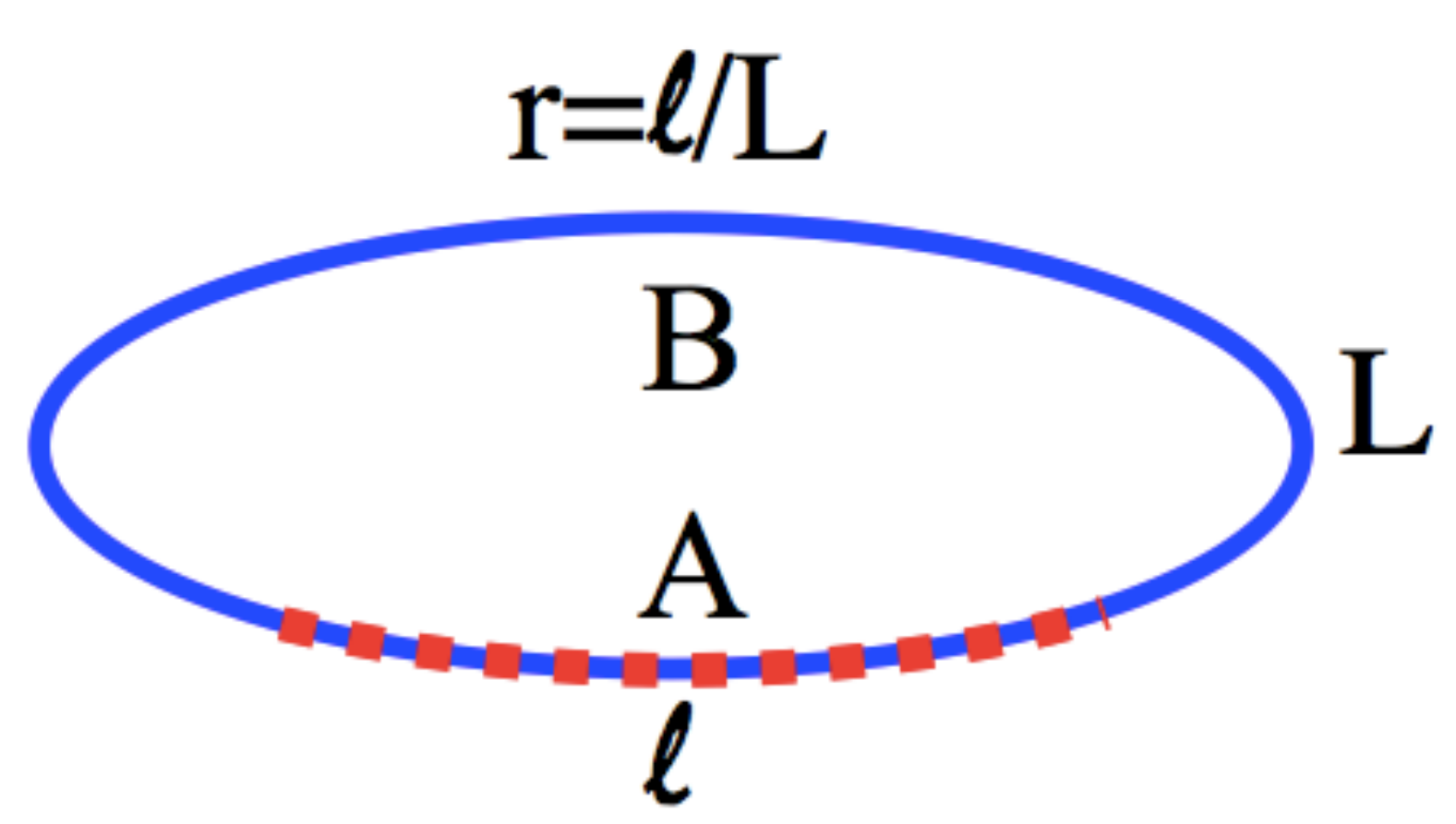} 
 \end{center} 
\end{minipage}

\noindent of a single-interval in one space dimension and its limits $n\rightarrow 1$ (von Neumann entropy) and $n\rightarrow \infty$ (single-copy entropy). 
Our work has focussed on a very particular class of QFTs and excited states $|\Psi\ket$: 
the former are massive free QFTs in 1+1 dimensions and the latter are zero-density states, populated by finite numbers of particles.  We have considered the particular limit $\ell, L \rightarrow \infty$ with $r:=\frac{\ell}{L}$ finite. 

It is well-known that the EE of finite-density excited states in gapped systems satisfies a volume law \cite{EERev2}. In the current work we have shown that for zero-density excited states in infinite volume the EE of one interval saturates to a value which, upon subtracting the ground state contribution, is a simple non-negative function of the ratio $r$. 
More precisely, for $0<r<1$ the excited state provides a net positive additive contribution to the saturation value of the entanglement entropy. For any zero-density states and entropies, this contribution is maximal for $r=1/2$. Moreover, for excited states consisting of $k$ excitations of distinct rapidities, the maximum is $k \log 2$, that is, every excitation ``adds" exactly $\log 2$ to the entanglement entropy of the ground state. The simple form of our results makes them amenable to a qubit interpretation in which each $k$-particle excited state is associated with an entangled qubit state with coefficients that are probabilities of finding $q$ excitations in region $A$ and $k-q$ in region $B$ (see figure above) for $q=0, \ldots, k$. 

Some of our results have previously appeared in the literature (see e.g. \cite{Vincenzo, ln2}) and have been described as semi-classical limits of the EE. Our work, together with the companion paper \cite{excitations}, strongly suggest that the results (\ref{ren})-(\ref{3}) apply much more generally, in fact, to any situations were one can reasonably speak of localized quantum excitations. It is also worth emphasizing that our derivation is the only analytic explicit computation we know of, leading to the formulae (\ref{ren})-(\ref{3}).
 
The domain of applicability of (\ref{ren})-(\ref{3}) may be formally characterized by the condition:
$$
\min(m^{-1},\frac{2\pi}{P})\ll\min(\ell, L-\ell)\,,
$$
where $P$ is the largest momentum of any of the excitations in the state $|\Psi\ket_L$ and $\frac{2\pi}{P}$ can be interpreted as the De Broglie wave length associated to that particular excitation, whereas $\xi=m^{-1}$ is the system's correlation length. Interestingly, this condition implies that we may have a situation where the correlation length of the system is very large and $P$ is also very large and yet still find the same results. Indeed, we provided numerical evidence of this in Fig.~\ref{numerics} and also for higher dimensions in \cite{excitations}.

This work offers ample scope for generalization and extension. It is reasonable to expect that the same results  should also hold for interacting integrable models of QFT. There are three main reasons for this expectation. First, technically, the key mathematical property leading to formulae (\ref{ren})-(\ref{3}) from the form factor calculations of Sections~4 and 5, is the kinematic pole structure of the form factors. However, this form is rather universal and not exclusive to free theories. Second, from \cite{excitations} and \cite{Vincenzo} there is evidence that the same results hold in gapped interacting quantum spin chains whose thermodynamic limit should be described by integrable QFT. Finally, the qubit interpretation is quite universal (at least, as long as there is no particle production) so that we see no reason why results should change in more general theories. However, it would be nice to have a rigorous derivation of this result and we hope to provide this in a future work.

Another interesting problem is the investigation of finite volume corrections to  (\ref{ren})-(\ref{3}). These can be computed both from the form factor expansion and numerically form the wave functional method presented earlier. Some numerical analysis of such corrections was presented in the supplementary material of \cite{excitations} but a more detailed analysis of how the corrections depend on the energy of excitations, the value of $r$ and the replica number $n$ would be very interesting. According to our general arguments in Appendix B we expect the next-to-leading order correction the entropy increment to be of order $1/L$ in the volume so that, for a generic state we should have
\beq
\Delta S_n^\Psi(rL,L)= \Delta S_n^\Psi(r)+ \frac{f(n,r, \{\theta_i\})}{mL}+O((mL)^{-2})\,,
\eeq
where $f(n,r,\{\theta_i\})$ is some function of $n$, the region size, and the rapidities of the excitations, which can be computed from a form factor expansion.
 It would also be interesting to extend the analysis to higher dimensions. For critical systems it has been shown that the EE contains information about the shape of the regions (e.g. the number of vertices) \cite{corners1,corners2,corners3,corners4,corners5,corners6,corners7} and we would like to investigate whether or not such information can also be red off from the finite volume corrections. At present we cannot compute these exactly in higher dimensional gapped QFT, but for free theories, we can use the wave functional method to investigate the problem numerically as in \cite{excitations}. 
 
 To conclude, our results provide further evidence that measures of entanglement encode universal information about quantum models, be it their universality class \cite{HolzheyLW94, Calabrese:2004eu, BCDLR}, operator content \cite{disco1,negativity1,german1,german2}, particle spectrum \cite{entropy, next, ourneg} or, as in this case, the number and nature of their excitations above the ground state. These results come at an exciting time in the understanding of entanglement measures as experimental results for particular R\'enyi entropies have recently become available \cite{Islam, Kaufman794}. It would be extremely interesting to connect our results to experiments and to understand their implications in the wider quantum information context. 
 

\paragraph{Acknowledgments:} The authors are grateful to EPSRC for providing funding through the standard proposal ``Entanglement Measures, Twist Fields, and Partition Functions in Quantum Field Theory" under reference numbers EP/P006108/1 and EP/P006132/1. Benjamin Doyon thanks the \'Ecole Normale Sup\'erieure de Paris for an invited professorship February 19th to March 20th 2018, where part of this work was carried out. Olalla Castro-Alvaredo and Benjamin Doyon acknowledge hospitality and funding from the Erwin Schr\"odinger Institute in Vienna (Austria) where part of this work was carried out and presented during the programme ``Quantum Paths" from April 9th to June 8th 2018. Likewise, all authors are indebted to  the Galileo Galilei Institute for financial support and hospitality during the ``Entanglement in Quantum Systems" workshop from May 21st to July 13th 2018, where the results were also presented. Cecilia De Fazio acknowledges financial support from the University of Bologna and from City, University of London. We are grateful to Vincenzo Alba for bringing reference \cite{Vincenzo} to our attention. Istv\'an M. Sz\'ecs\'enyi thanks Zolt\'an Bajnok for useful discussions. Cecilia De Fazio is grateful to Francesco Ravanini for his support and supervision during her MSc project. 
\appendix

\newpage

\section{Wave Functional Method}

In this appendix, we describe how to evaluate numerically traces of the $n^{\rm th}$ powers of reduced density matrices for few-particle excited states in the quantum free boson model. We use the  wave functional method, which is based on completely different principles than methods using form factors explained in the main text, thus offering an independent verification of our results. After discretizing the model to a finite chain of size $N$, the method reduces the problem to the inversion of a single $nN$ by $nN$ matrix, which can be performed numerically.

Consider the real free boson, with hamiltonian
\beq
	H = \frc12\int_0^L \dd x\lt( (\p_x\Phi(x))^2 + \Pi^2 + m^2\Phi^2\rt)\,,
\eeq
where $\Phi(x)$ and $\Pi(x)$ are hermitian canonically conjugate fields, $[\Phi(x),\Pi(x')] = i\delta(x-x')$. The wave functional of the ground state can be obtained by methods similar to those used for the ordinary harmonic oscillator in quantum mechanics. The annihilation and creation operators are $A_p$ and $A_p^\dag$ for $p\in (2\pi/L) \Z$ with
\beq
	A_p = \frc1{\sqrt{2LE_p}} \int_0^L \dd x\,e^{-\ri px}\lt(E_p \Phi(x) + \ri \Pi(x)\rt),\qquad E_p = \sqrt{p^2+m^2}\,,
\eeq
satisfying $[A_p, A_{p'}^\dag] = \delta_{p,p'}$. We use the representation of wave functionals $\Psi[\varphi] = \bra\varphi|\Psi\ket$, with wave functionals taking as arguments fields $\varphi:[0,L]\to\R$. In this representation,
\beq
	\Phi(x)\Psi[\varphi] = \varphi(x)\Psi[\varphi],\quad \ri \Pi\Psi[\varphi] = \frc{\delta \Psi[\varphi]}{\delta\varphi(x)}\,.
\eeq
The vacuum satisfies $A_p \Psi_{\rm vac} = 0$, which gives
\beq
	\Psi_{\rm vac}[\varphi] = {\cal N}\,\exp\lt[
	-\frc12\int_0^L \dd x\dd y\,K(x-y) \varphi(x)\varphi(y)\rt]\,,
	\quad K(x-y) = \frc1 L\sum_{p} E_p e^{\ri p(x-y)}\,, \label{gauswf}
\eeq
where ${\cal N}$ is a normalization factor.

Excited states are obtained by acting with the creation operator, giving for instance
\beqa
	A_p^\dag \Psi_{\rm vac}[\varphi] &=& \alpha_p[\varphi]
	\Psi_{\rm vac}[\varphi]\,, \n
	A_p^\dag A_q^\dag \Psi_{\rm vac}[\varphi] &=&
	\lt(\alpha_p[\varphi]\alpha_q[\varphi] - \delta_{p+q,0}\rt)
	\Psi_{\rm vac}[\varphi]\,,
\eeqa
where
\beq
	\alpha_p[\varphi] = \sqrt{\frc{2E_p}{L}}\int_0^L \dd x\,e^{\ri px}\varphi(x)\,.
\eeq
In general, for momenta $\{p_j\}$ with all partial sums $\sum_{i}p_{j_i}$  non-vanishing,
\beq
	\Psi_{\{p_j\}}[\varphi] := \prod_{j}A^\dag_{p_j}\Psi_{\rm vac}[\varphi] = \prod_j \alpha_{p_j}[\varphi]\Psi_{\rm vac}[\varphi]\qquad
	(\sum_{i}p_{j_i}\neq0)\,.\label{excwf}
\eeq

We now divide space into $A=[0,\ell)$ and $B=[\ell,L)$, and construct the reduced density matrix $\rho_B = \Tr_{{\cal H}_A} |\Psi\ket\bra\Psi|$. This acts on the space ${\cal H}_B$ of wave functionals taking as arguments fields $\varphi_B:B\to\R$. It has matrix elements
\beq
	\bra \varphi_B|\rho_B|\varphi_B'\ket
	= \int {\cal D}\varphi_A \Psi[\varphi_A,\varphi_B]
	\Psi[\varphi_A,\varphi_B']^*\,.
\eeq
Here we see $[\varphi_A,\varphi_B] = [\varphi]$ as a field on $[0,L]$, and $\Psi[\varphi_A,\varphi_B] = \Psi[\varphi] = \bra \varphi|\Psi\ket$ is the wave functional associated to the state $|\Psi\ket$. The trace of its $n^{\rm th}$ power is
\beqa
	\Tr (\rho_B^n) &=& \int {\cal D}\varphi_1\cdots{\cal D}\varphi_n\,
	\Psi[\varphi_{1A},\varphi_{1B}]\Psi[\varphi_{1A},\varphi_{2B}]^*
	\; \Psi[\varphi_{2A},\varphi_{2B}]\Psi[\varphi_{2A},\varphi_{3B}]^*\n &&
	\hspace{3cm}
	\cdots
	\Psi[\varphi_{nA},\varphi_{nB}]\Psi[\varphi_{nA},\varphi_{1B}]^*\,.
\eeqa
We denote the reduced density matrix of the vacuum state as $\rho_{B|\vac}$, and that of the excited state as $\rho_{B|\{p_j\}}$. We are interested in the ratio
\beq\label{ratiowf}
	\frc{\Tr (\rho_{B|\{p_j\}}^n)}{\Tr (\rho_{B|\vac}^n)}\,.
\eeq
By using the Gaussian form of the vacuum wave functional \eqref{gauswf} and the fact that excited states are obtained by multiplying by polynomial functionals of the fields, as in \eqref{excwf}, we see that \eqref{ratiowf} is the average in a Gaussian measure over the fields $\varphi_j$, of a product of the monomials $\alpha_{p_j}$.

In order to evaluate numerically this average, we discretize space. For this purpose, we choose
\beq
	\Delta x = L/N\,,
\eeq
for some $N\in\N$, restrict space and momentum variables to
\beq
	x = \b x \,\frc LN,\quad p = \b p\,\frc{2\pi}L\,,\qquad
	\b x,\,\b p \in \{0,1,2,\ldots,N-1\}\,,
\eeq
and make the replacement
\beq
	\int_0^L\dd x \mapsto \frc LN\sum_{x=0}^{L-\Delta x}\,.
\eeq
We also change the action to its discrete version, which gives rise to the following change in the equations of motion,
\beq
	\p^2_x\Phi(x) \mapsto \frc1{\Delta x^2}\big(\Phi(x+\Delta x)+\Phi(x-\Delta x)-2\Phi(x)\big)\,.
\eeq
This induces a change in the dispersion relation, the new energy function being
\beq
	E_p = \sqrt{m^2 + \lt(\frc{2N}L\sin\frc{pL}{2N}\rt)^2}\,.
	\label{ep}
\eeq

Putting these ingredients together, some calculations show that the final result can be expressed as follows. Define
\beqa
	K(x) &=& \frc1L\sum_{p=0}^{2\pi(N-1)/L} E_p e^{\ri px}\,, \\
	U_j(p) &=& \frc LN\sum_{x=0}^{L-\Delta x} e^{\ri px}\varphi_j(x)\,,\\
	V_j(p) &=& \frc LN\sum_{x=0}^{\ell-\Delta x} e^{-\ri px}\varphi_j(x)+
	\frc LN\sum_{x=\ell}^{L-\Delta x} e^{-\ri px}\varphi_{j+1}(x)\,.
\eeqa
Note that $K(x)$ with (\ref{ep}) is a real function. The ratio of interest is
\beq\label{aver}
	\frc{\Tr (\rho_{B|\{p_j\}}^n)}{\Tr (\rho_{B|\vac}^n)} =
	\lt(\prod_j \frc{2E_{p_j}}L\rt)^n\;
	\bra\bra \,\prod_{i=1}^n \prod_j U_i(p_j) V_i(p_j)\,\ket\ket\,.
\eeq
The average $\bra\bra\cdots\ket\ket$ is over the Gaussian measure given by the discretized vacuum wave functional,
\beq
	\bra\bra\Or[\varphi_1,\ldots,\varphi_n]\ket\ket
	= \frc{\int {\cal D}\varphi_1\cdots{\cal D}\varphi_n \,\Or[\varphi_1,\ldots,\varphi_n]\, \exp\lt[-\frc12{\cal M}\rt]}{
	\int {\cal D}\varphi_1\cdots{\cal D}\varphi_n \exp\lt[-\frc12{\cal M}\rt]} \, ,
\eeq
with
\beqa
	{\cal M} &=& \sum_{i,j=1}^n \sum_{x,y=0}^{L}
	\varphi_i(x)M_{i,x;j,y}\varphi_j(y) \n &=&
	2\lt(\frc LN\rt)^2\sum_{j=1}^n\Bigg[
	\Big(\sum_{x\in A,\,y\in A} + \sum_{x\in B,\,y\in B}\Big)
	K(x-y)\varphi_j(x)\varphi_j(y) \n && \qquad\qquad\qquad
	+ \sum_{x\in A,\,y\in B}K(x-y)\varphi_j(x)(\varphi_{j+1}(y) + \varphi_j(y)) 
	\Bigg] \,,
\eeqa
where $A$ is from $0$ to $\ell-\Delta x$, and $B$ from $\ell$ to $L-\Delta x$ (inclusively). The Gaussian average in \eqref{aver} is evaluated using Wick's theorem with the Wick contraction
\setlength\unitlength{0.5cm}
\beq
	\stackrel{\begin{picture}(2,0.3)\thicklines
\put(-0.3,0.5){\line(1,0){2}}\put(1.7,0.5){\line(0,-1){0.5}}\put(-0.3,0.5){\line(0,-1){0.5}}\end{picture}}{\;\varphi_i(x)\ \ \varphi_j(y)\;}
 = (M^{-1})_{i,x;j,y}\,.
\eeq
The matrix $M$ is an $nN$ by $nN$ matrix, and the inverse matrix $M^{-1}$ can easily be evaluated numerically. Schematically, the matrix $M$ has the following block structure

\renewcommand{\arraystretch}{1.4}

\begin{tabular}{ r r V{2} c : c V{2} c : c V{2} c V{2} c : c V{2} c : c V{2}}
\multicolumn{2}{c}{} & \multicolumn{2}{c}{$1$} & \multicolumn{2}{c}{$2$} & \multicolumn{1}{c}{}& \multicolumn{2}{c}{$N-1$} & \multicolumn{2}{c}{$N$} 
\\[-1.2em] 
\multicolumn{2}{c}{} & \multicolumn{2}{c}{\downbracefill} & \multicolumn{2}{c}{\downbracefill} & \multicolumn{1}{c}{}& \multicolumn{2}{c}{\downbracefill} & \multicolumn{2}{c}{\downbracefill} 
\\[-0.5em] 
\multicolumn{1}{c}{} &\multicolumn{1}{c}{} & \multicolumn{1}{c}{A} & \multicolumn{1}{c}{B} & \multicolumn{1}{c}{A} & \multicolumn{1}{c}{B} & \multicolumn{1}{c}{} & \multicolumn{1}{c}{A} & \multicolumn{1}{c}{B} & \multicolumn{1}{c}{A} & \multicolumn{1}{c}{B} 
\\[-1.2em] 
\multicolumn{1}{c}{} &\multicolumn{1}{c}{} & \multicolumn{1}{c}{\downbracefill} & \multicolumn{1}{c}{\downbracefill} & \multicolumn{1}{c}{\downbracefill} & \multicolumn{1}{c}{\downbracefill} & \multicolumn{1}{c}{} & \multicolumn{1}{c}{\downbracefill} & \multicolumn{1}{c}{\downbracefill} & \multicolumn{1}{c}{\downbracefill} & \multicolumn{1}{c}{\downbracefill} 
\\ \clineB{3-6}{2} \clineB{8-11}{2} 
\ldelim\{{2}{0.5em}[$1$] & \ldelim\{{1}{1.1em}[A]& $2 K_{AA}$ & $K_{AB}$ & $0$ & $K_{AB}$ & \multirow{2}{*}{ $\cdots$} & $0$ & $0$ & $0$ & $0$
\\ \cdashline{3-6} \cdashline{8-11}
& \ldelim\{{1}{1.1em}[B] & $K_{AB}^T$ & $2 K_{BB}$ & $0$ & $0$ & & $0$ & $0$ & $K_{AB}^T$ & $0$
\\ \clineB{3-6}{2} \clineB{8-11}{2} 
\ldelim\{{2}{0.5em}[$2$] &\ldelim\{{1}{1.1em}[A] & $0$ & $0$ & $2 K_{AA}$ & $K_{AB}$ & \multirow{2}{*}{ $\cdots$} & $0$ & $0$ & $0$ & $0$
\\ \cdashline{3-6} \cdashline{8-11}
& \ldelim\{{1}{1.1em}[B] & $K_{AB}^T$ & $0$ & $K_{AB}^T$ & $2 K_{BB}$ & & $0$ & $0$ & $0$ & $0$
\\ \clineB{3-6}{2} \clineB{8-11}{2} 
\multicolumn{2}{c}{} & \multicolumn{2}{c}{$\vdots$} & \multicolumn{2}{c}{$\vdots$} & \multicolumn{1}{c}{$\ddots$}& \multicolumn{2}{c}{$\vdots$} & \multicolumn{2}{c}{$\vdots$} 
\\ \clineB{3-6}{2} \clineB{8-11}{2} 
\ldelim\{{2}{2.6em}[$N-1$] &\ldelim\{{1}{1.1em}[A] & $0$ & $0$ & $0$ & $0$ &\multirow{2}{*}{ $\cdots$} & $2 K_{AA}$ & $ K_{AB}$ & $0$ & $K_{AB}$ 
\\ \cdashline{3-6} \cdashline{8-11}
& \ldelim\{{1}{1.1em}[B] & $0$ & $0$ & $0$ & $0$ & & $K_{AB}^T$& $2 K_{BB}$ & $0$ & $0$
\\ \clineB{3-6}{2} \clineB{8-11}{2} 
\ldelim\{{2}{0.9em}[$N$] &\ldelim\{{1}{1.1em}[A] & $0$ & $K_{AB}$ & $0$ & $0$ &\multirow{2}{*}{ $\cdots$} & $0$ & $0$ & $2 K_{AA}$ & $K_{AB}$
\\ \cdashline{3-6} \cdashline{8-11}
& \ldelim\{{1}{1.1em}[B] & $0$ & $0$ & $0$ & $0$ & &$K_{AB}^T$ & $0$ & $K_{AB}^T$ & $2 K_{BB}$
\\ \clineB{3-6}{2} \clineB{8-11}{2} \\[-10 pt]
\end{tabular}
\renewcommand{\arraystretch}{1}\\
where the matrices $K_{Q_1Q_2}$ have entries $(K_{Q_1Q_2})_{ij}:=\lt(L/N\rt)^2 K(x_i-x_j)$ with $x_i\in Q_1$ and $x_j \in Q_2$.

\section{Selection Rules for Leading Terms in the Form Factor Expansion}
\label{bens_rules}


In this appendix, we identify the terms in the form factor expansion that contribute in the limit of large system size $L$. We show that these terms contribute to order $L^0$ (that is, are finite and nonzero), and that all other terms contribute to orders $L^{-1}$ or less (that is, vanish as $L\rightarrow \infty$). The leading terms are analyzed in the main text, and give rise to the main results of this paper.

For simplicity, we will consider the case where the excited state depends on a single rapidity value: either it is a single particle state, or a  many-particle state, where all particles have the same rapidity $\theta$ (this is of course only possible in the free boson case). The general case, involving many distinct rapidities, can be understood along similar lines.

Consider a generic term in the form factor expansion (\ref{eq:2pt_sector_ff}). A generic term is characterized by a number $N$ of particles in the (bra) state on the left, a number $\t N$ of particles in the (ket) state on the right, the set $B=\{1,\ldots,M\}$ of rapidity labels in the intermediate state, and the subsets $A\subset B$ and $\t A\subset B$ of labels of the rapidities that are Wick contracted with those in the bra and ket states on the left and right, respectively.  A term is understood as a sum over the intermediate rapidities of the appropriate Wick contractions of products of finite-volume form factors,
\beq\label{eq:generictermproof}
	\sum_{\theta_B=\{\theta_1,\ldots,\theta_M\}} {}_L\bra 
	\stackrel{\begin{picture}(2,0.3)\thicklines
\put(-1,0.5){\line(1,0){1.8}}\put(0.8,0.5){\line(0,-1){0.5}}\put(-1,0.5){\line(0,-1){0.5}}\end{picture}}{
	N|\Or|\theta_A,\theta_{B\setminus A}}
	\ket\bra \theta_{B\setminus \t A},
	\stackrel{\begin{picture}(2,0.3)\thicklines
\put(-0.5,0.5){\line(1,0){2.2}}\put(1.7,0.5){\line(0,-1){0.5}}\put(-0.5,0.5){\line(0,-1){0.5}}\end{picture}}{
	\theta_{\t A}|\Or^\dag|\t N\ket_L}\,.
\eeq
In the calculation presented in Section~\ref{contours}, particles are additionally characterized by their sector as well as their $U(1)$ charge, the operators $\Or$ and $\Or^\dag$ are appropriate $U(1)$-twist fields and one must evaluate products of such terms over all sectors. However, these details are not important in the determination of the leading terms and their large-$L$ behaviour. Additional constraints, such as those from the $U(1)$ charges, can be assessed once the leading terms are identified.

We show that the generic term \eqref{eq:generictermproof} behaves as $O(L^0)$ if and only if $N=\t N$, $M\geq N$, and $A=\t A$ with $|A|=N$; and that otherwise it vanishes in the limit $L\to\infty$.

We first establish the leading power of $L$ corresponding to \eqref{eq:generictermproof}. Due to (\ref{ftofL}) a finite-volume form factor contributes a factor $1/\sqrt{L}$ for each rapidity:
\[
	L^{-\frc{N+\t N}2 - M}\,.
\]
Each particle in the intermediate state that is not contracted with a particle in left or right states (and is, each particle with label in $B\setminus(A\cup \t A)$) contributes a factor of $L$, as for such particles, the sum is evaluated by transforming it into an integral, $\sum_\theta \sim L\int \dd\theta$:
\[
	L^{M - |A\cup \t A|}\,.
\]
Finally, each element in $A$ contributes a factor $L$, and each element in $\t A$ also contributes a factor of $L$. This accounts for two situations. First, a particle may be contracted with one in the left (or right) state but not with any particle in the right (or left) state, $j\in A$ and $j\neq\in \t A$ (or vice versa). In this case, the contraction gives rise to a single pole. The sum over $
\theta_j$ can then be transformed into a converging, principal-value integral $L\,\prin\int\dd\theta_j$, giving a factor of $L$. Second, a particle may be contracted both with one in the state on the left, and one in the state on the right, $j\in A$ and $j\in\t A$. In this case, the leading contribution is obtained by ``zooming in" onto the second-order pole that develops, and summing the resulting second-order pole contribution without transforming the sum into an integral. This sum is convergent, and results in a factor $L^2$ using the fact that momenta are proportional to $1/L$. For instance $\sum_{\theta_j} 1/(\theta_j-\theta)^2 \sim \sum_{I_j\in\Z} L^2/(I_j-I-q)^2$ for some $I\in\Z$ and $q\in(0,1)$. The factor of $L^2$ indicates that we must count a factor of $L$ for the particle both as an element of $A$ and as an element of $\t A$. Thus, we have
\[
	L^{|A|+|\t A|}\,.
\]

In order to find the leading behaviour, we must therefore maximize
\beq\label{eq:tomaximizeproof}
	R = -\frc{N+\t N}2-|A\cup \t A|+|A|+|\t A|=-\frc{N+\t N}2+|A\cap \t A|\,.
\eeq
Thus $R$ will be maximized whenever the cardinality of $A\cap \t A$ is maximized. This occurs when either $A\subseteq \tilde{A}$ or $\tilde{A}\subseteq A$, giving
\beq
	R = -\frc{N+\t N}2+{\rm min}\,(|A|,|\t A|)\,.
\eeq
 Given $N$, $\t N$ and $M$, the number of contractions is constrained by the available particles, giving the bounds
\[
	0\leq |A|\leq {\rm min}\,(N,M)\,,\qquad
	0\leq |\t A|\leq {\rm min}\,(\t N,M)\,,
\]
and all possibilities within these ranges may occur. Let us now fix $N$, $\t N$ and $M$, and choose $A$ and $\t A$ in order to maximize $R$. We must take the maximal values for $|A|$ and $|\t A|$, and we obtain
\beq
	R = -\frc{N+\t N}2+{\rm min}\,(N,\t N,M)\,.
	\label{b4}
\eeq
Fixing $N$ and $\t N$, this is maximized by taking $M\geq {\rm max}\,(N,\t N)$. With this choice, $|A|$ and $|\tilde{A}|$ are maximized by $|A|=N$ and $|\t A| = \t N$, and
\beq
	R = -\frc{|N-\t N|}2\,.
\eeq
Finally, this is maximized by taking $N=\t N$. In this case, we have $|A|=|\t A|$ and thus $A=\t A$, and we find $R=0$.  This shows the claim at the beginning of this Appendix. Moreover, the argument can be easily generalized to states consisting of various particle types.


\section{The Functions $g_p^{n}(r)$}
\label{thefunctionsg}
Throughout this paper we have used the relations
\beq
g_p^{n}(r):=\frac{\sin^2\frac{\pi p}{n}}{\pi^2} \sum_{J\in \mathbb{Z}} \frac{e^{2\pi i r (J+\frac{p}{n})}}{(J+\frac{p}{n})^2}=1-(1-e^{\frac{2\pi i p}{n}})r\,.
\label{definitiong}
\eeq
The fact that the sum above is a simple polynomial in $r$ can be of course checked numerically. It can also be shown analytically, for instance, by showing that the second derivative with respect to $r$ is zero. We compute
\beqa
\partial^2_r g_p^{n}(r)&=&-4 \sin^2 \frac{\pi p}{n} \sum_{J \in \mathbb{Z}} {e^{2\pi i r(J+\frac{p}{n})}}=-4 \sin^2 \frac{\pi p}{n}  e^{\frac{2\pi i r p}{n}}\sum_{J \in \mathbb{Z}} {e^{2\pi i r J}}\nonumber\\
&=& -4 \sin^2 \frac{\pi p}{n}  e^{\frac{2\pi i r p}{n}}\left[-1+ \sum_{J =0}^\infty {e^{2\pi i r J}}+\sum_{J= 0}^\infty e^{-2\pi i r J}  \right]\,.
\eeqa 
The resulting sums are not convergent, but can be regularized by introducing a small parameter $\varepsilon \ll 1$ and computing instead
\beqa
\lim_{\varepsilon \rightarrow 0}\left[\sum_{J =0}^\infty {e^{2\pi i (r+ i\varepsilon) J}}+\sum_{J= 0}^\infty e^{-2\pi i (r-i \varepsilon) J}\right]=\lim_{\varepsilon \rightarrow 0} \left[\frac{1}{1-e^{2\pi i (r+ i\varepsilon) }}+\frac{1}{1-e^{-2\pi i (r- i\varepsilon) }}\right]=1\,.
\eeqa 
Taking the limit $\varepsilon \rightarrow 0$  we find he desired result $\partial_r^2 g_p^n(r)=0$.
Assuming that $g_p^n(r)$ is analytic for at least one value of $r$, we now know that 
\beq
g_p^n(r)=a_p^n+ b_p^n r\,,
\eeq
where $a_p^n, b_p^n$ are independent of the value of $r$. We can determine $a_p^n$ by setting $r=0$ which gives us the simple sum
\beqa
a_p^n=g_p^{n}(0)=\frac{\sin^2\frac{\pi p}{n}}{\pi^2} \sum_{J\in \mathbb{Z}} \frac{1 }{(J+\frac{p}{n})^2}=\frac{\sin^2\frac{\pi p}{n}}{\pi^2} \left[\Psi_1\left(\frac{p}{n}\right)+\Psi_1\left(1-\frac{p}{n}\right) \right]=1\,.
\eeqa
where $\Psi_1(z)=\frac{d^2}{dz^2} \ln \Gamma(z)$ and $\Gamma(z)$ is the Gamma-function. The equality above follows from the known reflection property \cite{gr}:
\beq
\Psi_1(1-z)+\Psi_1(z)=\frac{\pi^2}{\sin^2 \pi z}\,.
\eeq
Finally, we may fix the value of
\beq
b_p^n=\partial_r g_p^n(r)=\frac{2 i \sin^2\frac{\pi p}{n}}{\pi}\sum_{J\in \mathbb{Z}} \frac{e^{2\pi i r (J+\frac{p}{n})}}{J+\frac{p}{n}}\,.
\eeq
For $r=0, 1$ the sum above is singular, but for $r=\frac{1}{2}$ it can be computed to
\beq 
\sum_{J\in \mathbb{Z}} \frac{e^{\pi i  (J+\frac{p}{n})}}{J+\frac{p}{n}}=\frac{e^{\frac{i \pi p}{n}}}{2} \left[\Psi\left(\frac{1}{2}+\frac{p}{2n}\right) -
\Psi\left(\frac{1}{2}-\frac{p}{2n}\right)+\Psi\left(1-\frac{p}{2n} \right) -\Psi\left(\frac{p}{2n}\right) \right]\,,
\eeq 
where $\Psi(z)=\frac{d}{dz}\ln \Gamma(z)$. The $\Psi$-function also has a reflection property \cite{gr}, namely 
\beq
\Psi(1-z)-\Psi(z)=\pi \cot \pi z\,.
\eeq
Using this property, it is a simple matter to show that
\beq
b_p^n=e^{\frac{2\pi i p}{n}}-1\,.
\eeq
\subsection{Properties}
From the definition (\ref{definitiong}) it is also clear that 
\beq
g_p^n(r)=g_{p-jn}^n(r) \quad \mathrm{and} \quad g_p^n(r)=g_{jn-p}^n(r)^* \quad \forall \quad j\in \mathbb{Z}\,.
\eeq
An additional, not entirely obvious property, is that
\beq
\prod_{p=-\frac{n-1}{2}}^{\frac{n-1}{2}}  g_p^n(r)=r^n+(1-r)^n\,.
\label{polynomial}
\eeq 
For $n$ odd we have that
\beq
\prod_{p=-\frac{n-1}{2}}^{\frac{n-1}{2}} g_p^n(r)=\prod_{p=1}^{\frac{n-1}{2}} g_p^n(r) g_{-p}^n(r)=\prod_{p=1}^{\frac{n-1}{2}}\left[ r^2+ 2 r (1-r) \cos\frac{2\pi p}{n}+(1-r)^2\right]\,.
\label{this}
\eeq 
which follows simply from using (\ref{definitiong}). 
Then, the result (\ref{polynomial}) is a consequence of the more general identity \cite{gr}
\beq
 \prod_{p=0}^{n-1} \left[x^2-2 xy  \cos \left(\alpha+\frac{2\pi p}{n}\right)+y^2 \right]=x^{2n}-2x^n y^n \cos n\alpha + y^{2n}\,.
 \label{identity}
\eeq
For $n$ odd, $\alpha=\pi$, $x=r$ and $y=1-r$ (\ref{identity}) gives
\beq
 \prod_{p=1}^{n-1} \left[r^2+2 r (1-r)  \cos \frac{2\pi p}{n}+(1-r)^2 \right]=(r^n+(1-r)^n)^2\,.
\eeq
Note that the $p=0$ term is 1 in this case. We now simply need to observe that
\beq
\prod_{p=1}^{n-1} \left[r^2+2 r (1-r)  \cos \frac{2\pi p}{n}+(1-r)^2 \right]=\prod_{p=1}^{\frac{n-1}{2}} \left[r^2+2 r (1-r)  \cos \frac{2\pi p}{n}+(1-r)^2 \right]^2\,,
\eeq
which then proves (\ref{polynomial}). A similar argument also holds for $n$ even.


\end{document}